\documentclass[oneside,english,american,graybox,envcountchap,sectrefs,notitlepage]{scrbook}
\usepackage[T1]{fontenc}
\usepackage[utf8]{inputenc}
\usepackage{xcolor}
\usepackage{babel}
\usepackage{prettyref}
\usepackage{float}
\usepackage{calc}
\usepackage{mathtools}
\usepackage{amsbsy}
\usepackage{amstext}
\usepackage{graphicx}
\usepackage[numbers]{natbib}
\usepackage[bookmarks=true,bookmarksnumbered=false,bookmarksopen=false,
 breaklinks=false,pdfborder={0 0 1},backref=false,colorlinks=false]
 {hyperref}
\hypersetup{pdftitle={From Gaussian processes to feature learning},
 pdfauthor={Helias, Lindner, Schutzeichel, Ringel},
 pdfsubject={Lecture notes},
 pdfkeywords={statistical field theory, neural networks, feature learning, Gaussian processes},
 colorlinks=true,linkcolor=black,citecolor=black,urlcolor=black,filecolor=black}

\makeatletter

\let\pr@chap=\pr@cha
\providecommand{\tabularnewline}{\\}

\newsavebox{\@brx}
\newcommand{\llangle}[1][]{\savebox{\@brx}{\(\m@th{#1\langle}\)}%
  \mathopen{\copy\@brx\kern-0.5\wd\@brx\usebox{\@brx}}}
\newcommand{\rrangle}[1][]{\savebox{\@brx}{\(\m@th{#1\rangle}\)}%
  \mathclose{\copy\@brx\kern-0.5\wd\@brx\usebox{\@brx}}}
\makeatother

\usepackage{type1cm}         

\usepackage{makeidx}         
\usepackage{graphicx}        
\usepackage{multicol}        
\usepackage[bottom]{footmisc}

\usepackage{newtxtext}       %
\usepackage{newtxmath}       

\makeindex             

\usepackage[force]{feynmp-auto} 

\usepackage{amsmath}
\usepackage{ifthen}
\newboolean{includesolutions}
\newboolean{lecture}

\newrefformat{cap}{\hyperref[#1]{Figure~\ref{#1}}}
\newrefformat{fig}{\hyperref[#1]{Figure~\ref{#1}}}
\newrefformat{tab}{\hyperref[#1]{Table ~\ref{#1}}}
\newrefformat{sec}{\hyperref[#1]{Section~\ref{#1}}}
\newrefformat{sub}{\hyperref[#1]{Section~\ref{#1}}}
\newrefformat{chap}{\hyperref[#1]{Chapter~\ref{#1}}}

\newif\ifContLineOne
\newif\ifContLineTwo
\newif\ifContLineThree

\def\conC#1{\vbox{\ialign{##\crcr
  \ifContLineThree\hrulefill\else\vphantom{\hrulefill}\fi\crcr
  \noalign{\kern3.2pt\nointerlineskip}
  \ifContLineTwo\hrulefill\else\vphantom{\hrulefill}\fi\crcr
  \noalign{\kern3.2pt\nointerlineskip}
  \ifContLineOne\hrulefill\else\vphantom{\hrulefill}\fi\crcr
  \noalign{\nointerlineskip}
  $\hfil\textstyle{\vbox to 14pt{}#1}\hfil$\crcr}}}

\def\DrawLeg#1#2{
  \kern-.2pt              
  \dimen2 =#1             
  \advance\dimen2 by 2pt  
  \dimen3 = 10.6pt        
  \dimen4 =3.6pt          
  \advance\dimen3 by -\dimen2 
  \multiply\dimen4 by #2
  \advance\dimen3 by \dimen4
  \raise\dimen2 \hbox{\vrule height\dimen3 width .4pt} 
  \kern-.2pt}             

\def\begC#1#2{\setbox0 =\hbox{$\textstyle{#2}$}
  \dimen0=.5\wd0 \dimen1=\ht0
  \conC{\hskip\dimen0}
  \count255=#1
  \ifnum\count255 =1 \ContLineOnetrue\else
  \ifnum\count255 =2 \ContLineTwotrue\else
  \ifnum\count255 =3 \ContLineThreetrue\fi\fi\fi
  \DrawLeg{\dimen1}{\count255}
  \conC{\hskip\dimen0}
  \kern-\dimen0\kern-\dimen0 \box0}

\def\endC#1#2{\setbox0 =\hbox{$\textstyle{#2}$}
  \dimen0=.5\wd0 \dimen1=\ht0
  \conC{\hskip\dimen0}
  \count255=#1
  \ifnum\count255 =1 \ContLineOnefalse\else
  \ifnum\count255 =2 \ContLineTwofalse\else
  \ifnum\count255 =3 \ContLineThreefalse\fi\fi\fi
  \DrawLeg{\dimen1}{\count255}
  \conC{\hskip\dimen0}
  \kern-\dimen0\kern-\dimen0 \box0}

\makeatother

\begin{document}
\global\long\def\D{\mathcal{D}}%
\global\long\def\bx{\mathbf{x}}%
\global\long\def\bl{\mathbf{l}}%
\global\long\def\bh{\mathbf{h}}%
\global\long\def\bJ{\mathbf{J}}%
\global\long\def\N{\mathcal{N}}%
\global\long\def\hh{\hat{h}}%
\global\long\def\bhh{\mathbf{\hh}}%
\global\long\def\T{\mathrm{T}}%
\global\long\def\by{\mathrm{\mathbf{y}}}%
\global\long\def\diag{\mathrm{diag}}%
\global\long\def\Ftr#1#2{\mathcal{F}\left[#1\right]\left(#2\right)}%
\global\long\def\iFtr#1#2{\mathcal{F}^{-1}\left[#1\right]\left(#2\right)}%
\global\long\def\D{\mathcal{D}}%
\global\long\def\T{\mathrm{T}}%
\global\long\def\Gammafl{\Gamma_{\mathrm{fl}}}%
\global\long\def\gammafl{\gamma_{\mathrm{fl}}}%
\global\long\def\E#1{\left\langle #1\right\rangle }%

\global\long\def\D{\mathcal{D}}%
\global\long\def\J{\mathbf{J}}%
\global\long\def\one{\mathbf{1}}%
\global\long\def\e{\mathbf{e}}%
\global\long\def\Cpp{\mathcal{K}_{\phi^{\prime}\phi^{\prime}}^{(0)}}%
\global\long\def\CCpp{C_{\phi^{\prime}\phi^{\prime}}^{(0)}}%
\global\long\def\Cppj{\mathcal{K}_{\phi_{j}^{\prime}\phi_{j}^{\prime}}^{(0)}}%
\global\long\def\tx{\tilde{x}}%
\global\long\def\xo{x^{(0)}}%
\global\long\def\xii{x^{(1)}}%
\global\long\def\txi{\tilde{x}^{(1)}}%
\global\long\def\bx{\mathbf{x}}%
\global\long\def\tbx{\tilde{\mathbf{x}}}%
\global\long\def\bl{\mathbf{j}}%
\global\long\def\tbj{\tilde{\mathbf{j}}}%
\global\long\def\bk{\mathbf{k}}%
\global\long\def\tbk{\tilde{\mathbf{k}}}%
\global\long\def\bh{\mathbf{h}}%
\global\long\def\bJ{\mathbf{J}}%
\global\long\def\bN{\mathcal{N}}%
\global\long\def\bH{\mathbf{H}}%
\global\long\def\bK{\mathbf{K}}%
\global\long\def\bxo{\bx^{(0)}}%
\global\long\def\tbxo{\tilde{\bx}^{(0)}}%
\global\long\def\bxi{\bx^{(1)}}%
\global\long\def\tbxi{\tilde{\bx}^{(1)}}%
\global\long\def\tbxi{\tbx^{(1)}}%
\global\long\def\tpsi{\tilde{\psi}}%
\global\long\def\Cxi{C_{x^{(1)}x^{(1)}}}%
\global\long\def\bxxi{\mathbf{\xi}}%
\global\long\def\N{\mathcal{N}}%
\global\long\def\bW{\mathbf{W}}%
\global\long\def\bon{\mathbf{1}}%
\global\long\def\tj{\tilde{j}}%
\global\long\def\tJ{\tilde{J}}%
\global\long\def\Z{\mathcal{Z}}%
\global\long\def\SOM{S_{\mathrm{OM}}}%
\global\long\def\SMSR{S_{\mathrm{MSR}}}%

\global\long\def\cX{\mathcal{X}}%
\global\long\def\cK{\mathcal{K}}%
\global\long\def\cF{\mathcal{F}}%
\global\long\def\tr{\mathrm{tr}}%

\global\long\def\mcC{\mathcal{C}}%
\global\long\def\tX{\tilde{X}}%
\global\long\def\ty{\tilde{y}}%
\global\long\def\tQ{\tilde{Q}}%
\global\long\def\tk{\tilde{k}}%
\global\long\def\bR{\mathbf{R}}%
\global\long\def\bQ{\mathbf{Q}}%
\global\long\def\tbQ{\mathbf{\tilde{Q}}}%

\global\long\def\bX{\mathbf{\mathbf{X}}}%
\global\long\def\tbX{\tilde{\mathbf{\mathbf{X}}}}%
\global\long\def\bW{\mathbf{\mathbf{W}}}%
\global\long\def\bXi{\boldsymbol{\Xi}}%
\global\long\def\bC{\boldsymbol{C}}%

\selectlanguage{english}%

\selectlanguage{american}%
\global\long\def\order{\mathcal{O}}%
\global\long\def\bR{\mathbb{R}}%
\global\long\def\cL{\mathcal{L}}%
\global\long\def\I{\mathbb{I}}%

\global\long\def\bx{\mathbf{x}}%
\global\long\def\bth{\tilde{\mathbf{h}}}%
\global\long\def\order{\mathcal{O}}%
\global\long\def\tC{\tilde{C}}%
\global\long\def\cW{\mathcal{W}}%
\global\long\def\th{\tilde{h}}%
\global\long\def\I{\mathbb{I}}%
\global\long\def\bh{\mathbf{h}}%
\global\long\def\cR{\mathbb{R}}%
\global\long\def\ty{\tilde{y}}%
\global\long\def\bphi{\boldsymbol{\phi}}%
\global\long\def\bY{\mathbf{Y}}%
\global\long\def\bA{\mathbf{A}}%
\global\long\def\cZ{\mathcal{Z}}%
\global\long\def\cD{\mathcal{D}}%
\global\long\def\bI{\mathbf{I}}%
\global\long\def\tr{\mathrm{tr}}%
\global\long\def\hy{\hat{y}}%
\global\long\def\bj{\mathbf{j}}%
\global\long\def\const{\mathrm{const}}%
\global\long\def\tu{\tilde{u}}%
\global\long\def\erf{\mathrm{erf}}%
\global\long\def\tG{\tilde{G}}%

\title{Lecture notes: From Gaussian processes to feature learning in neural
networks}
\author{Moritz Helias$^{1,2}$, Javed Lindner$^{1,3}$, Lars Schutzeichel$^{1,3}$,
Zohar Ringel$^{4}$}

\maketitle
$^{1}$Institute for Advanced Simulation (IAS-6), Jülich Research
Centre, Jülich, Germany

$^{2}$Department of Physics, Faculty 1, RWTH Aachen University, Aachen,
Germany

$^{3}$RWTH Aachen University, Aachen, Germany

$^{4}$The Racah Institute of Physics, The Hebrew University of Jerusalem,
Jerusalem, Israel

\tableofcontents{}

\pagebreak{}

\chapter{Introduction\protect\label{sec:Introduction}}

Over the past decade, machine learning, in particular artificial
neural networks (ANNs), have significantly influenced all fields of
science, ranging from engineering and robotics to genomics, and has
also increasingly become a part of our daily lives. Despite their
widespread use, our understanding of the inner workings of neural
networks remains limited. Unlike traditional engineering, we lack
a solid understanding of first principles to guide design processes:
Whereas for example thermodynamics allows us to reliably simulate
processes within a combustion engine before constructing a costly
prototype, we lack a comparable comprehensive theory of artificial
neural networks to assess the performance of an architecture prior
to a potentially costly training process. This knowledge gap severely
impedes our ability to effectively guide the engineering process of
developing novel neural networks. Consequently, the design of artificial
intelligence systems has relied heavily on a mix of educated guesses,
intuition, and lastly also on trial and error. Facing the massive
costs and energy consumption of training contemporary networks clearly
underscores the urgent need for a solid theoretical understanding
of artificial neural networks.

Opening the black-box of artificial neural networks is an active field
of research today which has its roots in the original idea of the
perceptron \citet{Rosenblatt58} as the predecessor of artificial
neural networks. Since then, different avenues emerged to quantiatively
describe and analyze ANNs, focusing on diverse aspects and hence utilizing
different sets of tools to further understanding. As a matter of fact,
physics has shaped the theory of artificial neural networks. The Nobel
prize in physics 2024 awarded to John Hopfield and Geoffrey Hinton
prominently testifies some of these contributions. The reason why
statistical physics in particular has a successful track record in
contributing to the theory of AI is that complex systems and artificial
neural networks share many similarities, allowing one to borrow methods
from the one to solve problems in the other field.

The current set of lecture notes focuses on one facet of neural network
theory that investigates how neural networks learn features from finite
data and generalize to unseen data points. We treat deep feed-forward
and recurrent networks on the same footing, in a Bayesian framework.
This approach has a long and successful track record \citep{Seung92},
which maps the problem of learning to the study of a partition function,
where the learnable parameters play the role of the degrees of freedom;
an approach originally pioneered by Gardner to investigate the memory
capacity of the perceptron \citep{Gardner88_257}. Due to the large
number of degrees of freedom and their complex mutual interaction,
this partition function corresponds to a complex physical system,
which can be treated by suitable methods from statistical physics
such as field theory to study learning as an emergent collective phenomenon.
For example, phase transitions may occur in the training phase of
the network, where the number of training samples plays the role of
a control parameter and the transition marks the sudden onset of specialization
to the given data \citep{Rubin24_iclr}. Likewise, different forms
of information supplied to sequence processing networks, such as transformers
\citep{Vaswani17_nips}, induce phase transitions that either encode
the information contents or the position of an element in the sequence
\citep{Cui2024Phasetransition}. As usual in statistical physics,
such phase transitions arise from the competition between the energy
and the entropy, from the principle of minimal free energy.

We chose to focus on Bayesian network training to provide a general
framework agnostic of the learning rule. By Bayes' rule, one here
infers how the initially chosen \textbf{prior distribution} of trainable
parameters $\theta$ changes when the network is confronted with training
data. The resulting parameter distribution is known as the \textbf{posterior
distribution}. This approach only requires the specification of how
the network maps an input $x$ to an output $y$ given its parameters
$\theta$, the choice of a prior distribution $p(\theta)$ and the
specification of a loss function that measures departures of the network's
output from the desired value prescribed by the training data.

Some practical applications, however, in parts depend on the actual
implementation of network training. Hence there are branches of research
which in particular focus on network training dynamics and stochastic
optimization to get a deeper insight into possible constructive biases,
caveats and upgrades for stochastic gradient descent (SGD) as well
as studies on the choices of hyperparameters, such as the influence
of learning rates and modern SGD relatives like ADAM. Evaluating
networks in the Bayesian setting effectively translates to considering
networks that have been trained until their weights reach an equilibrium
distribution under a stochastic version of gradient descent known
as Langevin training. However, different aspects, such as on-line
learning, active learning as well as few-shot learning, transfer learning
and curriculum learning do depend on the learning dynamics and would
not be accessible by solely studying stationary states. Studying the
training dynamics in neural networks also allows investigations by
methods from dynamical systems analysis, such as Lyapunov exponents
or robustness measures.

\paragraph{}

Furthermore this set of notes primarily focuses on the \textbf{posterior
}of the network output distribution: Other works tend to focus on
the distribution of network weights, the loss landscapes as well as
the function spaces accessible to certain network architectures and
the internal representations and hidden manifold geometries generated
during network training to get a constructive insight on why certain
networks perform better than others. The focus on the network outputs
also neglects effects such as pruning or sparsification on the level
of the network weight distribution, which we do not study here.

\paragraph{}

Our focus on statistical field theory should not be confused with
the long-standing and rich field of statistical learning theory \citep{Vapnik98_learningtheory},
which, together with classical concepts such as the Vapnik-Chervonenkis
dimension \citep{VapnikChervonenkis1968} and Cramer-Rao bounds learning
bounds \citep{Cramer1946,Rao1947,Seroussi22}, provides rigorous worst
case estimates on learnability in neural networks and machine learning.
The methods from statistical field theory which we present here, on
the other hand, focus rather on average-case results than worst or
best case depictions. However, we do share the main assumption of
statistical learning theory that training and test data come from
the same distribution and hence, by definition, the current lecture
notes do not cover effects such as distributional shifts or transfer
learning.

\paragraph{}

Whereas there exists a whole zoo of different neural network architectures,
from convolutional neural networks (CNNs) for image processing \citep{Lecun1998,Krizhevsky2012},
to residual neural networks (ResNets) \citep{He2015}, graph neural
networks \citep{Scarselli2009} as well as attention-mechanism based
transformer architectures \citep{Vaswani17_nips}, which form the
basis of modern large language models, we focus on vanilla deep networks
(DNNs) and recurrent networks in discrete time (RNNs). Even though
the methods presented in this set of notes also cover the aforementioned
architectures and there already exists theoretical groundwork to study
those architectures \citep{Hron20_4376,Naveh21_NeurIPS,Epping2024}
we choose to focus on the more pedagogical examples of DNNs and RNNs.

\paragraph{}

The thermodynamic limit of large numbers of neurons but with a limited
number of training data points leads to a particularly simple theory
of Gaussian process regression. The most prominent of which are the
Neural Network Gaussian Process (NNGP \citep{Neal96,Williams98,Cho09})
and the Neural Tangent Kernel (NTK \citep{Jacot18_8580}), both of
which describe what is often termed ``lazy learning'' \citep{Chizat19_neurips},
a setting in which the distribution of weights after training does
not differ significantly from the one at initialization. We will recover
the NNGP in these lecture notes as the natural limit of the Bayesian
posterior when the number of training samples is small. We do not
cover the NTK, which would be obtained by studying, instead of the
stochastic gradient dynamics, the dynamics of deterministic gradient
flow. Yet, both approaches allow one to understand basic properties
of trainability and learning phases in ANNs, such as the inductive
bias towards implementing smooth functions \citep{Canatar2021} or
the emergence of neural scaling laws \citep{Bahri_2024}; the latter
are power laws that describe the decline of the loss as a function
of the number of training samples and network size \citep{kaplan2020scaling}.
The NNGP lacks what is known as ``feature learning''; the representations
of the data within the network in particular do not depend on the
training target. The NTK considers weak such effects, as it captures
faithfully only small changes of the weights from their initialization.

Understanding feature learning to its full extent is a field of active
research as understanding feature learning is central to understanding
the superior performance of ANNs and the reduction of sample complexity
compared to NNGP, which is the number of samples required to reach
a desired prediction accuracy. Within the realm of Bayesian networks,
there are currently two favored views of feature learning, one that
determines changes to the NNGP Gaussian process kernel described a
scaling parameter \foreignlanguage{english}{\citep{Li21_031059,Pacelli23_1497}}
and one that considers the adaptation of the kernel in a more flexible
way \citep{Naveh21_NeurIPS,seroussi23_908,Fischer24_10761}. These
notes also explain how these two views are connected, following \citep{Rubin25}.

These notes do not aim at reviewing the field of physics-inspired
theory of AI, but rather wants to introduce the reader into some of
the useful concepts to follow the literature with help of a minimal
set of examples explained in the main text, complemented by exercises.
We would therefore like to mention closely related works in the following
section on related works \prettyref{sec:related_works}.

The main goal of these notes is to provide an introduction into some
of the useful techniques from statistical physics, disordered systems,
and large deviation theory as far as they are needed to bring the
reader into the position to understand and extend the current state
of the literature on feature learning in neural networks. To achieve
this goal, the notes are self-contained and structured as follows:
The initial chapters \prettyref{chap:Probabilities-moments-cumulant}
and \prettyref{chap:Gaussian-distribution-Wick-theorem} introduce
basic notions of moment and cumulant-generating functions for probability
distributions as well as the Gaussian distribution as an important
example and Wick's theorem. These basic concepts are required to understand
the remainder of the notes. Chapter \ref{chap:Linear-regression-as-Bayesian-inference}
introduces the notion of supervised learning and its Bayesian formulation
on the simplest example of linear regression. Chapter \ref{chap:The-law-of-large-number}
introduces another technique, the law of large numbers and some notions
of its more powerful version – large deviation theory \citep{Touchette09},
which will be required to approximate probability distributions in
the limit of large numbers of degrees of freedom. Chapter \ref{chap:Dominant-behavior-of-DNNs}
studies how the NNGP arises in deep networks in the limit of large
width and small numbers of training samples. Chapter \ref{chap:Recurrent-networks}
performs the corresponding analysis for recurrent networks and draws
comparisons between NNGPs for recurrent and deep nets. Chapter \ref{chap:Non-equilibrium-statistical-mech}
derives the Fokker-Planck equation as a technique to study the time-evolution
the probability distribution of network parameters, resulting from
the training dynamics, which allows us to connect the Bayesian posterior
distribution to the stationary distribution of gradient-descent training
with stochastic Langevin dynamics. Finally, Chapter \ref{chap:Feature-learning}
considers feature learning as it arises in the limit where both, the
number of training data points and the width of the network tend to
infinity proportionally. We here cover both aforementioned views,
the scaling approach and the adaptive kernel approach and expose their
tight relation.

\section{Related works\protect\label{sec:related_works}}

Previous work has investigated deep networks within the Gaussian process
limit for infinite width $N\to\infty$ \citet{Schoenholz17_01232,Lee18}.
\citet{Schoenholz17_01232} found optimal backpropagation of signals
and gradients when initializing networks at the critical point, the
transition to chaos \citep{molgedey92_3717}, that we explain here.
The joint limit $N\to\infty$, $P\to\infty$ with $P/N=\alpha$ fixed
as well as and standard scaling of weights $w\propto1/\sqrt{N}$ has
been investigated with tools from statistical mechanics in deep linear
networks \citep{Li21_031059}, where kernels act as if only their
overall scale would change compared to the NNGP limit. We discuss
these results and present an alternative derivation in these notes.
A rigorous non-asymptotic solution for deep linear networks in terms
of Meijer-G functions \citep{Hanin23} has shown that the posterior
of infinitely deep linear networks with data-agnostic priors is the
same as that of shallow networks with evidence-maximizing data-dependent
priors. For a teacher-student setting, \citep{ZavatoneVeth22_064118}
show that in deep linear networks feature learning corrections to
the generalization error result from perturbation corrections only
at quadratic order or higher. For deep kernel machines, \citep{Yang23_39380}
find a trade-off between network prior and data term; their main results
can be obtained from these notes in the special case of deep linear
networks.

Previous theoretical work on non-linear networks of finite width $N<\infty$
has employed three different approximation techniques. First, a perturbative
approach that computes corrections where the non-linear terms constitute
the expansion parameter \citep{Halverson21_035002}. Second, a perturbative
approach based on the Edgeworth expansion that uses the strength of
the non-Gaussian cumulants as an expansion parameter. These corrections
are computed either in the framework of gradient-based training \citep{Dyer20_ICLR,Huang20_4542,Aitken20_06687,Roberts22,Bordelon23_114009}
or Bayesian inference \citep{Yaida20,Antognini19_arxiv,Naveh21_064301,Cohen21_023034,Roberts22}.
Ref. \citep{ZavatoneVeth21_NeurIPS_II} derive a general form of finite-width
corrections, resulting from the linear readout layer and the quadratic
loss function. Third, non-perturbative Bayesian approaches \citet{Naveh21_NeurIPS,seroussi23_908,Pacelli23_1497,Cui23_6468},
that derive self-consistency equations either by saddle-point integration
or by variational methods to obtain the Bayesian posterior. Ref. \citep{Cui23_6468}
exploits the Nishimori conditions that hold for Bayes-optimal inference,
where student and teacher have the same architecture and the student
uses the teacher's weight distribution as a prior; the latter is assumed
Gaussian i.i.d., which allows them to use the Gaussian equivalence
principle \citep{Goldt20_14709} to obtain closed-form solutions.
These notes are most closely related to these non-perturbative Bayesian
approaches. Ref. \citep{vanMeegen_24_16689} studies the limit of
very weak readout weights, so that readout weights concentrate and
different inner representations form, which depend on the employed
activation function.

The current presentation closely follows the following previous works:
For the Gaussian process limit, a unified derivation for deep and
recurrent networks presented in \citet{Segadlo22_accepted}. The scaling
approach to the feature learning theory has first been derived in
\citet{li2021statistical}; the derivation based on large deviation
theory that we follow here has first been presented in \citet{rubin2024a}.
The kernel adaptation theory of feature learning has been pioneered
in \citet{seroussi2023separation_main} and specific form of kernel
adaptation presented here closely follows \citet{Fischer24_10761}.
It has also been adapted by \citet{Lauditi2025} to the $\mu P$ parametrization
\citep{Yang2021_icml} of the readout weights. The unification of
the scaling and the kernel adaptation approach follows \citet{Rubin2025}.
As a complementary resource and presentation, the review \citet{Ringel2025}
presents a perspective on feature learning through the lens of field
theory, including approaches that make assertions on averages over
the data distribution; the current set of notes, instead, throughout
operates on one given train set.

\textbf{}

\chapter{Probabilities, moments, cumulants\protect\label{chap:Probabilities-moments-cumulant}}

This chapter introduces the fundamental notions to describe random
variables by a probability distribution, by the moment-generating
function, and by the cumulant-generating function. It, correspondingly,
introduces moments and cumulants and their mutual connections. These
definitions are key to the subsequent concepts, such as the perturbative
computation of statistics.

\section{Probabilities, observables, and moments\protect\label{sec:Probabilities}}

Assume we want to describe some physical system. Let us further assume
the state of the system is denoted as $x\in\mathbb{R}^{N}$. Imagine,
for example, the activity of $N$ neurons at a given time point. Or
the activity of a single neuron at $N$ different time points. We
can make observations of the system that are functions $f(x)\in\mathbb{R}$
of the state of the system. Often we are repeating our measurements,
either over different trials or we average the observable in a stationary
system over time. It is therefore useful to describe the system in
terms of the density
\begin{align*}
p(y) & =\lim_{\epsilon\to0}\,\frac{1}{\Pi_{i}\epsilon_{i}}\,\langle1_{\{x_{i}\in[y_{i},y_{i}+\epsilon_{i}]\}}\rangle_{x}\\
 & =\langle\delta(x-y)\rangle_{x},
\end{align*}
where the symbol $\langle\rangle$ denotes the average over many repetitions
of the experiment, over realizations for a stochastic model, or over
time. The indicator function $1_{x\in S}$ is $1$ if $x\in S$ and
zero otherwise, and the Dirac $\delta$-distribution acting on a vector
is understood as $\delta(x)=\Pi_{i=1}^{N}\delta(x_{i})$. The symbol
$p(x)$ can be regarded as a probability density, but we will here
use it in a more general sense, also applied to deterministic systems,
for example where the values of $x$ follow a deterministic equation
of motion. It holds that $p$ is normalized in the sense
\begin{align}
1 & =\int\,p(x)\,dx.\label{eq:normalization_p}
\end{align}
Evaluating for the observable function $f$ the expectation value
$\langle f(x)\rangle$, we may use the Taylor representation of $f$
to write
\begin{align}
\langle f(x)\rangle & :=\int\,p(x)\,f(x)\,dx\label{eq:f_Taylor}\\
 & =\sum_{n_{1},\ldots,n_{N}=0}^{\infty}\frac{f^{(n_{1},\ldots,n_{N})}(0)}{n_{1}!\cdots n_{N}!}\,\langle x_{1}^{n_{1}}\cdots x_{N}^{n_{N}}\rangle\nonumber \\
 & =\sum_{n=0}^{\infty}\sum_{i_{1},\ldots,i_{n}=1}^{N}\frac{f_{i_{1}\cdots i_{n}}^{(n)}(0)}{n!}\,\langle\prod_{l=1}^{n}x_{i_{l}}\rangle,\nonumber 
\end{align}
where we denoted by $f^{(n_{1},\ldots,n_{N})}(x):=\big(\frac{\partial}{\partial x_{1}}\big)^{n_{1}}\cdots\big(\frac{\partial}{\partial x_{N}}\big)^{n_{N}}\,f(x)$
the $n_{1}$-th to $n_{N}$-th derivative of $f$ by its arguments;
the alternative notation for the Taylor expansion denotes the $n$-th
derivative by $n$ (possibly) different $x$ as $f_{i_{1}\cdots i_{n}}^{(n)}(x):=\prod_{l=1}^{n}\frac{\partial}{\partial x_{i_{l}}}f(x)$.

We see that the two representations of the Taylor expansion are identical,
because each of the indices $i_{1},\ldots,i_{n}$ takes on any of
the values $1,\ldots,N$. Hence there are $\left(\begin{array}{c}
n\\
n_{k}
\end{array}\right)$ combinations that yield a term $x_{k}^{n_{k}}$, because this is
the number of ways by which any of the $n$ indices $i_{l}$ may take
on the particular value $i_{l}=k$. So we get a combinatorial factor
$\frac{1}{n!}\left(\begin{array}{c}
n\\
n_{k}
\end{array}\right)=\frac{1}{(n-n_{k})!n_{k}!}$. Performing the same consideration for the remaining $N-1$ coordinates
brings the third line of \eqref{eq:f_Taylor} into the second.

In \eqref{eq:f_Taylor} we defined\textbf{ }the \textbf{moments} as
\begin{align}
\langle x_{1}^{n_{1}}\cdots x_{N}^{n_{N}}\rangle & :=\int\,p(x)\,x_{1}^{n_{1}}\cdots x_{N}^{n_{N}}\,dx\label{eq:def_moments}
\end{align}
 of the system's state variables. Knowing only the latter, we are
hence able to evaluate the expectation value of arbitrary observables
that possess a Taylor expansion.

Alternatively, we may write our observable $f$ in its Fourier representation
$f(x)=\iFtr{\hat{f}}x=\frac{1}{\left(2\pi\right)^{N}}\,\int\,\hat{f}(\omega)\,e^{i\omega^{\T}x}\,d\omega$
so that we get for the expectation value
\begin{align}
\langle f(x)\rangle & =\frac{1}{\left(2\pi\right)^{N}}\int\,\hat{f}(\omega)\,\int\,p(x)\,e^{i\omega^{\T}x}\,dx\,d\omega\nonumber \\
 & =\frac{1}{\left(2\pi\right)^{N}}\int\,\hat{f}(\omega)\,\langle e^{i\omega^{\T}x}\rangle_{x}\,d\omega,\label{eq:exp_f_Fourier}
\end{align}
where $\omega^{\T}x=\sum_{i=1}^{N}\omega_{i}x_{i}$ denotes the Euclidean
scalar product.

We see that we may alternatively determine the function $\langle e^{i\omega^{\T}x}\rangle_{x}$
for all $\omega$ to characterize the distribution of $x$, motivating
the definition
\begin{align}
Z(j) & :=\langle e^{j^{\T}x}\rangle_{x}\nonumber \\
 & =\int\,p(x)\,e^{j^{\T}x}\,dx.\label{eq:def_char_fctn}
\end{align}
Note that we can express $Z$ as the Fourier transform of $p$, so
it is clear that it contains the same information as $p$ (for distributions
$p$ for which a Fourier transform exists). The function $Z$ is called
the \textbf{characteristic function} or \textbf{moment generating
function} \citep[p. 32]{Gardiner85}. The argument $j$ of the function
is sometimes called the ``source'', because in the context of quantum
field theory, these variables correspond to particle currents. We
will adapt this customary name here, but without any physical implication.
The moment generating function $Z$ is identical to the partition
function $\mathcal{Z}$ in statistical physics, apart from the lacking
normalization of the latter. The equivalence here refers to the fact
that in a partition function $\cZ=\int e^{-\beta H(x)}\,dx$ the integrand
$e^{-\beta H(x)}$ is proportional to the probability of the state
$x$ and often the energy $H(x)$ contains terms linear in $x$, such
as $j^{\T}x$, so that derivatives by $j$ allow one to measure moments
of $x$; here, however, the normalization $\Z^{-1}$ needs to be taken
into account.

From the normalization \eqref{eq:normalization_p} and the definition
\eqref{eq:def_char_fctn} follows that
\begin{align}
Z(0) & =1.\label{eq:normalization_Z}
\end{align}

We may wonder how the moments, defined in \eqref{eq:def_moments},
relate to the characteristic function \eqref{eq:def_char_fctn}. We
see that we may obtain the moments by a simple differentiation of
$Z$ as
\begin{align}
\langle x_{1}^{n_{1}}\cdots x_{N}^{n_{N}}\rangle & =\left.\left\{ \prod_{i=1}^{N}\partial_{i}^{n_{i}}\right\} \,Z(j)\right|_{j=0},\label{eq:moments_generation}
\end{align}
where we introduced the short hand notation $\partial_{i}^{n_{i}}=\frac{\partial^{n_{i}}}{\partial j_{i}^{n_{i}}}$
and set $j=0$ after differentiation. Conversely, we may say that
the moments are the Taylor coefficients of $Z$, from which follows
the identity
\begin{align*}
Z(j) & =\sum_{n_{1},\ldots,n_{N}}\frac{\langle x_{1}^{n_{1}}\ldots x_{N}^{n_{N}}\rangle}{n_{1}!\ldots n_{N}!}\,j_{1}^{n_{1}}\ldots j_{N}^{n_{N}}.
\end{align*}

\section{Transformation of random variables\protect\label{sec:Transformation-of-random}}

Often one knows the statistics of some random variable $x$ but would
like to know the statistics of $y$, a function of $x$
\begin{align*}
y & =f(x).
\end{align*}
The probability densities transform as
\begin{align*}
p_{y}(y) & =\int dx\,p_{x}(x)\,\delta(y-f(x)).
\end{align*}
It is obvious that the latter definition of $p_{y}$ is properly normalized:
integrating over all $y$, the Dirac distribution reduces to a unit
factor so that the normalization condition for $p_{x}$ remains. What
does the corresponding moment-generating function look like?

We obtain it directly from its definition \eqref{eq:def_char_fctn}
as
\begin{align*}
Z_{y}(j) & =\langle e^{j^{\T}y}\rangle_{y}\\
 & =\int dy\,p_{y}(y)\,e^{j^{\T}y}\\
 & =\int dy\,\int dx\,p_{x}(x)\,\delta(y-f(x))\,e^{j^{\T}y}\\
 & =\int dx\,p_{x}(x)\,e^{j^{\T}f(x)}\\
 & =\langle e^{j^{\T}f(x)}\rangle_{x},
\end{align*}
where we swapped the order of the integrals in the third line and
performed the integral over $y$ by employing the property of the
Dirac distribution. The dimension of the vector $y\in\mathbb{R}^{N^{\prime}}$
may in general be different from the dimension of the vector $x\in\mathbb{R}^{N}$.
In summary, we only need to replace the source term $j^{\T}x\to j^{\T}f(x)$
to obtain the transformed moment generating function.

\section{Joint distribution and conditional distribution}

A joint distribution is a distribution that depends on more than a
single random variable. The distributions we have introduced above
are such distributions, because they describe the joint distribution
of the $N$ elements of $x\in\bR^{N}$, written explicitly as $p(x_{1},\ldots,x_{N})$.
Consider the special case $p(x,y)$ of only two random variables $x$
and $y$. One defines the \textbf{marginal distributions} as 
\begin{eqnarray*}
p(x) & = & \int p(x,y)\,dy,\\
p(y) & = & \int p(x,y)\,dx,
\end{eqnarray*}
which describe the probability for a value $x$, independent of what
$y$ is ($p(x)$) or vice versa. Another object that frequently arises
are \textbf{conditional distributions}. These describe the distribution
of $y$ given that one has observed a value $x$ already: $p(y|x)$,
which is denoted as $y$ given $x$. It holds \textbf{Bayes' law
\begin{eqnarray}
p(y,x) & = & p(y|x)\,p(x).\label{eq:Bayes}
\end{eqnarray}
}Correspondingly we may define the moment-generating function for
the conditional distribution
\begin{eqnarray*}
Z(j|x) & = & \langle e^{jy}\rangle_{y\sim p(y|x)}.
\end{eqnarray*}

\section{Cumulants\protect\label{sec:Cumulants}}

For a set of independent variables the probability density factorizes
as $p^{\mathrm{indep.}}(x)=p_{1}(x_{1})\cdots p_{N}(x_{N})$. The
characteristic function, defined by \eqref{eq:def_char_fctn}, then
factorizes as well $Z^{\mathrm{indep.}}(j)=Z_{1}(j{}_{1})\cdots Z_{N}(j_{N})$.
Considering the $k$-point moment, the $k$-th ($k\le N$) moment
$\langle x_{1}\ldots x_{k}\rangle=\langle x_{1}\rangle\ldots\langle x_{k}\rangle$,
where individual variables only appear in single power, decomposes
into a product of $k$ first moments of the respective variables.
We see in this example that the higher order moments in this case
contain information which is already contained in the lower order
moments. In the general case of variables that are not statistically
independent, higher order moments still contain some information that
is already present at lower orders. Decomposing the statistical dependence
into one part that is implied by lower orders and one part that is
present only at order $n$, motivates the definition of cumulants
in the following.

One can therefore ask if it is possible to define an object that only
contains the dependence at a certain order and removes all dependencies
that are already contained in lower orders. The observation that the
moment-generating function in the independent case decomposes into
a product, leads to the idea to consider its logarithm
\begin{align}
W(j) & :=\ln\,Z(j),\label{eq:def_W}
\end{align}
Evidently for independent variables it consequently decomposes into
a sum $W(j)=\sum_{i}\ln\,Z_{i}(j_{i})$. The Taylor coefficients of
$W$ for independent variables therefore do not contain any mixed
terms, because $\partial_{k}\partial_{l}W\big|_{j=0}=0\quad\forall k\neq l$.
The same it obviously true for higher derivatives. This observation
motivates the definition of the \textbf{cumulants} as the Taylor coefficients
of $W$

\begin{align}
\llangle x_{1}^{n_{1}}\ldots x_{N}^{n_{N}}\rrangle & :=\left.\left\{ \prod_{i=1}^{N}\partial_{i}^{n_{i}}\right\} W(j)\right|_{j=0},\label{eq:def_cumulant}
\end{align}
which we here denote by double angular brackets $\llangle\circ\rrangle$.
For independent variables, as argued above, we have $\llangle x_{1}\ldots x_{N}\rrangle=0$.

The function $W$ defined by \eqref{eq:def_W} is called the \textbf{cumulant
generating function}. We may conversely express it as a Taylor series
\begin{align}
W(j) & =\ln\,Z(j)=\sum_{n_{1},\ldots,n_{N}}\frac{\llangle x_{1}^{n_{1}}\ldots x_{N}^{n_{N}}\rrangle}{n_{1}!\ldots n_{N}!}\,j_{1}^{n_{1}}\ldots j_{N}^{n_{N}}.\label{eq:cumulant_Taylor}
\end{align}
The cumulants are hence the Taylor coefficients of the cumulant-generating
function. The normalization \eqref{eq:normalization_Z} of $Z(0)=1$
implies 
\begin{align*}
W(0) & =0.
\end{align*}
For the cumulants this particular normalization is, however, not crucial,
because a different normalization $\tilde{Z}(j)=C\,Z(j)$ would give
an inconsequential additive constant $\tilde{W}(j)=\ln(C)+W(j)$.
The normalization therefore does not affect the cumulants, which contain
at least one derivative. The definition $W(j):=\ln\,\mathcal{Z}(j)$
for a partition function $\mathcal{Z}$ would hence lead to the same
cumulants. In statistical physics, this latter definition of $W$
corresponds to the free energy \citep{NegeleOrland98}.

Examples:
\begin{itemize}
\item The uniform distribution $x\sim U([-1,1])$ with the density $p(x)=\frac{1}{2}H(1+x)\,H(1-x)$
has the moment-generating function $Z(j)=\frac{1}{2}\int_{-1}^{1}dx\,e^{jx}=\frac{1}{2j}\,(e^{j}-e^{-j})=\frac{\sinh(j)}{j}$
and the cumulant-generating function $W(j)=\ln\,\sinh(j)-\ln j$.
\item An Ising spin $s\in\{-1,1\}$ with $p(s=1)=\frac{1}{2}$ has the moment-generating
function $Z(j)=\frac{1}{2}e^{j}+\frac{1}{2}e^{-j}=\cosh(j)$ and the
cumulant-generating function $W(j)=\ln Z(j)=\ln\,\cosh(j)$.
\end{itemize}

\section{Connection between moments and cumulants\protect\label{sec:Connection-moments-cumulants}}

Since both, moments and cumulants, characterize a probability distribution
one may wonder if and how these objects are related. The situation
up to this point is this:
\begin{center}
\includegraphics{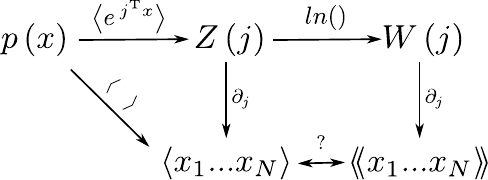}
\par\end{center}

We know how to obtain the moment generating function $Z$ from the
probability $p$, and the cumulant generating function from $Z$ by
the logarithm. The moments and cumulants then follow as Taylor coefficients
from their respective generating functions. Moreover, the moments
can also directly be obtained by the definition of the expectation
value. What is missing is a direct link between moments and cumulants.
This link is what we want to find now.

To this end we here consider the case of $N$ random variables $x_{1},\ldots,x_{N}$.
At first we restrict ourselves to the special case of the $k$-point
moment ($1\le k\le N$)
\begin{eqnarray}
\langle x_{1}\cdots x_{k}\rangle & = & \left.\partial_{1}\cdots\partial_{k}\,Z(j)\right|_{j=0},\label{eq:raw_corr_generated}
\end{eqnarray}
where individual variables only appear in single power.

It is sufficient to study this special case, because a power of $x^{n}$
with $n>1$ can be regarded by the left hand side of \eqref{eq:raw_corr_generated}
as the $n$-fold repeated occurrence of the same index. We therefore
obtain the expressions for repeated indices by first deriving the
results for all indices assumed different and setting indices indentical
in the final result. We will come back to this procedure at the end
of the section. 

Without loss of generality, we are here only interested in $k$-point
moments with consecutive indices from $1$ to $k$, which can always
be achieved by renaming the components $x_{i}$. We express the moment
generating function using \eqref{eq:def_W} as
\begin{align*}
Z(j) & =\exp(W(j)).
\end{align*}
Taking derivatives by $j$ as in \eqref{eq:raw_corr_generated}, we
anticipate due to the exponential function that the term $\exp(W(j))$
will be reproduced, but certain pre-factors will be generated. We
therefore define the function $f_{k}(j)$ as the prefactor appearing
in the $k$-fold derivative of $Z(j)$ as
\begin{align*}
\partial_{1}\cdots\partial_{k}\,Z(j) & =\partial_{1}\cdots\partial_{k}\,\exp(W(j))\\
 & =:f_{k}(j)\,\exp(W(j)).
\end{align*}
Obviously due to \eqref{eq:raw_corr_generated} and $\exp(W(0))=1$,
the function evaluated at zero is the $k$-th moment
\begin{align*}
f_{k}(0) & =\langle x_{1}\cdots x_{k}\rangle.
\end{align*}
We now want to obtain a recursion formula for $f_{k}$ by applying
the product rule as
\begin{align*}
\partial_{k}\underbrace{\big(f_{k-1}(j)\,\exp(W(j))\Big)}_{\partial_{1}\cdots\partial_{k-1}\,Z(j)} & \stackrel{\text{product rule}}{=}\underbrace{\left(\partial_{k}f_{k-1}+f_{k-1}\,\partial_{k}W\right)}_{f_{k}}\exp(W(j)),
\end{align*}
from which we obtain
\begin{align}
f_{k} & =\partial_{k}f_{k-1}+f_{k-1}\,\partial_{k}W.\label{eq:recursion_moments_from_cumulants_k_point}
\end{align}
The explicit first three steps lead to (starting from $f_{1}(j)\equiv\partial_{1}W(j)$)
\begin{align}
f_{1} & =\partial_{1}W\label{eq:correlators_moments_epxlicit_1_2_3}\\
f_{2} & =\partial_{1}\partial_{2}W+\left(\partial_{1}W\right)\left(\partial_{2}W\right)\nonumber \\
f_{3} & =\partial_{1}\partial_{2}\partial_{3}W\nonumber \\
 & +\left(\partial_{1}W\right)\left(\partial_{2}\partial_{3}W\right)+\left(\partial_{2}W\right)\left(\partial_{1}\partial_{3}W\right)+\left(\partial_{3}W\right)\left(\partial_{1}\partial_{2}W\right)\nonumber \\
 & +\left(\partial_{1}W\right)\left(\partial_{2}W\right)\left(\partial_{3}W\right).\nonumber 
\end{align}
The structure shows that the moments are composed of all combinations
of cumulants of all lower orders. More specifically, we see that
\begin{itemize}
\item the number of derivatives in each term is the same, here three
\item the three derivatives are partitioned in all possible ways to act
on $W$, from all derivatives acting on the same $W$ (first term
in last line) to each acting on a separate $W$ (last term).
\end{itemize}
Figuratively, we can imagine these combinations to be created by having
$k$ places and counting all ways of forming $n$ subgroups of sizes
$l_{1},\ldots,l_{n}$ each, so that $l_{1}+\ldots+l_{n}=k$. On the
example $k=3$ we would have
\begin{align*}
\langle1\,2\,3\rangle & =\underbrace{\llangle1\,2\,3\rrangle}_{n=1\;l_{1}=3}\\
 & +\underbrace{\llangle1\rrangle\llangle2\,3\rrangle+\llangle2\rrangle\llangle3\,1\rrangle+\llangle3\rrangle\llangle1\,2\rrangle}_{n=2;\,l_{1}=1\le l_{2}=2}\\
 & +\underbrace{\llangle1\rrangle\llangle2\rrangle\llangle3\rrangle}_{n=3;\,l_{1}=l_{2}=l_{3}=1}.
\end{align*}
We therefore suspect that the general form can be written as

\begin{align}
f_{k} & =\sum_{n=1}^{k}\sum_{\begin{array}[t]{c}
\{1\le l_{1}\le\ldots,\le l_{n}\le k\}\\
\sum_{i}l_{i}=k
\end{array}}\quad\times\label{eq:f_k_combinatorial_mom_corr}\\
 & \times\,\sum_{\sigma\in P(\{l_{i}\},k)}\left(\partial_{\sigma(1)}\cdots\partial_{\sigma(l_{1})}W\right)\ldots\left(\partial_{\sigma(k-l_{n}+1)}\cdots\partial_{\sigma(k)}W\right),\nonumber 
\end{align}
where the sum over $n$ goes over all numbers of subsets of the partition,
the sum 
\begin{eqnarray*}
\sum_{\begin{array}[t]{c}
\{1\le l_{1}\le\ldots,\le l_{n}\le k\}\\
\sum_{i}l_{i}=k
\end{array}}
\end{eqnarray*}
goes over all sizes $l_{1},\ldots,l_{n}$ of each subgroup, which
we can assume to be ordered by the size $l_{i}$, and $P(\{l_{i}\},k)$
is the set of all permutations of the numbers $1,\ldots,k$ that,
for a given partition $\{1\le l_{1}\le\ldots\le l_{n}\le k\}$, lead
to a different term: Obviously, the exchange of two indices within
a subset does not cause a new term, because the differentiation may
be performed in arbitrary order.

The proof of \eqref{eq:f_k_combinatorial_mom_corr} follows by induction.
Initially we have $f_{1}=\partial_{1}W$ which fulfills the assumption
\eqref{eq:f_k_combinatorial_mom_corr}, because there is only one
possible permutation. Assuming that in the $k$-th step \eqref{eq:f_k_combinatorial_mom_corr}
holds, the $k+1$-st step follows from the application of the product
rule for the first term on the right of \eqref{eq:recursion_moments_from_cumulants_k_point}
acting on one term of \eqref{eq:f_k_combinatorial_mom_corr} 
\begin{eqnarray*}
 &  & \partial_{k+1}\left(\partial_{\sigma(1)}\cdots\partial_{\sigma(l_{1})}W\right)\ldots\left(\partial_{\sigma(\sum_{i<n}l_{i}+1)}\cdots\partial_{\sigma(k)}W\right)\\
 & = & \sum_{j=1}^{n}\left(\partial_{\sigma(1)}\cdots\partial_{\sigma(l_{1})}W\right)\ldots\left(\partial_{k+1}\partial_{\sigma(\sum_{i<j}l_{i}+1)}\cdots\partial_{\sigma(\sum_{i\le j}l_{i})}W\right)\\
 &  & \phantom{\sum_{j=1}^{n}}\ldots\left(\partial_{\sigma(\sum_{i<n}l_{i}+1)}\cdots\partial_{\sigma(k)}W\right),
\end{eqnarray*}
which combines the additional derivative with each of the existing
terms in turn. Therefore, all terms together have $k+1$ derivatives
and no term exists that has a factor $\partial_{k+1}W$, because $f_{k}$
already contained only derivatives of $W$, not $W$ alone. The second
term in \eqref{eq:recursion_moments_from_cumulants_k_point} multiplies
$\partial_{k+1}W$ with $f_{k}$, containing all combinations of order
$k$. So the two terms together generate all combinations of the form
\eqref{eq:f_k_combinatorial_mom_corr}, proving the assumption.

Setting all sources to zero $j_{1}=\ldots=j_{k}=0$ leads to the expression
for the $k$-th moment by the $1\mathrm{st},\ldots,k$-point cumulants
\begin{align}
\langle x_{1}\cdots x_{k}\rangle & =\sum_{n=1}^{k}\sum_{\begin{array}[t]{c}
\{1\le l_{1}\le\ldots,\le l_{n}\le k\}\\
\sum_{i}l_{i}=k
\end{array}}\quad\times\label{eq:k_point_corr_from_cum}\\
 & \times\,\sum_{\sigma\in P(\{l_{i}\},k)}\llangle x_{\sigma(1)}\cdots x_{\sigma(l_{1})}\rrangle\cdots\llangle x_{\sigma(k-l_{n}+1)}\cdots x_{\sigma(k)}\rrangle.\nonumber 
\end{align}

\begin{itemize}
\item So the recipe to determine the $k$-th moment is: Draw a set of $k$
points, partition them in all possible ways into disjoint subsets
(using every point only once). Now assign, in all possible ways that
lead to a different composition of the subgroups, one variable to
each of the points in each of these combinations. The $i$-th subset
of size $l_{i}$ corresponds to a cumulant of order $l_{i}$. The
sum over all such partitions and all permutations yields the $k$-th
moment expressed in terms of cumulants of order $\le k$.
\end{itemize}
We can now return to the case of higher powers in the moments, the
case that $m\ge2$ of the $x_{i}$ are identical. Since the appearance
of two differentiations by the same variable in \eqref{eq:raw_corr_generated}
is handled in exactly the same way as for $k$ different variables,
we see that the entire procedure remains the same: In the final result
\eqref{eq:k_point_corr_from_cum} we just have $m$ identical variables
to assign to different places. All different assignments of these
variables to positions need to be counted separately.

\section{Recovering the probability density}

If one knows the moment-generating function or the cumulant-generating
function, we may recover the density by the inverse Fourier transform
\begin{eqnarray}
p(x) & = & \int\D j\,\exp\big(-j^{\T}x\big)\,Z(j)\label{eq:Fourier_W_Z_p}\\
 & = & \int\D j\,\exp\big(-j^{\T}x+W(j)\big),\nonumber 
\end{eqnarray}
where $\int\D j=\prod_{k=1}^{N}\int_{-i\infty}^{i\infty}\frac{dj_{k}}{2\pi i}$.

\section{Keypoints}

After reviewing the basics of probability theory we can summarize
the results as follows\\

\noindent\fcolorbox{black}{white}{\begin{minipage}[t]{1\textwidth - 2\fboxsep - 2\fboxrule}%
\textbf{Probabilities, Moments and Cumulants}
\begin{itemize}
\item The moments of a probability density are given by $\langle x^{n}\rangle=\int dx\,p(x)x^{n}$.
\item The moment generating function $Z(j)$ is equivalent to the Fourier
transform (for $j\in i\bR$) of $p(x)$: $Z(j)=\int dx\,\exp(j^{\T}x)\,p(x).$
\item The moments of a probability density are obtained from derivatives
of the moment generating function (MGF) $Z(j)$ w.r.t to the source
term $j$.
\item The cumulants are generated by the derivatives of cumulant generating
function (CGF) $W(j)=\ln\,Z(j).$
\item Moments can be reconstructed from the cumulants and vice versa.
\item CGF, MGF and the probability density contain the same information.
The CGF is less redundant than the MGF.
\end{itemize}
\end{minipage}}

\chapter{Gaussian distribution and Wick's theorem\protect\label{chap:Gaussian-distribution-Wick-theorem}}

We will now study a special case of a distribution that plays an
essential role in all further development, the Gaussian distribution.
In a way, field theory boils down to a clever reorganization of Gaussian
integrals. In this section we will therefore derive fundamental properties
of this distribution. In the limit of large numbers of neurons, neuronal
networks will exhibit Gaussian distributions in many of their quantities.

\section{Gaussian distribution\protect\label{sec:Gaussian-distribution}}

A Gaussian distribution of $N$ centered (mean value zero) variables
$x$ is defined for a positive definite symmetric matrix $A$ as

\begin{eqnarray}
p(x) & \propto & \exp\Big(-\frac{1}{2}x^{\T}Ax\Big).\label{eq:Gauss_unnormalized}
\end{eqnarray}
In the language of statistical field theory, the exponent on the right
hand side is often referred to as the ``action''; this nomenclature
is also used in cases that the expression is not a quadratic polynomial.
In the current case, one may also call it a quadratic or Gaussian
action.

A more general formulation for symmetry is that $A$ is self-adjoint
with respect to the Euclidean scalar product (see \prettyref{sec:Appendix-Self-adjoint-matrices}).
As usual, positive definite means that the bilinear form $x^{\T}\,A\,x>0\quad\forall x\neq0$.
Positivity equivalently means that all eigenvalues $\lambda_{i}$
of $A$ are positive. One may also define the Gaussian for a positive
semi-definite quadratic form, but we will here stick to the slightly
more specific case of a positive definite form. The properly normalized
distribution is
\begin{align}
p(x) & =\frac{\det(A)^{\frac{1}{2}}}{(2\pi)^{\frac{N}{2}}}\,\exp\left(-\frac{1}{2}x^{\T}Ax\right);\label{eq:N_dim_Gauss}
\end{align}
this normalization factor is derived in \prettyref{sec:Appendix_Normalization_Gaussian}.

\section{Moment and cumulant generating function of a Gaussian\protect\label{sec:Completion_of_square_Gaussian}}

The moment generating function $Z(j)$ follows from the definition
\eqref{eq:def_char_fctn} for the Gaussian distribution \eqref{eq:N_dim_Gauss}.
We utilize the substitution $y=x-A^{-1}j$ in the second line of \eqref{eq:Z_Gauss},
which is the $N$-dimensional version of the ``completion of the
square''. With the normalization $C=\det(A)^{1/2}/(2\pi)^{N/2}$
we get

\begin{align}
Z(j) & =\langle e^{j^{\T}x}\rangle_{x}\label{eq:Z_Gauss}\\
 & =C\,\int\,\Pi_{i}dx_{i}\exp\Big(-\frac{1}{2}x^{\T}Ax+\underbrace{j^{\T}x}_{\frac{1}{2}\,\big(A^{-1}\,j\big)^{\T}A\,x+\frac{1}{2}\,x^{\T}A\,\big(A^{-1}\,j\big)}\Big)\nonumber \\
 & =C\,\int\,\Pi_{i}dx_{i}\,\exp\Big(-\frac{1}{2}\underbrace{\left(x-A^{-1}j\right)^{\T}}_{y^{\T}}\,A\,\underbrace{\left(x-A^{-1}j\right)}_{y}+\frac{1}{2}\,j^{\T}\,A^{-1}\,j\Big)\nonumber \\
 & =\underbrace{C\,\int\,\Pi_{i}dy_{i}\,\exp\Big(-\frac{1}{2}y^{\T}A\,y\Big)}_{=1}\,\exp\Big(\frac{1}{2}\,j^{\T}\,A^{-1}\,j\Big)\nonumber \\
 & =\exp\,\Big(\frac{1}{2}\,j^{\T}\,A^{-1}\,j\Big).\nonumber 
\end{align}
 The integral measures do not change form the third to the fourth
line, because we only shifted the integration variables. We used from
the fourth to the fifth line that $p$ is normalized, which is not
affected by the shift, because the boundaries of the integral are
infinite. The cumulant generating function $W(j)$ defined by \prettyref{eq:def_W}
then is
\begin{align}
W(j) & =\ln\,Z(j)\nonumber \\
 & =\frac{1}{2}\,j^{\T}A^{-1}\,j.\label{eq:W_Gauss}
\end{align}
Hence the second order cumulants are 
\begin{align}
\llangle x_{i}x_{j}\rrangle & =\left.\partial_{i}\partial_{j}W\right|_{j=0}\label{eq:cumulants_Gauss}\\
 & =\big[A^{-1}\big]_{ij},\nonumber 
\end{align}
where the factor $\frac{1}{2}$ is canceled, because, by the product
rule, the derivative first acts on the first and then on the second
$j$ in \eqref{eq:W_Gauss}, both of which yield the same term due
to the symmetry of $A^{-1T}=A^{-1}$ (The symmetry of $A^{-1}$ follows
from the symmetry of $A$, because $\mathbf{1}=A^{-1}A=A^{\T}A^{-1\T}=A\,A^{-1\T}$;
because the inverse of $A$ is unique it follows that $A^{-1\T}=A^{-1}$).

All cumulants other than the second order \eqref{eq:cumulants_Gauss}
vanish, because \eqref{eq:W_Gauss} is already the Taylor expansion
of $W$, containing only second order terms and the Taylor expansion
is unique. This property of the Gaussian distribution will give rise
to the useful theorem by Wick in the following subsection.

Eq. \eqref{eq:cumulants_Gauss} is of course the covariance matrix,
the matrix of second cumulants. We therefore also write the Gaussian
distribution as
\begin{eqnarray*}
x & \sim & \N(0,A^{-1}),
\end{eqnarray*}
where the first argument $0$ refers to the vanishing mean value.

\section{Wick's theorem\protect\label{sec:Wicks-theorem}}

For the Gaussian distribution introduced in \prettyref{sec:Gaussian-distribution},
all moments can be expressed in terms of products of only second cumulants
of the Gaussian distribution. This relation is known as \textbf{Wick's
theorem} \citep{ZinnJustin96,Kleinert89}.

Formally this result is a special case of the general relation between
moments and cumulants \eqref{eq:k_point_corr_from_cum}: In the Gaussian
case only second cumulants \eqref{eq:cumulants_Gauss} are different
from zero. The only term that remains in \eqref{eq:k_point_corr_from_cum}
is hence a single partition in which all subgroups have size two,
i.e. $l_{1}=\ldots=l_{n}=2$; each such sub-group corresponds to a
second cumulant. In particular it follows that all moments with odd
power $k$ of $x$ vanish. For a given even $k$, the sum over all
$\sigma\in P[\{2,\ldots,2\}](k)$ includes only those permutations
$\sigma$ that lead to different terms

\begin{align}
\langle x_{1}\cdots x_{k}\rangle_{x\sim\N(0,A^{-1})} & =\sum_{\sigma\in P(\{2,\ldots,2\},k)}\quad\llangle x_{\sigma(1)}x_{\sigma(2)}\rrangle\cdots\llangle x_{\sigma(k-1)}x_{\sigma(k)}\rrangle\nonumber \\
 & \stackrel{(\ref{eq:cumulants_Gauss})}{=}\sum_{\sigma\in P(\{2,\ldots,2\},k)}A_{\sigma(1)\sigma(2)}^{-1}\cdots A_{\sigma(k-1)\sigma(k)}^{-1}.\label{eq:Wick_moments}
\end{align}

We can interpret the latter equation in a simple way: To calculate
the $k$-th moment of a Gaussian distribution, we need to combine
the $k$ variables in all possible, distinct pairs and replace each
pair $(i,j)$ by the corresponding second cumulant $\llangle x_{i}x_{j}\rrangle=A_{ij}^{-1}$.
Here ``distinct pairs'' means that we treat all $k$ variables as
different, even if they may in fact be the same variable, in accordance
to the note at the end of \prettyref{sec:Connection-moments-cumulants}.
In the case that a subset of $n$ variables of the $k$ are identical,
this gives rise to a \textbf{combinatorial factor}. Figuratively,
we may imagine the computation of the $k$-th moment as composed out
of so called \textbf{contractions}: Each pair of variables is contracted
by one Gaussian integral. This is often indicated by an angular bracket
that connects the two elements that are contracted. In this graphical
notation, the fourth moment $\langle x_{1}x_{2}x_{3}x_{4}\rangle$
of an $N$ dimensional Gaussian can be written as

\begin{align}
\langle x_{1}x_{2}x_{3}x_{4}\rangle_{x\sim\N(0,A^{-1})}= & \begC1{x_{1}}\endC1{x_{2}}\begC2{x_{3}}\endC2{x_{4}}+\begC1{x_{1}}\begC2{x_{2}}\endC1{x_{3}}\endC2{x_{4}}+\begC1{x_{1}}\begC2{x_{2}}\endC2{x_{3}}\endC1{x_{4}}\nonumber \\
= & \llangle x_{1}x_{2}\rrangle\llangle x_{3}x_{4}\rrangle+\llangle x_{1}x_{3}\rrangle\llangle x_{2}x_{4}\rrangle+\llangle x_{1}x_{4}\rrangle\llangle x_{2}x_{3}\rrangle\nonumber \\
= & A_{12}^{-1}\,A_{34}^{-1}+A_{13}^{-1}\,A_{24}^{-1}+A_{14}^{-1}\,A_{23}^{-1}.\label{eq:fourth_moment_example}
\end{align}
To illustrate the appearance of a combinatorial factor, we may imagine
the example that all $x_{1}=x_{2}=x_{3}=x_{4}=x$ in the previous
example are identical. We see from \prettyref{eq:fourth_moment_example}
by setting all indices to the same value that we get the same term
three times in this case, namely
\begin{eqnarray*}
\langle x^{4}\rangle & = & 3\,\llangle x^{2}\rrangle^{2}.
\end{eqnarray*}

\section{Appendix: Self-adjoint operators\protect\label{sec:Appendix-Self-adjoint-matrices}}

We denote as $(x,y)$ a scalar product. We may think of the Euclidean
scalar product $(x,y)=\sum_{i=1}^{N}x_{i}y_{i}$ as a concrete example.
The condition for symmetry of $A$ can more accurately be stated as
the operator $A$ being self-adjoint. In general, the adjoint operator
is defined with regard to a scalar product $(\cdot,\cdot)$ as
\begin{align*}
(x,A\,y) & \stackrel{\text{def. adjoint}}{=:}(A^{\T}\,x,y)\,\quad\forall x,y.
\end{align*}
An operator is self-adjoint, if $A^{\T}=A.$

If a matrix $A$ is self-adjoint with respect to the Euclidean scalar
product $(\cdot,\cdot)$, its diagonalizing matrix $U$ has orthogonal
column vectors with respect to the same scalar product, because from
the general form of a basis change into the eigenbasis $\diag(\{\lambda_{i}\})=U^{-1}\,A\,U$
follows that $(U^{-1\T},A\,U)\stackrel{\text{def. of adjoint}}{=}(A^{\T}\,U^{-1\T},U)\stackrel{\text{symm. of }(\cdot,\cdot)}{=}(U,A^{\T}\,U^{-1\T})\stackrel{A\text{ self. adj.}}{=}(U,A\,U^{-1\T})$.
So the column vectors of $U^{-1T}$ need to be parallel to the eigenvectors
of $A$, which are the column vectors of $U$, because eigenvectors
are unique up to normalization. If we assume them normalized we hence
have $U^{-1\T}=U$ or $U^{-1}=U^{\T}$. It follows that $(Uv,Uw)=(v,U^{\T}U\,w)=(v,w)$,
the condition for the matrix $U$ to be unitary with respect to $(\cdot,\cdot)$,
meaning its transformation conserves the scalar product.

\section{Appendix: Normalization of a Gaussian\protect\label{sec:Appendix_Normalization_Gaussian}}

The equivalence between positivity and all eigenvalues being positive
follows from diagonalizing $A$ by an orthogonal transform $U$
\begin{align*}
\diag(\{\lambda_{i}\}) & =U^{\T}\,A\,U,
\end{align*}
where the columns of $U$ are the eigenvectors of $A$ (see \prettyref{sec:Appendix-Self-adjoint-matrices}
for details). The determinant of the orthogonal transform, due to
$U^{-1}=U^{\T}$ is $|\det(U)|=1$, because $1=\det(\mathbf{1})=\det(U^{\T}U)=\det(U)^{2}$.
The orthogonal transform therefore does not affect the integration
measure. In the coordinate system of eigenvectors $v$ we can then
rewrite the normalization integral as

\begin{eqnarray*}
 &  & \int_{-\infty}^{\infty}\Pi_{i}dx_{i}\exp\Big(-\frac{1}{2}x^{\T}Ax\Big)\\
 & \stackrel{x=U\,v}{=} & \int_{-\infty}^{\infty}\Pi_{k}dv_{k}\exp\Big(-\frac{1}{2}v^{\T}U^{\T}AUv\Big)\\
 & = & \int_{-\infty}^{\infty}\Pi_{k}dv_{k}\exp\Big(-\frac{1}{2}\sum_{i}\lambda_{i}v_{i}^{2}\Big)\\
 & = & \Pi_{k}\sqrt{\frac{2\pi}{\lambda_{k}}}=(2\pi)^{\frac{N}{2}}\det(A)^{-\frac{1}{2}},
\end{eqnarray*}
where we used in the last step that the determinant of a matrix equals
the product of its eigenvalues.

\section{Keypoints}

In summary:

\noindent\fcolorbox{black}{white}{\begin{minipage}[t]{1\textwidth - 2\fboxsep - 2\fboxrule}%

\paragraph{Gaussian distribution and Wick's theorem}
\begin{itemize}
\item Gaussian distributions are characterized by only two cumulants: Mean
and variance (mean vector and covariance matrix for non-scalar Gaussian
variables). The CGF is a polynomial of degree 2 in the source terms
$j$.
\item The moments of a centered Gaussian can be obtained from the variance
using Wick's theorem, which considers all possible pairings.
\end{itemize}
\end{minipage}}

\section{Exercises\protect\label{sub:exercises_cumulants_moments}}

\subsection*{a) Cumulants}

Calculate the moment generating function and the cumulant generating
function for
\begin{enumerate}
\item the Gaussian distribution $p(x)=\frac{1}{\sqrt{2\pi}\sigma}\,e^{-\frac{(x-\mu)^{2}}{2\sigma^{2}}}$;
determine all cumulants of the distribution; (2 points)\textbf{}
\item the binary distribution $p(x)=(1-m)\,\delta(x)+m\,\delta(x-1)$ with
mean $m\in[0,1]$; determine the first three cumulants expressed in
$m$, verify that the first two correspond to the mean and the variance;
(2 points). Convince yourself that the link between moments and cumulants
holds on the example of the third moment, by once computing it directly
and once from the first three cumulants.
\end{enumerate}

\subsection*{b) Joint, marginal and conditional probability distribution}

Consider the joint Gaussian distribution of two random variables $x,y\in\mathbb{R}$

\[
p(x_{1},x_{2})=\frac{1}{2\pi\sqrt{\det\left(\Sigma\right)}}\exp\left(-\frac{1}{2}\left(\begin{array}{c}
x_{1}-\mu_{1}\\
x_{2}-\mu_{2}
\end{array}\right)^{\T}\Sigma^{-1}\left(\begin{array}{c}
x_{1}-\mu_{1}\\
x_{2}-\mu_{2}
\end{array}\right)\right)
\]
with $\Sigma=\left(\begin{array}{cc}
\Sigma_{11} & \Sigma_{12}\\
\Sigma_{12} & \Sigma_{22}
\end{array}\right)\in\mathbb{R}^{2\times2}$. State the cumulant generating function and show that it decomposes
into a sum for the hypothetical scenario of $\Sigma_{12}=0$ ($x_{1}$
and $x_{2}$ are independent)

\[
W_{x_{1},x_{2}}(\tilde{j}_{1},\tilde{j}_{2})=W_{x_{1}}(\tilde{j}_{1})+W_{x_{2}}(\tilde{j}_{2}).
\]
Now let us keep $\Sigma_{12}\neq0$ in the following so that the last
decomposition does not hold. Compute the marginal distribution $p(x_{1})=\int dx_{2}\,p(x_{1},x_{2})$.
We now want to compute the conditional probability $p(x_{1}|x_{2})=p(x_{1},x_{2})/p(x_{2})$.
In this formula, the denominator $p(x_{2})$ only serves as the correct
normalization; we will first ignore it and fix the normalization in
the end. Use that we can write the probability distribution in its
Fourier representation

\[
p(x_{1},x_{2})=\int_{-i\infty}^{i\infty}\frac{d\tilde{j}_{1}}{2\pi i}\int_{-i\infty}^{i\infty}\frac{d\tilde{j}_{2}}{2\pi i}\,\exp\left(-\tilde{j}_{1}x_{1}-\tilde{j}_{2}x_{2}+W_{x_{1},x_{2}}(\tilde{j}_{1},\tilde{j}_{2})\right).
\]
To compute the conditional probability distribution most easily, isolate
the $x_{1}$ dependency by integrating out $\tilde{j}_{2}$ and rewrite

\begin{align*}
p(x_{1}|x_{2})\propto p(x_{1},x_{2}) & =\underbrace{\int_{-i\infty}^{i\infty}\frac{d\tilde{j}_{1}}{2\pi i}\,\exp\left(-\tilde{j}_{1}x_{1}+W_{x_{1}}(\tilde{j}_{1})\right)}_{\text{independent of \ensuremath{\tilde{j}_{2}}}}\times\int_{-i\infty}^{i\infty}\frac{d\tilde{j}_{2}}{2\pi i}\,\dots,\\
 & =\int_{-i\infty}^{i\infty}\frac{d\tilde{j}_{1}}{2\pi i}\,\exp\left(-\tilde{j}_{1}x_{1}+W_{x_{1}|x_{2}}(\tilde{j}_{1})\right)
\end{align*}
with a new cumulant generating function $W_{x_{1}|x_{2}}(\tilde{j}_{1})$.
The Gaussian identity $\int_{-i\infty}^{i\infty}\frac{d\tilde{j}_{2}}{i}\,\exp\left(a\tilde{j}_{2}+\frac{b}{2}\tilde{j}_{2}^{2}\right)\propto\exp\left(-\frac{1}{2}a^{2}/b\right)$
may be useful in your calculations. Read off the cumulants $\llangle(x_{1}|x_{2})^{k}\rrangle$
from the cumulant generating function. How does this distribution
change if we had a larger uncertainty $\Sigma\to c\,\Sigma$ with
$c\in\bR$?

\subsection*{c) Pair of coupled spins}

Let us assume we have a pair of coupled Ising spins $s_{1,2}\in\{-1,1\}$.
Compute the partition function 
\begin{align*}
\mathcal{Z}(j_{1},j_{2},J) & =\sum_{s_{1},s_{2}=-1}^{1}\,\exp(Js_{1}s_{2}+j_{1}s_{1}+j_{2}s_{2}),
\end{align*}
Show that it holds that
\begin{align*}
\frac{\partial\mathcal{Z}}{\partial J} & =\frac{\partial^{2}\mathcal{Z}}{\partial j_{1}\partial j_{2}}
\end{align*}
and that for $W=\ln\mathcal{Z}$
\begin{align*}
\frac{\partial W}{\partial J} & =\frac{\partial^{2}W}{\partial j_{1}\partial j_{2}}+\frac{\partial W}{\partial j_{1}}\,\frac{\partial W}{\partial j_{2}}.
\end{align*}
Determine the first and second cumulants of the system for $j_{1}=j_{2}=h$.
Show that the second cumulant $\llangle s_{1}s_{2}\rrangle$ vanishes
in the case the that coupling $J=0$ vanishes between the spins.

Show that we may also obtain the same result by considering a pair
of spins $s_{1}$ and $s_{2}$ within a Gaussian fluctuating field
$h\sim\N(0,J)$, so
\begin{align}
\mathcal{Z} & =\sum_{s_{1},s_{2}=-1}^{1}\,\left\langle \exp\big((s_{1}+s_{2})\,h+j_{1}s_{1}+j_{2}s_{2}\big)\right\rangle _{h\sim\N(0,J)}.\label{eq:gaussian_fluctuating_field}
\end{align}

\subsection*{}

\chapter{Linear regression as Bayesian inference\protect\label{chap:Linear-regression-as-Bayesian-inference}}

The fundamental idea of machine learning, adapting parameters to
model and input-output relationship, goes back to ordinary linear
regression. Instead of dealing with a multitude of different non-linear
activation functions, network layers and parameters, we here first
focus on the problem of a linear setup, where the trainable parameters
consist of a single weight vector. Here we will reformulate the idea
of parameter selection from the point of view of Bayesian inference,
which will provide the starting point to analyze properties of shallow,
deep, and recurrent networks and will give an intuition on how to
relate classical approaches such as Maximum likelihood parameter estimations
to Bayesian statistics.

\section{Basics of supervised learning and generalization}

In machine learning we typically consider different settings where
learning takes place. Those settings usually depend on the task at
hand; all of which require input data sets, that we denote with $X$.
Single examples of this input data, e.g. a single image or measurement
point, are denoted in this manuscript using lowercase letters and
Greek indices $x_{\alpha}$.

In an \textbf{unsupervised setting}, we work solely with the input
data $X$ and without any additional information. A classical example
of this task setting is \textbf{clustering}, where we want to identify
different groups in the set of input data $X$. This is prominently
employed in natural language processing, where you might want to identify
different thematic clusters in texts. Another example might be the
analysis of clusters of participants in social networks. Classical
cluster algorithms are: $k$-means clustering, the principle component
analysis (PCA) or the Tf-idf score for natural language processing.

Here we will not go deeper into these aspects of learning but we rather
focus on learning in a\textbf{ supervised setting}. In addition to
input data we here have a corresponding set of labels or, more generally,
desired outputs $Y$. The labels can be a set of finite and discrete
values or continuous numbers. This supervised setting is closer to
what the reader might have already encountered so far in their studies:
The problem of regression, where you want to obtain a model of the
input-output relation $y(x)$ of a system, is a classical example
of supervised learning. You provide both the input samples $x_{\alpha}$
and the observed outputs $y_{\alpha}$ for some observations and want
to gain some knowledge of the relation between the two. During the
course of this lecture we focus on linear regression, Bayesian inference
and neural networks; all of which deal with learning in supervised
settings.

From a statistical viewpoint, the problem of supervised learning can
be formulated as follows:
\begin{itemize}
\item A core assumption is that the inputs $x$ and the outputs $y$ are
following a joint distribution $p(x,y)$; typically this distribution
is not known, but rather one has a set of $P$ tuples $\D:=\{(x_{\alpha},y_{\alpha})\}_{1\le\alpha\le P}$
drawn from this distribution. Note that this distribution does not
need to be stochastic but can also be deterministic and purely given
through empirical data. 
\item The task is to learn this joint distribution from the given set of
$P$ tuples.
\item Most of the time this is done with help of a \textbf{hypothesis class
$\phi_{\Theta}(x)$}; this is a set of functions of the input $x$
that depend on a set of parameters $\Theta$; in this lecture the
set of functions will be neuronal networks, which, for each provided
input $x$ produce an output $\phi_{\Theta}(x)$ that depends on the
parameters $\Theta$, which are here typically the connections between
the neurons (weights) and other parameters, such as biases.
\item The intermediate goal of learning is to adapt the parameters $\Theta$
such that the output $\phi_{\Theta}(x_{\alpha})$ matches the desired
output $y_{\alpha}$ for all $\alpha$; defining what ``matches''
means translates to defining a measure of distance between the obtained
output and the desired one, often called a \textbf{loss function}
$\cL(y;\phi_{\Theta}(X))$. The result of this step often is a single
value $\hat{\Theta}$ for all parameters (e.g., for standard training
of networks) or a distribution of such parameters in the case of Bayesian
inference (see below).
\item The ultimate goal of learning is to use the function $\phi_{\Theta}(x_{\alpha})$
to make \textbf{predictions }$y_{\ast}=\phi_{\Theta}(x_{\ast})$ for
the value $y_{\ast}$ that corresponds to a hitherto unseen input
$x_{\ast}$; such a task is denoted as generalization. In-distribution
generalization assumes that also the test input $x_{\ast}$ and the
test output $y_{\ast}$ follow the \textbf{same distribution} $p(x_{\ast},y_{\ast})$
that has been used to generate the training data.
\end{itemize}
Even though both the unsupervised and supervised setting differ in
their structure, they encounter similar problems that one needs to
address. A primary concern which appears in both settings is the
issue of \textbf{overfitting}. The intuition behind overfitting is
that the machine learning model takes the data at face-value and may
not be able to generalize the results to unknown data-points. This
is concerning, as we want the models to be flexible and somewhat indicative
of the underlying process. But how can we spot and mitigate this problem?

To spot overfitting we usually split the data that we have at our
hand $\D:=\{(x_{\alpha},y_{\alpha})\}_{1\le\alpha\le P}$ into a training-set
$\mathcal{P_{\mathrm{tr}}}$ and a test-set $\mathcal{P}_{\mathrm{test}}$.
We now train our model exclusively on the training set $\mathcal{P}_{\mathrm{tr}}$
and check the validity of the model on the test set $\mathcal{P}_{\mathrm{test}}$.
It is important, that the sets are distinct and do not share any instances,
$\mathcal{P}_{\mathrm{tr}}\cap\mathcal{P}_{\mathrm{test}}=\emptyset$.
If the performance of the model on the training set is good, whereas
it is bad on the test set, we know that the model overfit the training
set and did not learn the task properly and is not able to generalize.
To measure generalization one for example studies $\langle\cL\rangle_{\mathcal{P}_{\mathrm{test}}}=\sum_{(x,y)\in\mathcal{P}_{\mathrm{test}}}\,\cL(y;\phi_{\hat{\Theta}}(x))$.
There are different ways to mitigate this problem. One popular way
in neural networks is to introduce regularizers such as $L2$-regularizers
(which correspond to weight decay) or drop-out, where some of the
trainable network parameters are fixed for a training step at random
points during the training process.

In this course we further take assumptions made in statistical learning
theory: In order to make theoretical predictions within our frameworks
we make the assumption that all data points $\mathcal{P}$ are \textbf{independently
and identically distributed (i.i.d.)} according to $p(x,y)$. This
assumption is reasonable: Imagine that the task is to classify images
into cats and dogs. If, for some reason, there are also images of
toads present in your data-set this would correspond to a draw from
a different probability distribution, hence violating the ``identically''
in i.i.d. Likewise, for the case of independently distributed data-points
we simply require that there are no spurious correlations in the measurement
process between different samples; a particular example violating
this assumption are repeated data samples. Both assumptions are hence
reasonable and simply correspond to working with a cleaned and well
curated data-set. If you consider practical machine learning applications
a significant portion of time is indeed allocated to curate a clean
data-set before starting the training of a machine learning model. 

\section{Linear regression\protect\label{sec:Linear-regression}}

Consider the problem in a supervised learning setup with $P$ tuples
$\D:=\{(x_{\alpha},y_{\alpha})\}_{1\le\alpha\le P}$, where $x_{\alpha}\in\bR^{d}$
is a data point and $y_{\alpha}\in\mathbb{R}$ is the target, sometimes
also referred as a label. We wish to train a linear model of the data
of the form
\begin{align}
\phi_{w}(x_{\alpha}) & =w^{\T}x_{\alpha},\label{eq:family_lin_models}
\end{align}
where $w\in\bR^{d}$. If there are fewer data points than variables,
$P<d$, the problem is underdetermined. If there are more, $P>d$,
the problem is overdetermined. In both cases, one may formulate instead
an optimization problem, known as \textbf{linear regression},\textbf{
}which seeks to minimize the\textbf{ squared error loss (SE loss)
\begin{align}
\cL(w|\D) & :=\frac{1}{2}\,\sum_{\alpha=1}^{P}\,\big(y_{\alpha}-w^{\T}x_{\alpha}\big)^{2},\label{eq:mse_loss}
\end{align}
}which measures the sum of quadratic deviations. We may wish to minimize
this function to obtain the value $w^{\ast}$ which best fits the
data, so
\begin{align*}
\hat{w} & :=\arg\min_{w}\,\cL(w|\D),
\end{align*}
which leads to the stationarity condition $0\stackrel{!}{=}-\frac{\partial}{\partial w_{i}}\,\cL(w|\D)=\sum_{\alpha=1}^{P}\,\big(y_{\alpha}-w^{\T}x_{\alpha}\big)\,x_{\alpha i}$.
Defining the matrix $X\in\bR^{P\times d}$ as $X=\{x_{\alpha i}\}_{1\le\alpha\le P,1\le i\le d}$
one may write this in the form $0=y^{\T}X-X^{\T}X\,w^{\ast}$ which
is a linear equation for $w^{\ast}$ with the solution
\begin{align}
\hat{w} & =\big(X^{\T}X\big)^{-1}\,X^{\T}y.\label{eq:w_star_main}
\end{align}
The square matrix $X^{\T}X\in\bR^{d\times d}$ may not be invertible;
this happens in particular if the number of data points $P<d$, because
then the rank of the matrix is at most $P$. This problem will be
cured by what is known as regularization, adding a small diagonal
matrix $X^{\T}X+\kappa\I.$ Formally this may be seen as modifying
the loss function \eqref{eq:mse_loss} to read $\cL_{\kappa}(w|\D):=\frac{1}{2}\,\sum_{\alpha=1}^{P}\,\big(y_{\alpha}-w^{\T}x_{\alpha}\big)^{2}+\frac{1}{2}\kappa\,w^{\T}w$
which can be considered a term that favors solutions with small L2-norm
$\|w\|=w^{\T}w$. The limit $\kappa\searrow0$ yields the Moore-Penrose
pseudoinverse.

The \textbf{linear predictor} for a new test point $x^{\ast}$ is
then obtained as
\begin{align}
y^{\ast} & =w^{\ast\T}x^{\ast}\label{eq:lin_reg_pred}\\
 & =\big[\big(X^{\T}X\big)^{-1}\,X^{\T}y\big]^{\T}x^{\ast}\nonumber \\
 & =y^{\T}X\,\big(X^{\T}X\,\big)^{-1}\,x^{\ast}.\nonumber 
\end{align}
This predictor is linear in $x^{\ast}$, due to the linearity of the
model, but also linear in $y$. This calculation simply utilizes that
the output is a linear function of the network parameter, which allows
one to utilize the same equations for setups where the non-linear
transformations $\phi$ is applied to the data as $y=w^{\top}\phi(x)$.

\section{Bayesian formulation\protect\label{sec:Bayesian-formulation}}

We will now perform a reformulation of the problem of linear regression
in the framework of Bayesian inference. The idea of Bayesian inference
consists of two steps. First, one proposes a \textbf{family of models
}which is given by a probability distribution over models, called
the \textbf{prior}. Then one uses Bayes theorem \eqref{eq:Bayes}
for conditional distributions to compute the distribution of models
that one obtains when conditioning on the presented training data.
This approach is general and can be applied to any hypothesis class
$\phi_{\Theta}(x)$; it is easiest illustrated on the example of a
linear model. To this end, consider the family of linear models, parameterized
by $w$ given by \eqref{eq:family_lin_models}.

Now choose a prior distribution for the linear weights 
\begin{align}
w_{i} & \stackrel{\text{i.i.d.}}{\sim}\N(0,g\,d^{-1})\,,\label{eq:prior_w}
\end{align}
so all $w_{i}$ follow the same distribution and are independent of
one another. As we will see, both linear regression and Bayesian
inference share some similarities but let us focus on the key differences:
The aim of (linear) regression is to obtain a single value for each
of the parameters; in the example of linear regression the single
value $w^{\ast}$. One finds this value by demanding that the input-output
relation of the training dataset $\mathcal{P}$ is fullfilled as accurately
as possible. In short: Linear regression provides a point estimate
for all parameters. 

Opposed to this, the Bayesian approach follows a different, probabilistic,
reasoning: The set of network outputs on the training data $\{f_{\alpha}\}_{1\le\alpha\le P}$
produced by the linear model \eqref{eq:family_lin_models} with $f_{\alpha}:=\phi_{w}(x_{\alpha})$
may formally be written as a conditional probability distribution
(cf. \prettyref{sec:Transformation-of-random}) 
\begin{align}
p(f|X,w) & =\prod_{\alpha=1}^{P}\,\delta\big[f_{\alpha}-\phi_{w}(x_{\alpha})\big].\label{eq:conditional_output_given_w}
\end{align}
Here the product over all data points $\alpha$ appears, because we
want to enforce the same input-output relation \eqref{eq:family_lin_models}
for each data point $\alpha$, so that output $f_{\alpha}$ corresponds
to input $x_{\alpha}$ using the very same value $w$ for the weights
for all these points. The prior distribution of parameters \eqref{eq:prior_w},
by Bayes' law \eqref{eq:Bayes}, induces a joint distribution for
the set of outputs $\{y_{\alpha}\}_{1\le\alpha\le P}$ and the weights
via the chain rule of probabilities
\begin{align}
p(y,w|X) & =p(y|X,w)\,p(w).\label{eq:joint_dist_y_w}
\end{align}
The idea of Bayesian inference is to determine the \textbf{posterior
distribution} \textbf{for the weights} on the left hand side below
, which is the distribution of the $w$ given one fixes the outputs
$y$ to the known values of the training set, namely by Bayes' law
\eqref{eq:Bayes}
\begin{align}
p(w|y,X) & =\frac{p(y,w|X)}{p(y|X)}.\label{eq:weight_posterior}
\end{align}
Here $p(y|X)$ on the right hand side is the marginalization of \eqref{eq:joint_dist_y_w}
over $w$, namely $p(y|X)=\int dw\,p(y,w|X)$ such that the last expression
is properly normalized. Intuitively this means that we only allow
those parameters $w$ which correctly produce the output $y$, because
$p(y|X,w)$ vanishes for all other parameters. One may then use the
conditional distribution \eqref{eq:conditional_output_given_w} and
the posterior for the weights \eqref{eq:weight_posterior} to obtain
the \textbf{predictive distribution} for the output corresponding
to a new (so far unseen) data point $x_{\ast}$, namely
\begin{align}
p(y_{\ast}|y,X,x_{\ast}) & =\int dw\,p(y_{\ast}|x_{\ast},w)\,p(w|y,X).\label{eq:posterior_output}
\end{align}
One hence obtains the \textbf{posterior distribution for the output
$y_{\ast}$}. To illustrate the logic of Bayesian inference, we here
went the following way: 
\begin{enumerate}
\item prior for weights $w$ \eqref{eq:prior_w}
\item condition on training outputs $y$ to obtain posterior for weights
\eqref{eq:weight_posterior}
\item use posterior for weights to obtain posterior for the output \eqref{eq:posterior_output}
for a new input to make a prediction
\end{enumerate}
We will now investigate a shortcut that omits the intermediate step
of computing the weight posterior. This shortcut can be appreciated
by inserting \eqref{eq:joint_dist_y_w} and \eqref{eq:weight_posterior}
into \eqref{eq:posterior_output} with the result
\begin{align}
p(y_{\ast}|y,X,x_{\ast}) & =\frac{\int dw\,p(y|X,w)\,p(y_{\ast}|x_{\ast},w)\,p(w)}{p(y|X)}\label{eq:posterior_y_star_inserted}\\
 & =\frac{p(y,y_{\ast}|X,x_{\ast})}{p(y|X)},\nonumber 
\end{align}
where in the second step we used that the numerator is the joint distribution
of the training outputs $y_{1\le\alpha\le P}$ and the test output
$y_{\ast}$ under the prior distribution for the weights $w$; the
numerator can hence be considered as the \textbf{prior distribution
for the outputs}. This prior distribution is here directly conditioned
onto the training data, which is given by Bayes formula for a conditional
probability distribution. We have hence circumvented the intermedate
step of computing the posterior for the weights and instead directly
obtained the posterior for the test output $y_{\ast}$. The denominator
in \eqref{eq:posterior_y_star_inserted}, called the \textbf{model
evidence} because it quantifies how likely the training data $y$
came from the model, does not depend on $y_{\ast}$, so it does not
change the shape of the distribution for $y_{\ast}$; it only affects
its normalization. The interesting structure of the posterior for
$y_{\ast}$ must hence be contained in the numerator of \eqref{eq:posterior_y_star_inserted},
corresponding to the joint distribution \eqref{eq:joint_dist_y_y_star}. 

Considering the test input $x_{\ast}$ and the test output $y_{\ast}$
as the sample $\alpha=P+1$, the numerator in \eqref{eq:posterior_y_star_inserted}
may be written with \eqref{eq:conditional_output_given_w} as

\begin{align}
p(y,y_{\ast}|X,x_{\ast}) & =\Big\langle\prod_{\alpha=1}^{P+1}\,\delta\big[y_{\alpha}-\phi_{w}(x_{\alpha})\big]\Big\rangle_{w_{i}\stackrel{\text{i.i.d.}}{\sim}\N(0,gd^{-1})}\label{eq:joint_dist_y_y_star}\\
 & =\Big\langle\prod_{\alpha=1}^{P+1}\,\delta\big[y_{\alpha}-w^{\T}x_{\alpha}\big]\Big\rangle_{w_{i}\stackrel{\text{i.i.d.}}{\sim}\N(0,gd^{-1})}.\nonumber 
\end{align}
Due to the linearity of the output $\phi_{w}$ in $w$, for the linear
model this is a joint Gaussian distribution on the set of $\{y_{\alpha}\}_{1\le\alpha\le P+1}$.
Note that the $X$ are given and fixed here and the randomness is
entirely caused by the random weights $w_{i}$. In the following we
will write $\bar{y}:=(y,y_{\ast})$ and $\bar{X}=(X,x_{\ast})$ for
the outputs and inputs jointly for all training points and the one
test point.

We know by the linear appearance of $w_{i}$ that the distribution
$p(\bar{y}|\bar{X})$ is Gaussian. We here introduce a helpful technique
we will use throughout to see this explicitly which in particular
allows us to conveniently compute the conditional \prettyref{eq:posterior_y_star_inserted}.
To this end one uses the Fourier transform of the Dirac distribution
\begin{align*}
\cF[\delta](\omega) & =\int\delta(y)\,e^{-i\omega y}\,dy=1\\
\delta(y) & =\frac{1}{2\pi}\,\int_{-\infty}^{\infty}\,e^{i\omega y}\,1\,d\omega\\
 & =\int_{-i\infty}^{i\infty}\frac{d\ty}{2\pi i}\,e^{\ty y}
\end{align*}
to express the Dirac distribution with $\delta(y)=\delta(-y)$ in
\eqref{eq:joint_dist_y_y_star} as
\begin{align}
p(\bar{y}|\bar{X}) & =\Big\langle\prod_{\alpha=1}^{P+1}\int_{-i\infty}^{i\infty}\frac{d\ty_{\alpha}}{2\pi i}\,\exp\big(\ty_{\alpha}\,\big[w^{\T}x_{\alpha}-y_{\alpha}\big]\big)\Big\rangle_{w_{i}\stackrel{\text{i.i.d.}}{\sim}\N(0,gd^{-1})}\label{eq:p_y_X_Dirac}\\
 & =\int\D\ty\,\exp\big(-\sum_{\alpha=1}^{P+1}\,\ty_{\alpha}\,y_{\alpha}\big)\,\Big\langle\exp\big(\sum_{\alpha=1}^{P+1}\ty_{\alpha}\,w^{\T}x_{\alpha}\big)\Big\rangle_{w_{i}\stackrel{\text{i.i.d.}}{\sim}\N(0,gd^{-1})},\nonumber 
\end{align}
where 
\begin{align}
\int\D\ty & =\prod_{\alpha=1}^{P+1}\int_{-i\infty}^{i\infty}\frac{d\ty_{\alpha}}{2\pi i}.\label{eq:measure_dtily}
\end{align}
The latter expectation value takes the form of the moment-generating
function for the independently drawn Gaussian weights $w_{i}$ with
a source $j_{i}=\sum_{\alpha=1}^{P+1}\,\ty_{\alpha}x_{\alpha i}$
\begin{align*}
\Big\langle\exp\big(\ty_{\alpha}\,w^{\T}x_{\alpha}\big)\Big\rangle_{w_{i}\stackrel{\text{i.i.d.}}{\sim}\N(0,gd^{-1})} & =\prod_{i=1}^{d}\,\Big\langle\exp\big(w_{i}\,\sum_{\alpha=1}^{P+1}\ty_{\alpha}x_{\alpha i}\big)\Big\rangle_{w_{i}\stackrel{\text{i.i.d.}}{\sim}\N(0,gd^{-1})}\\
 & =\prod_{i=1}^{d}\,\exp\Big(\frac{g}{2d}\,\Big[\sum_{\alpha=1}^{P+1}\ty_{\alpha}x_{\alpha i}\Big]^{2}\Big)\\
 & =\exp\Big(\frac{1}{2}\sum_{\alpha,\beta=1}^{P+1}\ty_{\alpha}\ty_{\beta}\,\frac{g}{d}\,\sum_{i=1}^{d}x_{\alpha i}x_{\beta i}\Big),
\end{align*}
where we used that the cumulant-generating function of the Gaussian
$w_{i}\sim\N(0,g/d)$ by \prettyref{eq:W_Gauss} is $W_{w_{i}}(j_{i})=\frac{g}{2d}j_{i}^{2}$.
Inserted back into \eqref{eq:p_y_X_Dirac} one has
\begin{align}
p(\bar{y}|\bar{X}) & =\int\D\ty\,\exp\big(-\ty^{\T}y+W_{y}(\ty)\big),\label{eq:p_y_X_final}
\end{align}
where 
\begin{align}
W_{y}(\ty) & =\frac{1}{2}\,\ty^{\T}C^{(xx)}\ty,\label{eq:cum_gen_prior}\\
C^{(xx)} & =\frac{g}{d}\,\bar{X}\bar{X}^{\T}\in\bR^{(P+1)\times(P+1)}.\nonumber 
\end{align}
The covariance matrix $C^{(xx)}$ is obviously symmetric, as it has
to be (see \prettyref{sec:Gaussian-distribution}). Note that the
data here only enters in the form of an inner product across indices
$i$, which is also called a \textbf{dot-product kernel}. In linear
regression, we had the matrix $X^{\T}X$, so an inner product over
the sample indices $\alpha$ instead. In the case of non-linear regression
$y=w^{\T}\phi(x)$, instead, $\phi(X)^{\T}\phi(X)$ would appear.

\begin{figure}
\begin{centering}
\includegraphics[width=0.8\textwidth]{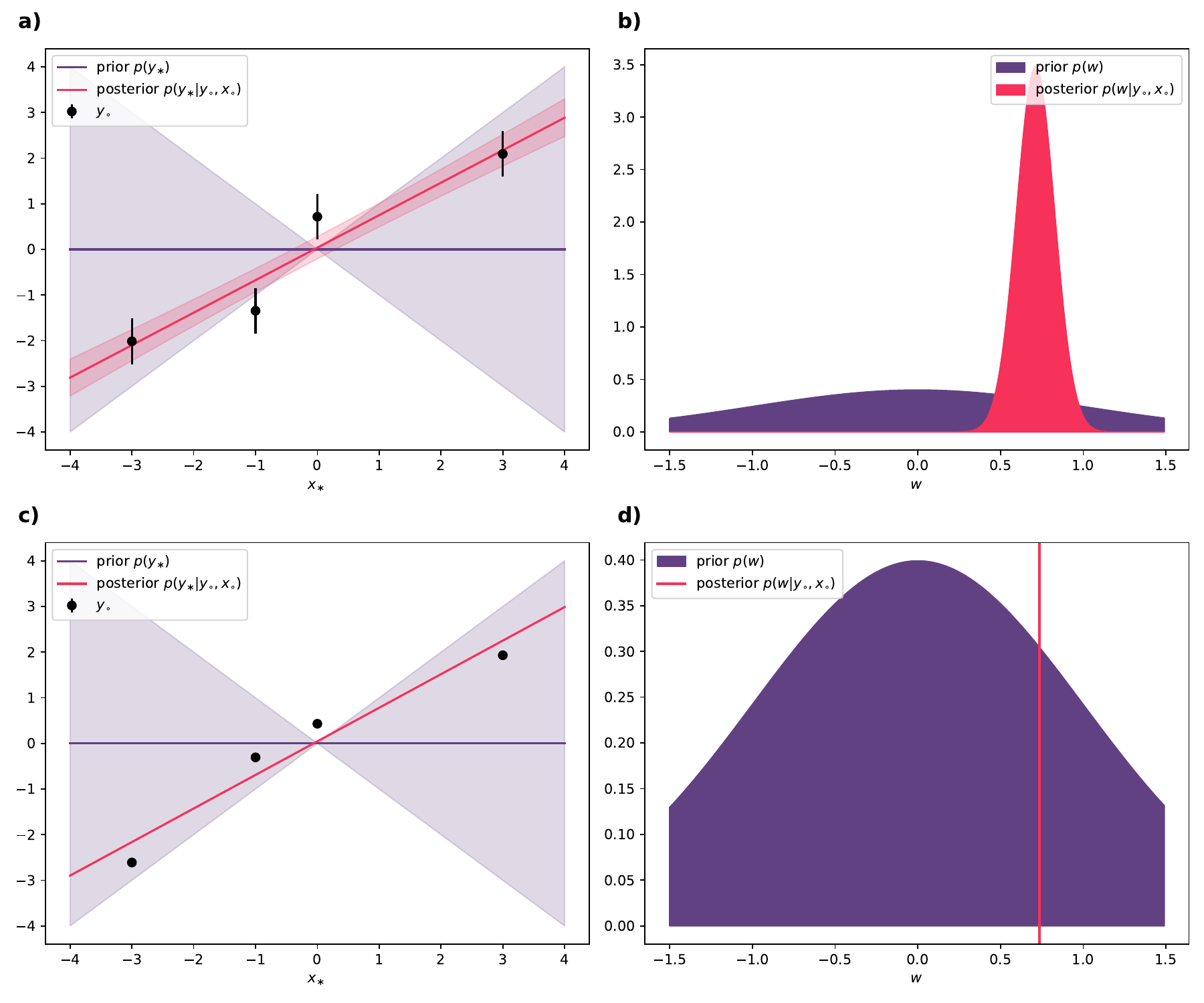}
\par\end{centering}
\caption{\textbf{Linear regression in Bayesian framework.} Comparison between
prior and posterior distributions for the linear model. Here the output
$p(y|X,w)$ of the linear regression is assumed to be stochastic with
a Gaussian regularization noise, namely instead of \eqref{eq:conditional_output_given_w}
we here use $p(y|X,w)=\protect\N(y|w^{\protect\T}x,\kappa\,\protect\I)$
which corresponds to adding Gaussian noise $\xi_{\alpha}\stackrel{\text{i.i.d.}}{\sim}\protect\N(0,\kappa)$,
i.e. $y_{\alpha}\to y_{\alpha}+\xi_{\alpha}$; this is often done
as a means of regularization: it forces the outputs to be close to
the training points, but allows for some wiggle room.\textbf{ a)}
Prior and posterior of labels $y_{\ast}$ shown as mean and standard
deviation from \eqref{eq:posterior_Gaussi}. The posterior is obtained
by conditioning on the training labels $y_{\circ}$. \textbf{b)} Prior
and posterior distributions of the slope of the linear model. \textbf{c)}
Same as \textbf{a)} but for zero noise ($\kappa=0$). \textbf{d)}
Same as \textbf{b)} but for zero noise ($\kappa=0$). (Adapted from
Bachelor thesis by Bastian Epping, 2020.)\protect\label{fig:Prior-vs-posterior-Bastian-inspired}}
\end{figure}

\prettyref{eq:p_y_X_final} is the Fourier representation of the distribution
for $y$ (cf. \prettyref{eq:Fourier_W_Z_p}) and \eqref{eq:cum_gen_prior}
is a quadratic polynomial, so it describes a zero mean Gaussian distribution
with covariance given by $C^{(xx)}$, so
\begin{align}
\{y_{\alpha}\} & \sim\N(0,C^{(xx)}).\label{eq:y_Gauss_BI}
\end{align}
 To obtain the form of the posterior for the test output \eqref{eq:posterior_y_star_inserted}
it is sufficient to consider the numerator explicitly, because the
denominator only supplies the correct normalization, which we may
fix post hoc. This conditional distribution is obtained most easily
from the Fourier representation \eqref{eq:p_y_X_final} by inserting
the set of training outputs $y_{\circ}:=y$ (using the subscript $\circ$
here to denote all sample indices $\circ=\{\alpha\in\mathbb{d}:1\le\alpha\le P\}$)
\begin{align*}
p(y_{\circ},y_{\ast}|X,x_{\ast}) & =\int\D(\ty_{\circ},\ty_{\ast})\,\exp\big(-\ty_{\circ}^{\T}y_{\circ}-\ty_{\ast}y_{\ast}+\frac{1}{2}\,(\ty_{\circ},\ty_{\ast})^{\T}C^{(xx)}(\ty_{\circ},\ty_{\ast})\big)\\
 & =\int\D\ty_{\ast}\,\exp\big(-\ty_{\ast}y_{\ast}+\frac{1}{2}\,\ty_{\ast}C_{\ast\ast}^{(xx)}\ty_{\ast}\big)\\
 & \phantom{=}\times\int\D\ty_{\circ}\,\exp\big(\big[\ty_{\ast}C_{\ast\circ}^{(xx)}-y_{\circ}\big]^{\T}\ty_{\circ}+\frac{1}{2}\,\ty_{\circ}^{\T}C_{\circ\circ}^{(xx)}\ty_{\circ}\big),
\end{align*}
where we split the matrix $C^{(xx)}=\left[\begin{array}{cc}
C_{\circ\circ}^{(xx)} & C_{\circ\ast}^{(xx)}\\
C_{\ast\circ}^{(xx)} & C_{\ast\ast}^{(xx)}
\end{array}\right]$ into four blocks and used the symmetry $C_{\ast\circ}^{(xx)}=C_{\circ\ast}^{(xx)}$.
Performing the integration over $\ty_{\circ}$ only affects the latter
line, using the Gaussian identity \eqref{eq:Z_Gauss} to obtain
\begin{align}
p(y_{\circ},y_{\ast}|X,x_{\ast}) & \propto\int\D\ty_{\ast}\,\exp\big(-\ty_{\ast}y_{\ast}+W_{\ast}(\ty_{\ast})\big),\label{eq:conditioning_Gauss_Fourier}\\
W_{\ast}(\ty_{\ast}) & =\frac{1}{2}\,\ty_{\ast}C_{\ast\ast}^{(xx)}\ty_{\ast}-\frac{1}{2}\,\big[\ty_{\ast}C_{\ast\circ}^{(xx)}-y_{\circ}\big]^{\T}\,[C_{\circ\circ}^{(xx)}]^{-1}\,\big[C_{\circ\ast}^{(xx)}\ty_{\ast}-y_{\circ}\big]\nonumber \\
 & =\ty_{\ast}C_{\ast\circ}^{(xx)}[C_{\circ\circ}^{(xx)}]^{-1}y_{\circ}+\frac{1}{2}\,\ty_{\ast}\big[C_{\ast\ast}^{(xx)}-C_{\ast\circ}^{(xx)}[C_{\circ\circ}^{(xx)}]^{-1}C_{\circ\ast}^{(xx)}\big]\,\ty_{\ast}+\mathrm{const.},\nonumber 
\end{align}
where we dropped terms independent of $\ty$, because they only affect
the normalization. The form \eqref{eq:conditioning_Gauss_Fourier}
is the Fourier representation of a Gaussian distribution in $y_{\ast}$
with a mean given by the linear coefficient of $\ty$ and a covariance
given by the quadratic coefficient, so that one reads off
\begin{align}
y_{\ast} & \sim\N\Big(C_{\ast\circ}^{(xx)}[C_{\circ\circ}^{(xx)}]^{-1}y_{\circ},\label{eq:posterior_Gaussi}\\
 & \phantom{\sim\N\Big(}C_{\ast\ast}^{(xx)}-C_{\ast\circ}^{(xx)}[C_{\circ\circ}^{(xx)}]^{-1}C_{\circ\ast}^{(xx)}\Big).\nonumber 
\end{align}
The meaning of this distribution is the distribution for the output
$y_{\ast}$ if one conditions the distribution of models on the presented
training data. The mean value can be considered the mean output, the
variance represents a form of uncertainty which is still present despite
the presented training data. The knowledge of the training data and
its labels leads to a mean-predictor $\langle y_{\ast}\rangle$ which
is in general non-zero, in contrast to the prior in \eqref{eq:y_Gauss_BI}.
The variance consists of the initial uncertainty of the prior and
a term which reduces this initial uncertainty due to the conditioning.
The change of the distribution of the weights and the predicted labels
by conditioning on training data is displayed in \prettyref{fig:Prior-vs-posterior-Bastian-inspired}.

\subsection{Bias-variance decomposition}

The Gaussian distribution of the output for the test point allows
us to measure how well the test point matches a potentially given
ground truth $y_{\ast}^{0}$. If we use the squared error loss, we
may be interested in the mean of this loss under the distribution
of outputs \prettyref{eq:posterior_Gaussi}, whose two cumulants we
here denote as $\mu$ and $\Sigma$, so $y_{\ast}\sim\N(\mu,\Sigma)$
\begin{align}
\langle\cL(y_{\ast},y_{\ast}^{0})\rangle & =\frac{1}{2}\langle(y_{\ast}-y_{\ast}^{0})^{2}\rangle_{y_{\ast}\sim\N(\mu,\Sigma)}\label{eq:bias_variance_decomp}\\
 & =\frac{1}{2}\langle(\mu+\delta y_{\ast}-y_{\ast}^{0})^{2}\rangle_{\delta y_{\ast}\sim\N(0,\Sigma)}\nonumber \\
 & =\frac{1}{2}(\mu-y_{\ast}^{0})^{2}+(\mu-y_{\ast}^{0})\,\underbrace{\langle\delta y_{\ast}\rangle}_{=0}+\frac{1}{2}\underbrace{\langle\delta y_{\ast}^{2}\rangle}_{\Sigma}\nonumber \\
 & =\frac{1}{2}(\mu-y_{\ast}^{0})^{2}+\frac{1}{2}\Sigma,\nonumber 
\end{align}
which is called a \textbf{bias-variance decomposition}. The bias describes
the average output of the ensemble of posterior networks, the variance
the variability of these outputs around this average. Both terms enter
the expected loss; the mean output enters in relation to the desired
output, $y_{\ast}^{0}$.

\subsection{Connection to linear regression}

To make connection to the result from linear regression, \eqref{eq:lin_reg_pred},
we may compare it to the mean of the predictor in \eqref{eq:posterior_Gaussi}
written as (writing again $y\equiv y_{\circ}$ here)
\begin{align*}
\langle y_{\ast}\rangle & =C_{\ast\circ}^{(xx)}[C_{\circ\circ}^{(xx)}]^{-1}y\\
 & \stackrel{(\ref{eq:cum_gen_prior})}{=}x_{\ast}^{\T}\,X^{\T}\,\big(XX^{\T}\big)^{-1}\,y,
\end{align*}
while the predictor for linear regression is
\begin{align*}
y_{\ast} & \stackrel{(\ref{eq:posterior_Gaussi})}{=}y^{\T}X\,\big(X^{\T}X\big)^{-1}x_{\ast}\\
 & =x_{\ast}^{\T}\,\big(X^{\T}X\big)^{-1}X^{\T}\,y.
\end{align*}
The two expressions are in fact identical because by the associativity
of matrix multiplication
\begin{align*}
(X^{\T}\,X)\,X^{\T} & =X^{\T}\,(X\,X^{\T}),
\end{align*}
from which follows by multiplying with $(X^{\T}X)^{-1}$ from left
and by $(XX^{\T})^{-1}$  from right
\begin{align*}
X^{\T}\,(XX^{\T})^{-1} & =(X^{\T}X)^{-1}\,X^{\T},
\end{align*}
showing that the two expressions for the linear predictor and the
mean of the predictive distribution agree. So linear regression yields
the mean predictor of Bayesian inference, while the latter, in addition,
yields the variance that can be used to quantify the uncertainty of
the prediction.

\section{Keypoints}

In summary

\noindent\fcolorbox{black}{white}{\begin{minipage}[t]{1\textwidth - 2\fboxsep - 2\fboxrule}%
\textbf{Linear regression as Bayesian inference}
\begin{itemize}
\item Linear regression is a type of supervised learning.
\item Bayesian inference in a linear regression setup produces Gaussian
processes with dot-product kernels.
\item The results of linear regression for inferred labels matches the mean
prediction of the Gaussian process in Bayesian approach. Linear regression
provides a point estimate for model parameters whereas Bayesian inference
produces a posterior distribution of parameters.
\item Supervised learning in a Bayesian setting corresponds to computing
the probability of inferred network outputs at the inference points
conditioned on the training data.
\item One can reframe the problem of supervised learning in Bayesian setting
in a field theoretic language by enforcing the regression model using
Dirac-constraints and by introducing conjugate fields $\tilde{y}$.
\item Regularization mitigates ill-conditioned matrices and can be introduced
by observation noise or by constraining the weight norm using L2 regularizers
(in the exercises)
\end{itemize}
\end{minipage}}

\selectlanguage{english}%

\section{Exercises}

\subsection*{Exercise a) Addition of independent variables (2p)}

Show that the cumulant-generating function of a sum $z=x+y$ of two
independently distributed multivariate variables $x\in\bR^{N}$ and
$y\in\bR^{N}$ decomposes into a sum itself:
\begin{align*}
W_{z}(j) & =W_{x}(j)+W_{y}(j).
\end{align*}
Apply this result to a pair of Gaussian variables $x\sim\N(\mu_{x},\Sigma_{x})$
and $y\sim\N(\mu_{y},\Sigma_{y})$ to show that the sum is again a
Gaussian with distribution
\begin{align*}
z & \sim\N(\mu_{x}+\mu_{y},\Sigma_{x}+\Sigma_{y}).
\end{align*}

\subsection*{Exercise b) Useful derivatives (4p)}
\begin{enumerate}
\item Show
\begin{equation}
\frac{\partial\ln\left(\det\left(C\right)\right)}{\partial C_{ij}}=C_{ij}^{-1},
\end{equation}
for $\det\left(C\right)>0$. Hint: Express $\det\left(C\right)$ through
a Gaussian integral using the normalization condition of the Gaussian
\begin{align*}
1 & =\frac{1}{\sqrt{2\pi}^{D}\det\left(C^{-1}\right)^{1/2}}\int dx\,\exp\left(-\frac{1}{2}x^{\T}Cx\right).
\end{align*}
 (2p)
\item Compute the derivative
\begin{equation}
\frac{\partial C^{-1}}{\partial C_{ij}}
\end{equation}
Hint: take the derivative of $\I=CC^{-1}$ (2p)
\end{enumerate}

\subsection*{Exercise c) Linear regression and Bayesian inference (9p)}

As we have seen in the main text there are inherent similarities
but also important differences between Bayesian inference and linear
regression, both of which we will explore a bit further in this exercise.
First we want to understand what happens to the inferred label mean
and variance

\begin{align}
\langle y_{*}\rangle & =C_{\ast\circ}^{(xx)}[C_{\circ\circ}^{(xx)}]^{-1}y_{\circ}\\
\langle\langle y_{*}^{2}\rangle\rangle & =C_{\ast\ast}^{(xx)}-C_{\ast\circ}^{(xx)}[C_{\circ\circ}^{(xx)}]^{-1}C_{\circ\ast}^{(xx)}
\end{align}
if the test point $*$ is part of the training data-set $\circ$ ?
Does the result match your intuition? (2p)We now want to understand
what happens, when we make the labels stochastic; alternatively, this
can be seen as adding random noise on-top of our network output. Consider
the following construction for labels $y_{\alpha}$ according to a
simple linear function

\begin{align}
f_{\alpha} & :=w^{\T}x_{\alpha},\label{eq:LinearBayesian_Model1_Regression}\\
y_{\alpha} & =f_{\alpha}+\xi_{\alpha},\\
\xi_{\alpha} & \stackrel{\text{i.i.d. over }\alpha}{\sim}\mathcal{N}(0,\kappa),
\end{align}
where we assume that both the scalar inputs $x_{\alpha}$ and the
scalar noise $\xi_{\alpha}\sim\N(0,\kappa)$ are both i.i.d. and we
assume a prior distribution on $w_{i}\stackrel{\text{i.i.d.}}{\sim}\mathcal{N}(0,g\,d^{-1})$.
The data is given by pairs of inputs $x$ and labels $z$ as $\D:=\{(x_{\alpha},z_{\alpha})_{1\le\alpha\le P}$.
Show that we can write the distribution $p(y,f\vert X)$ as

\begin{align}
p(y,f\vert X) & =\N(y|f,\kappa)\,\int\D\tilde{f}\,\big\langle\exp\big(\sum_{\alpha}\,\tilde{f}_{\alpha}\big[f_{\alpha}-w^{\T}x_{\alpha}\big]\,\big)\big\rangle_{w},\label{eq:p_y_z_joint_exc}\\
\text{where}\nonumber \\
\int\mathcal{D}\tilde{f} & =\prod_{\alpha}\int_{-i\infty}^{i\infty}\frac{1}{2\pi i}d\tilde{f}_{\alpha}\,,\\
\N(\xi|\mu,\kappa) & :=\prod_{\alpha=1}^{D}\,\N(\xi_{\alpha}|\mu_{\alpha},\kappa)\,,\\
\N(\xi_{\alpha}|\mu_{\alpha},\kappa) & :=\frac{1}{\sqrt{2\pi\kappa}}\,e^{-\frac{(\xi_{\alpha}-\mu_{\alpha})^{2}}{2\kappa}}.
\end{align}
(2p). Start by enforcing the relations $f_{\alpha}=w^{\T}x_{\alpha}$
and $y_{\alpha}=f_{\alpha}+\xi_{\alpha}$ using Dirac-delta constraints
and average over the noise $\xi$ in addition to the weights $w$.
Following the arguments in the main text (1p) show, that the distribution
of $y$, given by $p(y|X)=\int df\,p(y,f|X)$, is a Gaussian with
covariance (kernel)

\begin{equation}
C_{\alpha\beta}^{(zz)}=\frac{g}{d}\,x_{\alpha}^{\T}x_{\beta}+\kappa\,\delta_{\alpha\beta}.\label{eq:DotProductKernel}
\end{equation}

Why could the additional term on the diagonal of \eqref{eq:DotProductKernel}
become relevant in numerical implementation? Hint: Consider the formula
for the inferred label (0.5p). Consider now setting with regularization
$\kappa>0$: What are the mean and the variance of the inferred label
if the test point $*$ is now of the training data-set $\circ$ ?

We want to connect this result to linear regression: To this end marginalize
the distribution \eqref{eq:p_y_z_joint_exc} as

\begin{align*}
p(y\vert X) & =\int df\,p(y,f|X).
\end{align*}
Perform the integral over $y$ and write the result as $p(y|X)=\int dw\,\exp\big(S(w)\big)$.
Then compute the $w$ which maximizes the exponent $S$ (2p) and compare
to the result \eqref{eq:w_star_main}. What happens, if instead
of \eqref{eq:LinearBayesian_Model1_Regression} we take a model of
the structure

\begin{equation}
y_{\alpha}=w^{\T}\phi(x_{\alpha})+\xi_{\alpha}\,.
\end{equation}
What changes and what is the new value for $w$ which maximizes the
exponent (0.5p)?

\subsection*{Exercise d) Bayesian regression; Numerical exercise (4p)}

For the linear model \eqref{eq:LinearBayesian_Model1_Regression}
we provided some source code to make you familiar with the numerical
implementation of the concept. The program contains gaps with the
note ``TODO'' where you will need to add code. In the program you
will
\begin{enumerate}
\item Implement the computation of the mean and the variance of Bayesian
inference using a Gaussian process with a kernel from a linear function
such as \eqref{eq:DotProductKernel}. Do you need explicit regularization?
If yes how big should the regularizer be? What happens if you make
the regularization very large? (1.5p)
\item Implement sampling from the posterior of the Bayesian inference and
plot samples from it (0.5p)
\end{enumerate}
The kernels produced by a linear model are so called dot-product kernels.
Nevertheless there are also different versions of kernels like the
Radial Basis Function kernel (or short RBF kernel). In the program
you will also implement the RBF kernel with the correlation length
named $\zeta$

\begin{equation}
K_{\alpha\beta}(\zeta)=\exp(-\frac{1}{2\zeta^{2}}\vert x_{\alpha}-x_{\beta}\vert^{2})\,.
\end{equation}
 What are similarities and what are differences between the results?
Investigate what happens in the following cases:
\begin{enumerate}
\item Small amount of training data and large amount of training data (0.5p).
\item What happens when you increase or decrease the correlation length
$\zeta$ (0.5p)?
\item What happens when you increase or decrease the regularization parameter
? Are there differences compared to the dot-product kernel ? What
happens if $\zeta=0$. (1p)
\end{enumerate}
Produce appropriate plots to answer the questions.\selectlanguage{american}%

\chapter{The law of large numbers – large deviation principle\protect\label{chap:The-law-of-large-number}}

In this section we visit an important principle that helps us to
describe systems with large numbers of degrees of freedom – the law
of large numbers and its formal underpinning, large deviation theory.
We will see that the former describes small deviations of a random
quantity away from its typical value, while the latter also captures
deviations far away from the expected value. We will subsequently
use these results to quantify properties of networks with large numbers
of neurons.

\section{G\"artner-Ellis theorem}

We here follow \citep[Appendic C]{Touchette09}. Consider a sequence
of random variables $(x_{1},\ldots,x_{N})\sim p$, which are jointly
distributed with a certain distribution $p$. Assume we are interested
in a random variable $S_{N}(x_{1},\ldots,x_{N})=S_{N}(x)$. For example,
we may want to known the sample mean $S_{N}(x)=N^{-1}\,\sum_{i=1}^{N}x_{i}$.
In particular, we would like to know the distribution of $S_{N}$
in the limit of large $N$. This distribution is, with \prettyref{sec:Transformation-of-random},
\begin{align}
p_{N}(s) & =\big\langle\delta\big[s-S_{N}(x)\big]\big\rangle_{x\sim p}.\label{eq:p_N_delta}
\end{align}
We will now employ the Laplace representation of the Dirac distribution
\begin{align}
\delta(x) & =\int_{a-i\infty}^{a+i\infty}\,\frac{dj}{2\pi i}\,e^{jx},\label{eq:Laplace_delta}
\end{align}
where $a\in\bR$ is arbitrary. The integration contour is called a
\textbf{Bromwich contour}. The integration runs parallel to the imaginary
axis. One can show that this representation is correct by considering
how the Dirac distribution acts on a test function $f$
\begin{align*}
 & \int_{-\infty}^{\infty}\,dx\,f(x)\,\int_{a-i\infty}^{a+i\infty}\,\frac{dj}{2\pi i}\,e^{jx}\\
= & \int_{-\infty}^{\infty}dx\,e^{ax}\,f(x)\,\underbrace{\int_{-i\infty}^{i\infty}\,\frac{dj}{2\pi i}\,e^{jx}}_{\delta(x)}=f(0),
\end{align*}
where we used the Fourier representation of the Dirac distribution
$\delta(x)=\int_{-i\infty}^{i\infty}\,\frac{dj}{2\pi i}\,e^{jx}=1$
and note that the result is independent of $a$, the position where
the Bromwich contour intersects the real axis.

Using the representation \prettyref{eq:Laplace_delta} in \prettyref{eq:p_N_delta}
one has
\begin{align}
p_{N}(s) & =\int_{a-i\infty}^{a+i\infty}\,\frac{dj}{2\pi i}\,\big\langle\exp\big(j\,(S_{N}(x)-s)\big)\big\rangle_{x}\nonumber \\
 & =\int_{a-i\infty}^{a+i\infty}\,\frac{dj}{2\pi i}\,\exp\big(-j\,s+W_{N}(j)),\label{eq:p_N_Laplace}
\end{align}
where $W_{N}(j)=\ln\,\big\langle\exp\big(j\,S_{N}(x)\big)\big\rangle_{x}$
is the cumulant-generating function of $S_{N}$. In the case that
the $x_{i}$ were independent, one would obtain $W_{N}\propto N$
due to the $N$ independent variables (see exercises). In the general
case this motivates the definition of the \textbf{scaled cumulant-generating
function} as
\begin{align*}
\lambda_{N}(k) & :=N^{-1}W_{N}(N\,k),
\end{align*}
where the factor $N$ in the argument can be thought of in the example
of the sample mean $S_{N}(x)=N^{-1}\sum_{i=1}^{N}x_{i}$ to compensate
the prefactor $N^{-1}$: In this example, we would have $W_{N}(j)=N\,W_{1}(j/N)$,
where $W_{1}(j)$ is the cumulant-generating function of a single
$x_{i}$ (see also example below); so in this case $\lambda_{N}(k)=W_{1}(k)$,
which is independent of $N$.

In the general case that the limit $N\to\infty$ of this function
$\lambda_{N}$ exists, one defines
\begin{align*}
\lambda(k) & :=\lim_{N\to\infty}N^{-1}W_{N}(N\,k).
\end{align*}
In this case, we may replace $W_{N}(j)\stackrel{N\to\infty}{\simeq}N\,\lambda(j/N)$
to get for the distribution \eqref{eq:p_N_Laplace}
\begin{align}
p_{N}(s) & \stackrel{N\gg1}{\simeq}\int_{a-i\infty}^{a+i\infty}\,\frac{dj}{2\pi i}\,\exp\big(-j\,s+N\,\lambda(j/N)\big)\label{eq:back_trafo}\\
 & \stackrel{\text{subst. }N\,k=j}{=}\int_{a/N-i\infty}^{a/N+i\infty}\,\frac{d(Nk)}{2\pi i}\,\exp\big(-N\,\big[k\,s-\lambda(k)\big]\,\big).\nonumber 
\end{align}
Since the parameter $a$ was arbitrary in the Laplace representation
\eqref{eq:Laplace_delta} we may choose it conveniently. We here choose
it such that the Bromwich contour passes through the saddle point
of the integrand; this is the real-valued point $a/N=k^{\ast}$, where
$\frac{\partial}{\partial k}\big[k\,s-\lambda(k)\big]=0$. Due to
the imaginary unit in the integration variable, all other points along
the Bromwich contour produce oscillatory contributions that cancel
each other, so that the dominant contribution comes from this saddle
point $k^{\ast}(s)\in\bR$, so one approximates
\begin{align}
p_{N}(s) & \stackrel{N\gg1}{\simeq}\exp\big(-N\,\big[k^{\ast}\,s-\lambda(k^{\ast})\big]\,\big)\nonumber \\
\lim_{N\to\infty}\,-\frac{1}{N}\,\ln\,p_{N}(s) & =\sup_{k}\big[k\,s-\lambda(k)\big]=:\gamma(s),\label{eq:def_rate_function-1}
\end{align}
where we could write $\sup$ here, because we know that $\lambda^{\prime\prime}$
is the variance, which must be $\ge0$, so that $\lambda$ has positive
curvature, so $-\lambda$ has negative, so the stationary point must
be a local maximum. This is the G\"artner-Ellis theorem. The function
$\gamma(s)$ is called the \textbf{rate function} and we have
\begin{align*}
p_{N}(s) & \stackrel{N\gg1}{\simeq}e^{-N\,\gamma(s)}
\end{align*}
up to proportionality.

\section{Example\protect\label{sec:Example_ldp}}

Consider a set of $N$ random numbers $(x_{1},\ldots,x_{N})\stackrel{\text{i.i.d.}}{\sim}p$.
Assume that we are interested in another random variable, the \textbf{sample
mean} or \textbf{empirical average}
\begin{align}
S_{N}(N) & :=N^{-1}\,\sum_{i=1}^{N}x_{i}.\label{eq:sum_of_iid}
\end{align}
Let us check that the scaled cumulant generating function exists.
The cumulant-generating function for $S_{N}$ is
\begin{align}
W_{N}(j) & :=\ln\big\langle e^{j\,S_{N}(x)}\big\rangle_{x}=\ln\,\langle e^{j\,N^{-1}\,\sum_{i=1}^{N}x_{i}}\rangle_{x\stackrel{\text{i.i.d.}}{\sim}p}\label{eq:indep_W_S-1}\\
 & =\ln\,\prod_{i=1}^{N}\langle e^{j\,N^{-1}\,x_{i}}\rangle_{x_{i}\sim p}\nonumber \\
 & =N\,W_{1}(j/N),\nonumber 
\end{align}
where $W_{1}(k):=\ln\,\langle e^{j\,x}\rangle_{x\sim p}$ is the cumulant-generating
function for a single variable $x_{i}$. So the scaled cumulant-generating
function is
\begin{align}
\lambda_{N}(k) & :=N^{-1}\,W_{N}(N\,k)\nonumber \\
 & =W_{1}(k).\label{eq:lambda_example}
\end{align}
The limit $N\to\infty$ here exists trivially $\lambda(k)=\lim_{N\to\infty}\,\lambda_{N}(k)=W_{1}(k)$,
as the expression is independent of $N$. So the rate function is
\begin{align}
\gamma(s) & =\sup_{k}\,sk-W_{1}(k).\label{eq:rate_function_gamma}
\end{align}
This result shows that the large deviation approach is more general
than the law of large numbers: It is not restricted to the summation,
as we have seen in the general derivation and it also holds for arbitrary
values of $s$, so also for $s$ that are far away from the expected
mean value. Also we note that the function $\gamma$ is in general
not a quadratic polynomial; only in that case one obtains the law
of large numbers, when $e^{-N\gamma(s)}$ reduces to a Gaussian distribution.

The law of large numbers can be obtained from here by expanding $W_{1}(k)=\mu\,k+\frac{1}{2}\Sigma\,k^{2}+\order(k^{3})$,
which is valid for small $k$. In the rate function \eqref{eq:rate_function_gamma}
small values of $k$ correspond to values of $s$ close the mean value,
because a supremum at $k^{\ast}=0$ is obtained precisely when $0=s-W_{1}^{\prime}(0)$
and $W_{1}^{\prime}(0)=\mu$, the mean value of $x_{i}$ and hence
of $S_{N}$. Using this quadratic expansion one gets
\begin{align*}
\gamma(s) & =\sup_{k}\,(s-\mu)\,k-\frac{1}{2}\Sigma\,k^{2}+\order(k^{3}),
\end{align*}
whose supremum is at $0=(s-\mu)-\Sigma\,k^{\ast}$, which, inserted
back, yields
\begin{align*}
\gamma(s) & =\frac{(s-\mu)^{2}}{\Sigma}-\frac{1}{2}\frac{\Sigma\,(s-\mu)^{2}}{\Sigma^{2}}\\
 & =\frac{1}{2}\,\frac{(s-\mu)^{2}}{\Sigma},
\end{align*}
so for $s\simeq\mu$ we get the law of large numbers
\begin{align*}
p_{N}(s) & \simeq e^{-N\gamma(s)}\simeq e^{-\frac{1}{2}\,\frac{(s-\mu)^{2}}{\Sigma/N}}.
\end{align*}

\section{Legendre transform and equation of state}

The operation in \prettyref{eq:def_rate_function-1} is called a \textbf{Legendre-Fenchel
transform}. It may in general be described for any cumulant-generating
function $W(j)=\ln\langle e^{j^{\T}x}\rangle_{x}$ as
\begin{align*}
\Gamma(x^{\ast}) & :=\sup_{j}\,j^{\T}x^{\ast}-W(j).
\end{align*}
Evaluating the supremum condition (assuming the r.h.s. be differentiable
in $j$) one gets the condition
\begin{align*}
0 & \stackrel{!}{=}x^{\ast}-\underbrace{W^{(1)}(j)}_{\langle x(j)\rangle},
\end{align*}
which shows that $j$ is chosen such that $x^{\ast}$ becomes the
mean value of $x$. This condition always has a solution because one
can show that $W(j)$ is a convex function.

So the Legendre transform can be considered the tool to define an
ensemble with a fixed mean value given by $x^{\ast}$. The large deviation
principle that we have seen before simply states that, because the
empirical average $S=N^{-1}\sum_{i}x_{i}$ concentrates around a typical
value, it is sufficient to know this value to (approximately) know
the number of (microscopic) states of the $x_{i}$ that contribute
and thus provide the probability for $p(s)$.

The Legendre transform has the general property
\begin{align}
\frac{d\Gamma}{dx^{\ast}}(x^{\ast})= & j+\frac{\partial j^{\T}}{\partial x^{\ast}}x^{\ast}-\underbrace{\frac{\partial W^{\T}}{\partial j}}_{x^{\ast\T}}\frac{\partial j}{\partial x^{\ast}}\label{eq:equation_of_state-2}\\
= & j.\nonumber 
\end{align}
The latter equation is also called \textbf{equation of state}, as
its solution for $x^{\ast}$ allows us to determine the mean value
for a given source $j$. In statistical physics this mean value is
typically an order parameter, an observable that characterizes the
state of the system. One therefore has the reciprocity between the
pair of functions $W$ and $\Gamma$
\begin{align*}
W^{(1)}(j) & =x^{\ast},\\
\Gamma^{(1)}(x^{\ast}) & =j.
\end{align*}

\section{Keypoints}

We summarize the key points on large deviations

\noindent\fcolorbox{black}{white}{\begin{minipage}[t]{1\textwidth - 2\fboxsep - 2\fboxrule}%
\textbf{The law of large numbers - The large deviation principle}
\begin{itemize}
\item The G\"artner Ellis theorem allows us to estimate probabilities using
the rate function $\gamma$: $p(s)\propto\exp(-N\gamma(s))$, if we
have a scaled CGF and large $N$.
\item The rate function is the Legendre-Fenchel transform of the scaled
cumulant generating function.
\item The large deviation principle is more general than the central limit
theorem, because it is also valid for deviations far away from from
the mean.
\item Source and mean in the rate function are related via the equation
of state, which is a general property of Legendre-Fenchel transforms;
first derivatives $\Gamma^{\prime}$ and $W^{\prime}$ form a pair
of inverse functions of one another.
\end{itemize}
\end{minipage}}

\selectlanguage{english}%

\section{Exercises}

\subsection*{a) Sums of random variables: Central limit theorem vs large deviation
theory}

Let \textbf{$x_{i}$} be distributed according to the Bernoulli distribution

\[
x_{i}\stackrel{\text{i.i.d.}}{\sim}\begin{cases}
1 & \text{with probability }p\\
0 & \text{with probability }1-p
\end{cases}
\]
with the probability $0\le p\le1$. The $x_{i}$ are independently
and identically distributed (i.i.d.). You can imagine $x_{i}=1$ being
heads and $x_{i}=0$ being tails of a (rigged) coin. and $i=1,\ldots,N$.
Let us consider the empirical average

\[
S_{N}(N)=\frac{1}{N}\sum_{i=1}^{N}\,x_{i}
\]

\begin{enumerate}
\item Calculate the cumulants $\kappa_{1}\coloneqq\llangle x\rrangle$ and
$\kappa_{2}\coloneqq\llangle x^{2}\rrangle$.
\item What is the average value of $S_{N}$?
\item To obtain the higher cumulants of \textbf{$S_{N}$}, first show that
$Z_{N}(j)=Z_{x}(\frac{j}{N},\ldots,\frac{j}{N})\stackrel{\text{i.i.d}}{=}\left[Z_{1}(\frac{j}{N})\right]^{N}$,
where $Z_{x}(j_{1},\ldots,j_{N})$ is the moment generating function
of the vector $x$ in the general case and $Z_{1}(j)$ is the moment
generating function of a single variable $x_{i}$ in the i.i.d. case.
Derive the corresponding relation for $W_{N}(j)$ and $W_{1}(j)$.
\item Using this relation, show that the $n$-th cumulant of $S_{N}$ is
given by $\llangle S_{N}^{n}\rrangle=\frac{\kappa_{n}}{N^{n-1}}$.
\item We now employ the large deviation approach: Compute $p(s)\simeq\exp(-N\gamma(s))$
by computing the rate function $\gamma(s)$ using the Legendre-Fenchel
transform $\gamma(s)=\sup_{k}\,k\,s-W_{x}(k)$. ($k\coloneqq j/N$,
see \eqref{eq:rate_function_gamma}) \\
Check: What symmetries does $\gamma(s)$ have? What happens if one
sends $s\to0$ or $s\to1$?
\item Pretend that you do not trust the result that the approximation $p(s)\simeq\exp(-N\gamma(s))$
and instead you want to compute $p(s)$ directly. To this end, convince
yourself that $S_{N}$ can take on values $S_{N}\in[0,1]$ and that
the value $S_{N}=\frac{n}{N}\quad\mathrm{with}\,n=0,1,\ldots,N$ is
assumed with the probability $p(s=N^{-1}n)=\frac{N!}{n!(N-n)!}p^{n}(1-p)^{N-n}$.
Compute $p(s)$ for large $N$. Make use of Stirling's approximation
$\ln\left(M!\right)\approx M\ln M-M+\frac{1}{2}\ln\left(2\pi M\right)$
for large $M$, which you may remember from the statistical physics
calculation for the microcanonical ensemble. Hint: Do not forget to
insert $n=Ns$ in the binomial coefficient. Compute $I(s)$ with $\ln\left(p(s=N^{-1}n)\right)\eqqcolon-N\,I(s)+\ln(ds)$.
As $s$ is discrete, we introduce the probability density function
$p(s=N^{-1}n)=\exp\left(-N\,I(s)\right)ds$ with $ds=1/N$. Now compare
this expression to the approximated result using the rate function.
What happens in the limit $N\to\infty$?
\item Consider now the same setup but from a different (central limit theorem)
perspective:\\
From 4. we know that for large $N$, the probability distribution
becomes Gaussian $p(s)=\N\left(\kappa_{1},\frac{\kappa_{2}}{N}\right)\equiv\frac{1}{\sqrt{2\pi\kappa_{2}/N}}\,\exp\left(-\frac{1}{2}\frac{\left(s-\kappa_{1}\right){{}^2}}{\kappa_{2}/N}\right)$.\\
We now want to compare the results from the CLT and large deviation
theory.\\
What happens if one wanted to send $s$ to a smaller number than $0$
or a bigger number than $1$? Why is this so and does this seem reasonable
within the setup of the exercise?\\
\item Taylor expand $N\,I(s)$ from 6. with $s=p+\delta p$ for $\frac{\delta p}{p}\ll1$
up to second order in $\frac{\delta p}{p}$ and compare the result
to $\ln\left(p(s)\right)$ from 7.. Hint: $\ln(1+x)\simeq x-\frac{1}{2}x^{2}$
around $x=0$. Comment on similarities and discrepancies between the
two results and the applicability of both the central limit theorem
approach and the large deviation approach.
\end{enumerate}

\subsection*{Optional:: Probability bounds and Large deviation theory (9p)}

We want to use the results from the main text for some applications
in probability theory. Especially in computer science so called ``probability
bounds'' for random variables $x$ are used quite often. The use-case
for the bounds is to produce upper/lower limits on the probabilities
such as $P(x>a)$ with limited information about the distribution
of $x$. The most basic relation is the so called Markov bound on
positive random numbers $x$ which only utilizes the mean of the distribution
$\mathbb{E}(x)$ to create a bound on the probability $p(x\geq a)$

\begin{equation}
p(x\geq a)\leq\frac{\mathbb{E}(x)}{a}\quad\mathrm{for\quad}x\geq0\quad.
\end{equation}
Show that the Markov bound is valid for any $a\geq0$ (1p). Hint:
Start from the definition of the expectation value $\mathbb{E}(x)$
and follow successive inequalities to reach the result. As it turns
out, the Markov bound is a reasonable first guess but only a rough
heuristic. One can however refine the statement using more information
about the probability distribution of $x$. One such way is to utilize
the Chebychev bound (1p):

\begin{equation}
p(\vert x-\mathbb{E}(x)\vert\geq a)\leq\frac{\mathrm{Var}(x)}{a^{2}}\quad\mathrm{with\quad}a\geq0.
\end{equation}
Use the Markov bound to prove this statement. Hint: Use an auxiliary
variable $y:=(x-\mathbb{E}(x))^{2}$. As you can see, the Chebychev
bound uses knowledge of both the mean $\mathbb{E}(x)$ and the variance
$\mathrm{Var}(x)$. In a similar spirit we can now ask whether there
is a way to include even more knowledge about the distribution of
$x$ into the bounds? One way to do this is the Chernoff bound. This
bound relates the cumulant generating function $W(s)$ of the variable
$x$ to the probabibility $p(x\geq a)$ in the following way:

\begin{equation}
p(x\geq a)\leq\frac{\mathbb{E}(\exp(sx))}{\exp(sa)}=\frac{\exp(W(s))}{\exp(sa)}\quad\mathrm{for\quad}s,a\geq0\quad.\label{eq:Chernoff}
\end{equation}
Use the Markov bound to show this inequality. Hint: Use the fact that
$\exp(x)$ is a convex function (2p). As \eqref{eq:Chernoff} is
a bound for arbitrary $s$, we want to find the $s$ which provides
the best possible bound on $p(x\geq a)$. Relate this to the rate
function in the main text and show that the minimal Chernoff bound
in terms of the rate function $I(a)$ reads (1p)

\begin{equation}
p(x\geq a)\leq\exp(-I(a))\quad\mathrm{for\quad}x,a\geq0\quad.
\end{equation}

Let us investigate the validity of the results by using the example
of $n$ i.i.d. coin tosses, where the coin yields the value $+1$
(heads) with probability $p$ and $0$ (tails) with probability $(1-p)$.
What are the Markov, Chebychev and Chernoff bounds on the probability
to observe at least $\frac{3}{4}n$ times heads when $p=0.6$. Comment
on the scaling of the three bounds with $n$ and the consequences
to estimate the probability of large deviations from the mean. (3p+1p)
. Bonus: Assume you know the mean $\kappa_{1}$ and the variance $\kappa_{2}$
of the random variables $x_{i}$ and you want to obtain the bound
on

\[
p(y\geq\delta n)\quad\mathrm{with\quad}y=\sum_{i}x_{i}\,\quad\mathrm{and}\quad\delta\in[0,1].
\]
What is the best guess that you can make using any of the above bounds?\selectlanguage{american}%

\chapter{Neural network Gaussian processes}

We will now apply the idea of Bayesian inference to networks that
include hidden layers. We begin with the single hidden layer network
and subsequently extend the theory to networks with multiple hidden
layers known as deep networks or multi-layer perceptrons. We begin
with the limit where we take the layer width to infinity and will
see that this limit yields to Gaussian processes, very similar to
linear regression. This limit will serve as the starting point to
study networks of finite width.

\section{Single hidden layer network}

We here consider a network with a single hidden layer whose activations
are called $h\in\bR^{N}$

\begin{eqnarray}
h & = & V\,x,\label{eq:single_hidden_net}\\
y & = & w^{\T}\phi(h),\nonumber \\
z & = & y+\xi,\nonumber 
\end{eqnarray}
where $\phi$ is a point-wise applied activation function and $\xi$
is a Gaussian readout noise $\xi_{\alpha}\stackrel{\text{i.i.d.}}{\sim}\N(0,\kappa)$.
We will see that this noise acts as a regulator. As in \prettyref{sec:Linear-regression}
$x\in\cR^{d}$ is the data and $y\in\cR$ is the scalar output and
we consider $P$ tuples of training data $\cD=\{(x_{\alpha},z_{\alpha})\}_{1\le\alpha\le P}$.
We again consider the matrix $\{\bR^{P\times d}\ni X\}_{\alpha i}=x_{\alpha i}$.
The readout $w\in\cR^{N}$ and the matrix $V\in\cR^{N\times d}$ are
the trainable weights. The noise is introduced to implement regularization,
as we have seen it in the case of linear regression of Bayesian inference,
so we will condition on $z$ being fixed to the desired output, but
will be interested also in the distribution of the readout $y$ of
the hidden layer.

\section{Intuitive approach to neural network Gaussian processes\protect\label{sec:Intuitive-approach-to-NNGP}}

Before embarking on a more formal approach to derive the behavior
of networks in the limit of infinite width, we here present a heuristic
derivation. Let us consider the single hidden layer example \prettyref{eq:single_hidden_net}.
For simplicity we here set the regularization noise $\xi=0$; an extension
to $\xi\neq0$ is straight forward.

Due to the sum appearing in $y_{\alpha}=\sum_{j=1}^{N}w_{i}\,\phi(h_{\alpha i})$
and the independence of the $w_{i}$, we expect the distribution of
the $\{y_{\alpha}\}_{1\le\alpha\le P}$ to be jointly Gaussian. We
also note that due to $\langle w_{i}\rangle=0$, the mean $\langle y\rangle=0$
vanishes. It is therefore sufficient to characterize the statistics
by its second moment $\langle y_{\alpha}y_{\beta}\rangle$, where
the average is with regard to the weights $w$ and $V$. We start
from

\begin{align*}
\langle y_{\alpha}y_{\beta}\rangle & =\left\langle \sum_{i,j=1}^{N}w_{i}w_{j}\phi(h_{\alpha i})\phi(h_{\beta j})\right\rangle _{W,V}\\
 & =\sum_{i,j=1}^{N}\left\langle w_{i}w_{j}\right\rangle _{w}\left\langle \phi(h_{\alpha i})\phi(h_{\beta j})\right\rangle _{V}=\sum_{i=1}^{N}\,\frac{g_{w}}{N}\,\left\langle \phi(h_{\alpha i})\phi(h_{\beta i})\right\rangle _{V},
\end{align*}
where we have split the averages, as $W$ and $V$ are independent
and we utilized that the $w_{i}$ are pairwise independent $w_{i}\stackrel{\text{i.i.d.}}{\sim}\N(0,g/N)$
to write $\langle w_{i}w_{j}\rangle=\delta_{ij}\,g_{w}/N$, eliminating
one of the sums. As $h_{\alpha i}$ is the only dependence on $V$,
we can replace 

\[
\left\langle \phi(h_{\alpha i})\phi(h_{\beta i})\right\rangle _{V}=\left\langle \phi(h_{\alpha i})\phi(h_{\beta i})\right\rangle _{h_{\alpha i},h_{\beta i}}.
\]
The distribution of the $h_{\alpha i}=\sum_{j=1}^{d}V_{ij}\,x_{\alpha j}$
is Gaussian, because the $x_{\alpha j}$ are fixed and $h$ depends
linearly on the Gaussian $V_{ij}\stackrel{\text{i.i.d.}}{\sim}\N(0,g/d)$;
due to the vanishing mean of $V$, also the mean of $h$ vanishes,
so it is sufficient to compute the second moment
\begin{align*}
\big\langle h_{\alpha i}h_{\alpha j}\big\rangle & =\sum_{k,l=1}^{d}\,\underbrace{\big\langle V_{ik}\,V_{jl}\big\rangle}_{\delta_{ij}\,\delta_{kl}\,g_{V}/d}\,x_{\alpha k}x_{\beta l}=\delta_{ij}\,\frac{g_{V}}{d}\sum_{k=1}^{d}x_{\alpha k}x_{\beta k}=:\delta_{ij}C_{\alpha\beta}^{(xx)},
\end{align*}
so $h_{\alpha i}\stackrel{\text{i.i.d. over }i}{\sim}\N(0,C^{(xx)})$
in particular, they are independent across $i$. Hence we obtain 
\begin{align*}
\langle y_{\alpha}y_{\beta}\rangle_{W,V} & =\frac{g_{w}}{N}\,\sum_{i=1}^{N}\,\left\langle \phi(h_{\alpha i})\phi(h_{\beta i})\right\rangle _{h_{\alpha i}\stackrel{\text{i.i.d. over }i}{\sim}\N(0,C^{(xx)})}\\
 & =g_{w}\,\left\langle \phi(h_{\alpha})\phi(h_{\beta})\right\rangle _{h_{\alpha}\sim\N(0,C^{(xx)})},
\end{align*}
where we used in the last step that all expectation values in the
sum yield the same result.

This Gaussian distribution for $y$ is the prior distribution of the
network outputs. Conditioning on the training data, we may therefore
use the same expressions for the mean and covariance of the predictive
distribution as derived for linear regression \eqref{eq:posterior_Gaussi},
only replacing $C^{(xx)}$ (linear regression) by $g_{w}C^{(\phi\phi)}$
(NNGP).

The derivation above can be made precise by employing the central
limit theorem. However, it prevents us from studying finite $N$ corrections
and the interplay between the amount of training data $P$ and $N$.
From the derivation above it is also clear that the procedure may
be iterated across multiple layers: One proceeds by induction to show,
layer by layer, that the central limit theorem assures a Gaussian
distribution.

\section{Network field theory\protect\label{sec:Network-field-theory}}

To obtain the network prior, for an i.i.d. prior $V_{ij}\stackrel{\text{i.i.d.}}{\sim}\N(0,g_{V}/d)$,
it is easy to see that also the $h_{\alpha i}=\sum_{j}V_{ij}\,x_{\alpha j}$
are Gaussian random variables that are independent across $i$ but
correlated across different $\alpha$ following the distribution
\begin{align}
h_{\alpha i} & \stackrel{\text{i.i.d. over }i}{\sim}\N(0,C^{(xx)}),\label{eq:hidden_first}\\
\bR^{P\times P}\ni C^{(xx)} & =\frac{g_{V}}{d}\,X\,X^{\T}\nonumber 
\end{align}
by the same arguments as in \prettyref{sec:Bayesian-formulation}.
The probability distribution of the readout $y$ is only a function
of $C^{(xx)}$ and given with the Gaussian distribution $\N(z|y,\kappa)=\frac{1}{(2\pi\kappa)^{\frac{P}{2}}}\,e^{-\frac{\|z-y\|^{2}}{2\kappa}}$
a Gaussian distribution in $z$ with mean $y$

\begin{align}
p(z,y|C^{(xx)}) & =\N(z|y,\kappa)\,\big\langle\,\prod_{\alpha=1}^{P}\delta\,\big[y_{\alpha}-\sum_{i=1}^{N}w_{i}\,\phi(h_{\alpha i})\big]\big\rangle_{w_{i},h_{\alpha i}}\label{eq:pre_disorder_W}\\
 & =\N(z|y,\kappa)\,\int\D\ty\,\big\langle\exp\big(\,\ty_{\alpha}\big[-y_{\alpha}+w_{i}\,\phi(h_{\alpha i})\big]\,\big)\big\rangle_{w_{i},h_{\alpha i}},\nonumber 
\end{align}
where $w_{i}\stackrel{\text{i.i.d.}}{\sim}\N(0,\frac{g_{w}}{N})$
and $h_{\alpha i}\stackrel{\text{i.i.d. over }i}{\sim}\N(0,C^{(xx)})$
and $\int\D\ty$ is given by \prettyref{eq:measure_dtily} and we
use Einstein's summation convention, summing over repeated indices
$\alpha$ and $i$ on the right from the second line on. The normal
distribution $\N(z|y,\kappa)=p(z|y)=p_{\xi}(\xi=z-y)$ arises as the
conditional of $z$ given $y$, which is the probability to have the
right realization of the readout noise $\xi_{\alpha}=z_{\alpha}-y_{\alpha}$.
Now take the expectation over $w_{i}\stackrel{\text{i.i.d.}}{\sim}\N(0,g_{w}/N)$,
which yields 
\begin{align}
\big\langle\exp\big(\sum_{\alpha=1}^{P}\ty_{\alpha}\sum_{i=1}^{N}w_{i}\phi_{\alpha i}\big)\big\rangle_{W} & =\exp\big(\frac{1}{2}\,\sum_{\alpha,\beta=1}^{P}\,\ty_{\alpha}\ty_{\beta}\,\frac{g_{W}}{N}\,\sum_{i=1}^{N}\phi{}_{\alpha i}\,\phi_{\beta i}\big),\label{eq:disorder_avg}
\end{align}
where we wrote for short $\phi_{\alpha i}=\phi(h_{\alpha i})$.

The latter term shows by the appearance of the $\sum_{i=1}^{N}$ that
all neurons in the hidden layer contribute in a similar manner, so
the neuron identity has been lost here. We note that the terms 
\begin{align}
C_{\alpha\beta}= & \frac{g_{w}}{N}\,\sum_{i=1}^{N}\phi{}_{\alpha i}\,\phi_{\beta i}=:\frac{g_{w}}{N}\,\phi_{\alpha}\phi_{\beta}^{\T},\label{eq:def_C_phiphi}
\end{align}
have a form as in \prettyref{eq:sum_of_iid} in \prettyref{chap:The-law-of-large-number}:
We have a sum over $i$ and the distribution of $\phi_{\alpha i}\phi_{\beta i}$
is i.i.d. over $i$. We hence expect that for large $N$ a large deviation
principle holds for each of the $P^{2}$ terms $1\le\alpha,\beta\le P$.
To exploit this principle, it is convenient to introduce $C_{\alpha\beta}^{(\phi\phi)}\quad\forall1\le\alpha,\beta\le P$
as what are called \textbf{auxiliary fields} by enforcing the definition
\prettyref{eq:def_C_phiphi} in \prettyref{eq:pre_disorder_W} as
\begin{align*}
\int\D C\,\ldots\prod_{\alpha,\beta=1}^{P}\delta\big[-C_{\alpha\beta}+\frac{g_{w}}{N}\phi_{\alpha}\phi_{\beta}^{\T}\big] & =\int\D C\,\int\D\tC\,\ldots\exp\big(\sum_{\alpha,\beta=1}^{P}\,-\tC_{\alpha\beta}C_{\alpha\beta}+\tC_{\alpha\beta}\frac{g_{w}}{N}\phi_{\alpha}\phi_{\beta}^{\T}\big),
\end{align*}
where $\int\D C=\prod_{\alpha,\beta=1}^{P}\,\int_{-\infty}^{\infty}C_{\alpha\beta}$
and $\int\D\tC=\prod_{\alpha,\beta=1}^{P}\,\int_{-i\infty}^{i\infty}\,\frac{d\tC_{\alpha\beta}}{2\pi i}$
and $\ldots$ may be any function of the matrix $C$, namely

\begin{align}
p(z,y|C^{(xx)}) & =\N(z|y,\kappa)\,\int\D\ty\,\int\D C\label{eq:joint_outputs-2}\\
 & \phantom{=\N(y,\kappa;z)\,}\times\exp\Big(\ty^{\T}y+\frac{1}{2}\,\ty^{\T}C\ty\Big)\nonumber \\
 & \phantom{=\N(y,\kappa;z)\,}\times\int\D\tC\,\exp\big(-\tr\,\tC^{\T}C+W\,\big(\tC|C^{(xx)}\big)\Big),\nonumber 
\end{align}
where we defined the cumulant-generating function
\begin{align}
W(\tC|C^{(xx)}) & =\ln\,\Big\langle\exp\,\big(\frac{g_{w}}{N}\,\sum_{i=1}^{N}\phi_{\alpha i}\tC_{\alpha\beta}\,\phi_{\beta i}\big)\Big\rangle_{h_{\alpha i}\stackrel{\text{i.i.d. over }i}{\sim}\N(0,C^{(xx)})}\label{eq:W_C_tilde}\\
 & =\ln\,\prod_{i=1}^{N}\,\Big\langle\exp\,\big(\frac{g_{w}}{N}\,\phi_{\alpha i}\tC_{\alpha\beta}\,\phi_{\beta i}\big)\Big\rangle_{h_{\alpha i}\stackrel{\text{i.i.d. over }i}{\sim}\N(0,C^{(xx)})}\nonumber \\
 & =N\,\ln\,\Big\langle\exp\,\big(\frac{g_{w}}{N}\,\phi_{\alpha}\tC_{\alpha\beta}\,\phi_{\beta}\big)\Big\rangle_{h_{\alpha}\sim\N(0,C^{(xx)})}.\nonumber 
\end{align}
The independence across the $N$ neuron indices of the hidden layer
here shows up as the prefactor $N$, analogous to the simple example
in \prettyref{sec:Example_ldp} where we considered a sum of independent
variables.

The last line in \prettyref{eq:joint_outputs-2} is the Fourier-representation
of the probability distribution for $C$, so we may write it as
\begin{align}
p(C) & =\int\D\tC\,\exp\big(-\tr\,\tC^{\T}C+W\,\big(\tC|C^{(xx)}\big)\Big).\label{eq:p_C}
\end{align}
The second line in \prettyref{eq:joint_outputs-2}, for $C$ given
and fixed, describes a joint Gaussian distribution of $\{y_{\alpha}\}$
with covariance $C$, so that in total we get
\begin{align}
p(z,y|C^{(xx)}) & =\N(z|y,\kappa)\,\int\D C\,\N(y|0,C)\,p(C),\label{eq:p_y_C}
\end{align}
which is a weighted sum of Gaussian distributions for $y$ with different
variances $C$ and the explicit form of the Gaussian including the
normalization condition $1/\big[(2\pi)^{N/2}\det(C)^{\frac{1}{2}}\big]$
from \eqref{eq:N_dim_Gauss}.

\section{Dominant behavior at large width\protect\label{sec:Dominant-behavior-single-hidden}}

The form of the cumulant-generating function \eqref{eq:W_C_tilde}
has the form of a scaled cumulant-generating function. So we expect
a large deviation principle to apply in the limit $N\to\infty$. To
be more precise, we consider the case where the number of training
samples $P$ is kept finite, but the width of the layers $N\to\infty$.
Since $W$ in \eqref{eq:W_C_tilde} has the scaling form, which allows
us to define the scaled cumulant-generating function $\lambda_{N}(k):=N^{-1}\,W(N\,k|C^{(xx)})$,
which is independent of $N$ and hence its limit $\lambda(k):=\lim_{N\to\infty}\lambda_{N}(k)$
exists trivially. So from the G\"artner-Ellis theorem in \prettyref{chap:The-law-of-large-number}
we know that we may approximate the distribution $p(C)$ defined by
\prettyref{eq:p_C} with help of the rate function
\begin{align}
\Gamma(C) & :=\sup_{\tC}\,\tr\,\tC^{\T}C-W\,\big(\tC|C^{(xx)}\big)\label{eq:gamma_C}
\end{align}
as
\begin{align}
\ln\,p(C) & \simeq-\Gamma(C)\label{eq:approx_P_C}
\end{align}
so that one obtains from \eqref{eq:joint_outputs-2}

\begin{align}
p(z,y|C^{(xx)}) & \simeq\N(z|y,\kappa)\,\int\D C\,\N(y|0,C)\,e^{-\Gamma(C)}.\label{eq:p_y_given_Cxx_as_Gauss-1}
\end{align}
Writing all terms in the exponent one has
\begin{align}
p(z,y|C^{(xx)}) & \simeq\N(z|y,\kappa)\,\int\D C\,\exp\big(-\frac{1}{2}y^{\T}C^{-1}y-\frac{1}{2}\ln\det(C)-\Gamma(C)\big).\label{eq:p_z_y}
\end{align}

The matrix $C\in\bR^{P\times P}$ has a fixed dimension $P<\infty$.
Assuming its eigenvalues $\lambda_{1\le\mu\le P}$, the bilinear form
$-\frac{1}{2}y^{\T}C^{-1}y=-\frac{1}{2}\sum_{\mu=1}^{P}\lambda_{\mu}(v_{\mu}^{\T}y)^{2}=\order(P)$,
because for the orthogonal eigenvectors $v_{\mu}^{\T}v_{\nu}=\delta_{\mu\nu}$
of $C$ and for $y_{\alpha}=\order(1)$ it is $\|y\|^{2}=\order(P)$,
so one has $\order(P)=\|y\|^{2}=\sum_{\mu=1}^{P}\,(v_{\mu}^{\T}y)^{2}$.
The determinant is $\ln\det(C)=\prod_{\mu=1}^{P}\ln(\lambda_{\mu})=\order(P)$.
The rate function \eqref{eq:gamma_C}, however, is $\Gamma\propto N$,
because, with the scaled cumulant-generating function $\lambda(k)$
we may write $W(\tC)=N\,\lambda(\tC/N)$. So the rate function
\begin{align*}
\Gamma(C) & =\sup_{\tC}\,\tr\,\tC^{\T}C-N\,\lambda(\tC/N)\\
 & \stackrel{k=\tC/N}{=}N\,\sup_{k}\,\tr\,k^{\T}C-\lambda(k)=N\,\gamma(C)=\order(N)
\end{align*}
is a product of a function $\gamma$ that is independent of $N$ and
a prefactor $N$.

Taking a saddle point approximation of the integral $\int\D C$, in
the exponent \eqref{eq:p_z_y} and in the limit $N\to\infty$, the
rate function $\Gamma$ dominates its stationary point for $C^{\ast}$,
namely
\begin{align}
0 & \stackrel{!}{=}\Gamma^{\prime}(C)=\tC,\label{eq:eq_of_state_GammaC}
\end{align}
where the last equal sign follows from the equation of state \eqref{eq:equation_of_state-2}
so the explicit supremum condition in \eqref{eq:gamma_C} reads
\begin{align}
0 & \stackrel{!}{=}C-W^{\prime}(0|C^{(xx)}),\nonumber \\
C_{\alpha\beta} & =g_{w}\,\langle\phi_{\alpha}\,\phi_{\beta}\rangle.\label{eq:NNGP}
\end{align}
We do not have to check the second derivative for the stationary point
because $\Gamma$ is convex. We may therefore determine the network
prior, so the distribution of $y$ and $z$, as
\begin{align}
p(z,y|C^{(xx)}) & \simeq\N(z|y,\kappa)\,\N(y|0,C^{\ast}).\label{eq:GP_final}
\end{align}
 We thus obtain what is known as the \textbf{neural network Gaussian
process} (NNGP), because the outputs $\{y_{\alpha}\}$ follow a joint
Gaussian distribution.

We recapitulate the important steps which led to this result, because
in the following we will repeat the very same steps for different
architectures:
\begin{itemize}
\item enforce the equations defining the network input-output mapping with
help of Dirac distributions
\item compute the expectation value over the prior of the weights to obtain
an expression for the joint distribution of $z$ and $y$
\item quantities appear which are summed over all neurons in the hidden
layer (in this case the matrix $C_{\alpha\beta}$); introduce these
as auxiliary variables, because we expect these to show a concentration
phenomenon, namely they will be dominated by their mean and fluctuations
will be small in the large $N$ limit
\item write the network outputs $y$ and $z$ as a conditional distribution,
conditioned in $C$, multiplied by $p(C)$
\item approximate $p(C)$ for large $N$ by the G\"artner-Ellis theorem
\item compute the integral over $C$ in saddle point approximation; in the
NNGP limit $N\to\infty$, $P=\const.$ this amounts to neglecting
the data term in favor of the term $\Gamma(C)$ from the prior
\end{itemize}

\section{Mean and covariance of the NNGP predictor}

Since we have obtained a Gaussian process \eqref{eq:GP_final}, as
in the case of the Bayesian treatment of linear regression, when conditioning
on the training data by fixing the $z$, the mean and covariance for
a test point is the same as in linear regression \eqref{eq:posterior_Gaussi}

\begin{align}
y_{\ast}\sim & \N\Big(C_{\ast\circ}[C_{\circ\circ}+\kappa\I]^{-1}z_{\circ},\label{eq:stat_pred_NNGP}\\
 & \phantom{\N(}C_{\ast\ast}-C_{\ast\circ}[C_{\circ\circ}+\kappa\I]^{-1}C_{\circ\ast}\Big).\nonumber 
\end{align}
The covariance matrices arising here result from the $y$ and $z$
following a joint Gaussian distribution, with $z_{\alpha}=y_{\alpha}+\xi_{\alpha}$,
so that
\begin{align*}
\langle y_{\ast}z_{\circ}\rangle & =\langle y_{\ast}(y_{\circ}+\xi_{\circ})\rangle=\langle y_{\ast}y_{\circ}\rangle=C_{\ast\circ},\\
\langle y_{\ast}y_{\ast}\rangle & =C_{\ast\ast},\\
\langle z_{\circ}z_{\circ}\rangle & =\langle(y_{\circ}+\xi_{\circ})\,(y_{\circ}+\xi_{\circ})\rangle=C_{\circ\circ}+\kappa\,\I,
\end{align*}
where we used that the readout noise $\xi_{\alpha}$ is independent
of $y_{\alpha}$ and independent across different $\alpha$.

\section{Multi-layer perceptrons – deep networks\protect\label{sec:Multi-layer-perceptrons}}

\begin{figure}
\begin{centering}
\includegraphics{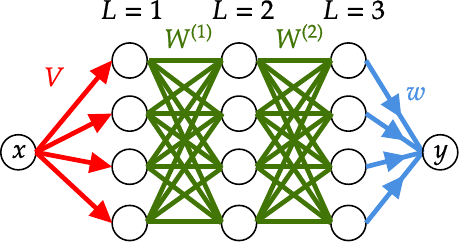}
\par\end{centering}
\caption{Sketch of a deep network with input $x$, $L+1$ hidden layers $h^{(0)},\ldots,h^{(L)}$
and a scalar output $y$.}
\end{figure}

The treatment shown for a network with a single hidden layer can be
extended straight forwardly to deep networks (DNN), which have many
intermediate hidden layers. To keep the notation simple we here treat
the case of identical width in each layer; the extension to different
widths is easy and can be looked up in \citep{Segadlo21_arxiv}. Likewise,
the treatment of bias terms, additive contributions to the pre-activations
is treated there and will be considered in the exercises.

Consider a DNN with $L$ inner layers as specified in \citep{Segadlo21_arxiv}
as

\begin{eqnarray}
h^{(0)} & = & V\,x,\label{eq:eq_of_motion}\\
h^{(a)} & = & W^{(a)}\phi(h^{(a-1)})\quad1\le a\le L,\nonumber \\
y & = & w^{\T}\phi(h^{(L)}),\nonumber \\
z & = & y+\xi,\nonumber 
\end{eqnarray}
where $x\in\bR^{d}$, $h\in\bR^{N}$ and $y\in\bR$. We again assume
Gaussian priors $V_{ij}\stackrel{\text{i.i.d.}}{\sim}\N(0,g_{V}/d)$
and $W_{ij}\stackrel{\text{i.i.d.}}{\sim}\N(0,g_{w}/N)$. Due to the
Gaussian i.i.d. prior on $V$ one again has that 
\begin{align*}
h_{\alpha i}^{(0)} & \stackrel{\text{i.i.d. in }i}{\sim}\N(0,C_{\alpha\beta}^{(xx)}),
\end{align*}
where $C^{(xx)}$ is given as before by \eqref{eq:hidden_first}.

The distribution of outputs $z$ and readouts $y$, given the inputs
$X$, is computed completely analogously as in the single hidden case
by enforcing \prettyref{eq:eq_of_motion} for each layer as

\begin{align}
p(z,y|C^{(xx)}) & =\N(z|y;\kappa)\,\int\D h^{(1\le a\le L)}\,\big\langle\prod_{\alpha=1}^{P}\delta\big[y_{\alpha}-w^{\T}\phi(h_{\alpha}^{(L)})\big]\big\rangle_{w}\label{eq:p_z_y_deep}\\
 & \times\prod_{a=1}^{L}\,\big\langle\prod_{\alpha=1}^{P}\,\delta\big[h_{\alpha}^{(a)}-W^{(a)}\,\phi(h_{\alpha}^{(a-1)})\big]\big\rangle_{W^{(a)},h_{\alpha i}^{(0)}\stackrel{\text{i.i.d. in }i}{\sim}\N(0,C_{\alpha\beta}^{(xx)})}.\nonumber 
\end{align}
We here used that the weights are drawn i.i.d. across layers, so that
the expectation values factorize over layers.

Each factor 
\begin{align*}
p(h^{(a)}|h^{(a-1)})= & \big\langle\prod_{\alpha=1}^{P}\,\delta\big[h_{\alpha}^{(a)}-W^{(a)}\,\phi(h_{\alpha}^{(a-1)})\big]\big\rangle_{W^{(a)}}
\end{align*}
has the meaning of a conditional distribution, so that one may interpret
\eqref{eq:p_z_y_deep} as a chain of conditional probabilities, marginalized
(integrated) over all hidden layers
\begin{align*}
p(z,y|C^{(xx)}) & =\N(z|y;\kappa)\,\int\D h^{(0\le a\le L)}\,p(y|h^{(L)})\,p(h^{(L)}|h^{(L-1)})\cdots p(h^{(1)}|h^{(0)})\,p(h^{(0)}|C^{(xx)}).
\end{align*}
Resolving the Dirac distributions with their Fourier representation
as in \prettyref{eq:pre_disorder_W}, we compute the expectations
over all weights. Consider some pair of intermediate layers, coupled
by the matrix $W^{(a)}$
\begin{align}
p(h^{(a)}|h^{(a-1)}) & =\int\D\th\,\big\langle\exp\big(-\sum_{\alpha,i}\th_{\alpha i}h_{\alpha i}^{(a)}+\sum_{\alpha,i,j}\,\th_{\alpha i}W_{ij}^{(a)}\phi_{\alpha j}^{(a-1)}\big)\big\rangle_{W_{ij}^{(a)}\stackrel{\text{i.i.d. in }ij}{\sim}\N(0,g_{w}/N)}.\label{eq:p_h_a_given_h_a_1}
\end{align}
We hence need

\begin{align}
 & \big\langle\exp\big(\sum_{\alpha ij}\,\th_{\alpha i}W_{ij}^{(a)}\phi_{\alpha j}^{(a-1)}\big)\big\rangle_{W_{ij}^{(a)}\stackrel{\text{i.i.d. in }ij}{\sim}\N(0,g_{w}/N)}\label{eq:disorder_avg-1}\\
 & =\prod_{ij}\,\Big\{\,\big\langle\exp\big(\sum_{\alpha}\,\th_{\alpha i}w\,\phi_{\alpha j}^{(a-1)}\big)\big\rangle_{w\sim\N(0,g_{w}/N)}\Big\}, & \big\langle\exp\big(\sum_{\alpha,i,j}\,\th_{\alpha i}W_{ij}^{(a)}\phi_{\alpha j}^{(a-1)}\big)\big\rangle_{W_{ij}^{(a)}\stackrel{\text{i.i.d. in }ij}{\sim}\N(0,g_{w}/N)}\nonumber 
\end{align}
where we used the independence of $W_{ij}$ over both indices to factorize
the expectation value into $N^{2}$ expectations over a univariate
Gaussian $w\sim\N(0,g_{w}/N)$. Taking this Gaussian integral one
again notices that this computation corresponds to the one obtaining
the moment-generating function of the univariate Gaussian variable
$w$, so

\begin{align*}
 & =\prod_{i,j}\,\Big\{\,\exp\big(\frac{1}{2}\,\sum_{\alpha\beta}\th_{\alpha i}\th_{\beta i}\,\frac{g_{w}}{N}\,\phi_{\alpha j}^{(a-1)}\phi_{\beta j}^{(a-1)}\big)\Big\}\\
 & =\exp\big(\frac{1}{2}\,\sum_{\alpha\beta}\,\sum_{i}\,\th_{\alpha i}\th_{\beta i}\,\frac{g_{w}}{N}\,\sum_{j}\phi{}_{\alpha j}^{(a-1)}\,\phi_{\beta j}^{(a-1)}\big).
\end{align*}
The appearance of the sum over $i$ shows that the statistics of $h_{\alpha i}^{(a)}$
is independent across $i$. The appearance of the sum $j$ shows that
the problem becomes symmetric in the index of the sendind neuron indices.
To deal with the term coupling four variables $\th\th\phi\phi$, we
introduce auxiliary fields, analogous to \eqref{eq:def_C_phiphi}

\begin{align}
C_{\alpha\beta}^{(a)}:= & \frac{g_{w}}{N}\,\sum_{j=1}^{N}\phi{}_{\alpha j}^{(a-1)}\,\phi_{\beta j}^{(a-1)}=:\frac{g_{w}}{N}\,\phi_{\alpha}^{(a-1)}\cdot\phi_{\beta}^{(a-1)}\,\quad1\le a\le L,\;1\le\alpha,\beta\le P,\label{eq:def_C_phiphi-1}
\end{align}
where we use $\cdot$ to denote the summation over the $N$ neuron
indices. This allows us to write \eqref{eq:p_h_a_given_h_a_1} as
\begin{align*}
p(h^{(a)}|C^{(a)}) & =\int\D\th\,\exp\big(-\sum_{\alpha i}\th_{\alpha i}h_{\alpha i}^{(a)}+\frac{1}{2}\sum_{\alpha\beta}\,\th_{\alpha}C^{(a)}\th_{\beta}\big)\\
 & =\N(\{h_{\alpha i}^{(a)}\}|0,\delta_{ij}C_{\alpha\beta}^{(a)}).
\end{align*}
So conditioned on the value of the auxiliary field $C^{(a)}$, the
$h_{\alpha i}^{(a)}$ are Gaussian, which are independent across neuron
indices $i$, but correlated across sample indices $\alpha$ with
$C_{\alpha\beta}^{(a)}$.

The definition of the field \eqref{eq:def_C_phiphi-1}is enforced
by conjugate fields $\tC_{1\le\alpha\beta\le P}^{(1\le a\le L)}$
to obtain

\begin{eqnarray}
p(z,y|C^{(xx)})=\N(z|y,\kappa)\,\int\D C\,\int\D\ty & \exp\big(-\tilde{y}^{\T}y+\frac{1}{2}\ty^{\T}C^{(L+1)}\ty\big)\,p(C|C^{(xx)}) & ,\label{eq:p_y_prefinal}
\end{eqnarray}
where the distribution of the $C$ is of the form of a chain of conditional
distribution is given by
\begin{align}
p(C^{(0\le a\le L+1)}|C^{(xx)}) & =\prod_{a=1}^{L+1}\,P(C^{(a)}|C^{(a-1)})\Big|_{C^{(0)}=C^{(xx)}}\label{eq:P_C_given_C_1}\\
P(C^{(a)}|C^{(a-1)}) & =\int\D\tC^{(a)}\,\exp\Big(-\tr\tilde{C}^{(a)}C^{(a)}+W(\tC^{(a)}|C^{(a-1)})\Big)\nonumber \\
W(\tC^{(a)}|C^{(a-1)}) & =N\,\ln\Big\langle\exp\big(\frac{g_{w}}{N}\phi^{(a-1)\T}\tC\phi^{(a-1)}\big)\Big\rangle_{h^{(a-1)}\sim\N(0,C^{(a-1)})},\label{eq:W_Ctilde_DNN}
\end{align}
where the independence of the $h_{\alpha i}^{(a)}$ in the index $i$
yields the factor $N$ in the cumulant-generation function $W$, in
the same way as in \eqref{eq:W_C_tilde}.

\section{Behavior of deep networks at large width\protect\label{chap:Dominant-behavior-of-DNNs}}

\begin{figure}
\begin{centering}
\includegraphics{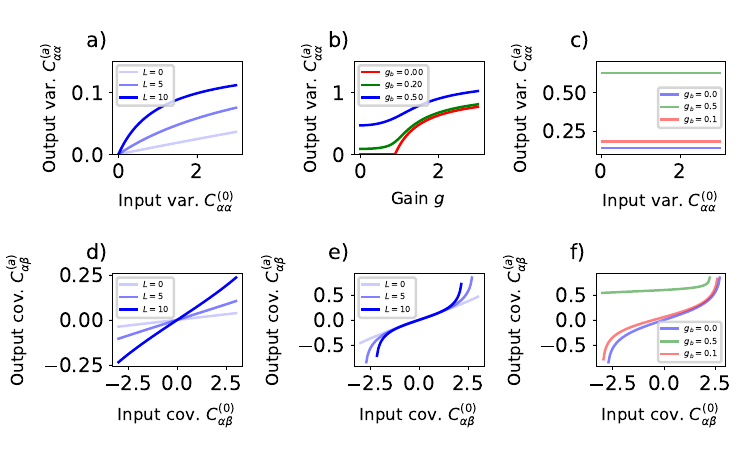}
\par\end{centering}
\caption{\textbf{Neural Network Gaussian Process (NNGP) for erf-activation
function. }Display of the diagonal $C_{\alpha\alpha}$ and off-diagonal
$C_{\alpha\beta}$ elements of NNGP kernel \textbf{a)} Dependence
of output variance $C_{\alpha\alpha}^{(a)}$ for different depths.
\textbf{b)} Fixpoint values of variance for different hidden gain
values $g$. \textbf{c)} Input weight variance initialization determined
such that one obtains a fixed point $C_{\alpha\alpha}^{(a)}$ for
different bias values.\textbf{ d)} Output covariance $C_{\alpha\beta}^{(a)}$
as a function of input covariance $C_{\alpha\beta}^{(0)}$ with $g_{v}=0.1$.
\textbf{e) }Output covariance $C_{\alpha\beta}^{(a)}$ as a function
of input covariance $C_{\alpha\beta}^{(0)}$ with $g_{v}$ set so
that $C_{\alpha\alpha}^{(a)}$ is initialized at the fixpoint. \textbf{f)
}Same setting as in e) for different values of the bias variance $g_{b}$
and for $5$ layers. All results are produced for $\phi=\mathrm{erf}$
and with a regularization variance of $\kappa=1$.\protect\label{fig:Neural-Network-Gaussian}}
\end{figure}

At large network width, due to the independence across neuron indices
in each layer we have obtained the scaling form of the cumulant-generating
function in \eqref{eq:W_Ctilde_DNN}, so the scaled cumulant-generating
function $\lambda_{N}(k):=N^{-1}W(N\,k|C^{(a-1)})=\ln\Big\langle\exp\big(g_{w}\phi^{(a-1)\T}\tC\phi^{(a-1)}\big)\Big\rangle_{h^{(a-1)}\sim\N(0,C^{(a-1)})}$
trivially possesses the limit $N\to\infty$ (because it is independent
of $N$), so one may compute $\ln\,P(C^{(a)}|C^{(a-1)})$ on exponential
scales in large deviation theory to obtain
\begin{align*}
\ln\,P(C^{(a)}|C^{(a-1)}) & \simeq-\Gamma(C^{(a)}|C^{(a-1)})\\
 & =\sup_{\tC^{(a)}}-\tr\tilde{C}^{(a)}C^{(a)}+W(\tC^{(a)}|C^{(a-1)}).
\end{align*}
Performing the same approximation as in \eqref{sec:Dominant-behavior-single-hidden}
one then obtains the NNGP result for the deep network from the above
supremum condition
\begin{align}
\bar{C}^{(a)} & =W^{(1)}(0|\bar{C}^{(a-1)})\label{eq:iteration_DNN}\\
 & =g_{w}\langle\phi^{(a-1)}\phi^{(a-1)}\rangle_{h^{(a-1)}\sim\N(0,\bar{C}^{(a-1)})}\quad\forall1\le a\le L+1\nonumber 
\end{align}
with the initial condition $C^{(0)}=C^{(xx)}$ \citep{Cho09}. The
mean of the predictor is then given by \eqref{eq:stat_pred_NNGP},
only replacing $\bar{C}=C^{(L+1)}$. The iteration of these kernels
is shown in an example in \prettyref{fig:Neural-Network-Gaussian}.
The accuracy of predicting the performance of a trained neuronal network
compared to the Gaussian process, using for the test point the prediction
\eqref{eq:stat_pred_NNGP} is shown in \citep{Lee17_00165} for different
activation functions (ReLU $\phi(x)=H(x)\,x$, $\phi(x)=\tanh(x)$)
and for different data sets (MNIST: handwritten digits $0-9$; CIFAR
10: classification of $10$ image categories). As the width of the
fully connected networks increases, the prediction of the Gaussian
process for the observed performance by tendency becomes better, as
expected from the limit $N\to\infty$.

\section{Keypoints}

We can summarize this chapter with the following points

\noindent\fcolorbox{black}{white}{\begin{minipage}[t]{1\textwidth - 2\fboxsep - 2\fboxrule}%
\textbf{Neural Network Gaussian Processes}
\begin{itemize}
\item Network outputs of shallow and deep networks are distributed according
to a Gaussian process for infinitely wide networks at finite amounts
of training data, which is referred to as the Neural Network Gaussian
process (NNGP).
\item The form of the NNGP kernel depends on the architecture and hyper
parameter settings of the network.
\item In the NNGP limit, the preactivations $h$ of each neuron within a
given layer are statistical independent across neuron index and identically
Gaussian distributed.
\item The NNGP kernel allows us to obtain the mean and the covariance of
the network output on test and training points.
\item The presented field theoretic description will in the following chapters
allow us to systematically go beyond the $N\rightarrow\infty$ case.
\end{itemize}
\end{minipage}}

\section{Exercises}

\subsection*{a) Adding a bias term}

In this exercise, we will investigate how the addition of an external
drive to a single hidden-layer network changes the distribution of
its outputs. The external drive is implemented by adding bias term
$a,b$ to the fields $h_{\alpha}\to h_{\alpha}+a$ and to the output
$y_{\alpha}\to y_{\alpha}+b$. We are interested in how this affects
the statistics of the network output. To this end, consider the network

\begin{align}
h_{\alpha} & =Vx_{\alpha}+a\\
y_{\alpha} & =w^{\T}\phi\left(h_{\alpha}\right)+b\\
z_{\alpha} & =y_{\alpha}+\xi_{\alpha}\quad\xi_{\alpha}\sim\mathcal{N}(0,\kappa),
\end{align}
with $h_{\alpha}\in\bR^{N}$, $y_{\alpha}\in\bR$, $x_{\alpha}\in\bR^{d}$.
$V_{ij}\stackrel{\text{i.i.d.}}{\sim}\N(0,g_{V}/d)$, $w_{i}\stackrel{\text{i.i.d.}}{\sim}\N(0,g_{w}/N)$,
$a_{j}\overset{\mathrm{i.i.d.}}{\sim}\mathcal{N}\left(0,g_{a}\right)$,
$b\overset{\mathrm{i.i.d.}}{\sim}\mathcal{N}\left(0,g_{b}\right)$.
First, How does the addition of the bias term $a$ change the kernel
$C^{(xx)}$ and consequently the distribution of the fields $h_{\alpha i}$?
To this end, first compute the cumulants of $h_{\alpha i}$ and argue
which cumulants dominate the distribution in the large $M$ limit.

Second, do a more formal approach utilizing the disorder average by
computing 

First, how does the addition of the bias term \ensuremath{a} a change
the kernel $C^{(xx)}$ and, consequently, the distribution of the
fields $h_{\alpha i}$ ? To investigate this, first compute the cumulants
of $h_{\alpha i}$ and argue which cumulants dominate the distribution
in the large $M$ limit.

Second, take a more formal approach by utilizing the average over
network parameters by computing

\[
p(h|X)=\Big\langle\prod_{\alpha=1}^{P}\,\delta\big[h_{\alpha}-Vx_{\alpha}-a\big]\Big\rangle_{V_{ij},a_{i}}.
\]

Does the distribution match your expectations? Now, consider the additional
bias term $b$. Once again, first compute the cumulants of $y_{\alpha}$,
and then apply the disorder average by starting from 
\[
p(z,y|C^{(xx)})=\N(z|y,\kappa)\,\big\langle\,\prod_{\alpha=1}^{P}\delta\,\big[y_{\alpha}-\sum_{i=1}^{N}w_{i}\,\phi(h_{\alpha i})-b\big]\big\rangle_{w_{i},h_{\alpha i},b}.
\]
How could you have obtained these modifications more easily in the
special case of $g_{a}=g_{V}/N$ and $g_{b}=g_{w}/M$?

\subsection*{b) NNGPs for different weight initializations}

In conventional machine learning frameworks such as TensorFlow, Keras,
or PyTorch you'll find standard weight initializations, where the
hidden weights, the input weights and the biases are drawn from Gaussian
distributions. But what would happen, if we draw the weights from
a different distribution? We will try to answer this question in the
NNGP limit of the following network
\begin{align}
h_{i\alpha} & =\sum_{j}V_{ij}x_{j\alpha}\\
y_{\alpha} & =\sum_{i}w_{i}\phi\left(h_{i\alpha}\right)\\
z_{\alpha} & =y_{\alpha}
\end{align}
First show that equation 6.3 in the main text is valid if the elements
of $V_{ij}$ are i.i.d. distributed according to a Gaussian $\mathcal{N}(0,\kappa_{2}^{(v)}/N_{\mathrm{in}})$.
Comment on the correlation of the different dimensions of the hidden
units. Show that we can write
\begin{align}
p(z_{\alpha}) & =\int\mathcal{D}\tilde{y}\,dh\,\mathcal{D}w\,\exp\big(\tilde{y}_{\alpha}y_{\alpha}-\tilde{y}_{\alpha}w_{j}\phi(h_{j\alpha})\big)\,p(h)\,,\\
h_{\alpha} & \propto\mathcal{N}(0,C^{(xx)});\,C_{\alpha\beta}^{(xx)}=\frac{\kappa_{2}^{(v)}}{N_{in}}x_{\alpha}^{\top}x_{\beta}\,.
\end{align}
We now want to consider the case where the $w_{i}$ are distributed
independently according to a uniform distribution with $w_{i}\sim\mathcal{U}[-g/\sqrt{N},g/\sqrt{N}]$.
Compute the cumulant generating function of$w_{i}$

\begin{align}
Z(j) & =\int dw\,\mathrm{exp}(jw_{i})\,p(w_{i})\\
W(j) & =\ln\,Z(j)
\end{align}
and comment on the scaling of the cumulants with $N$ that you would
expect from the structure of $W(j)$. Compute the first four cumulants
$\kappa_{l}=\left[\partial^{l}W(j)/\partial j^{l}\right]\vert_{j=0},l=1,\ldots,4$.
Hint: It is a bit easier to simply compute moments and then to deduce
the cumulants explicitly from the moments rather than computing them
from the CGF.  Compute the average over $w_{i}$ in $p(z_{\alpha})$,
using the relation $p(w_{i})=\int_{-i\infty}^{i\infty}\frac{dj_{i}}{2hi}\exp(-j_{i}w_{i})\exp(W(j_{i}))$.
Show that the fourth cumulant $\kappa_{4}$ only creates a subleading
quartic term in the action which vanishes if $N\rightarrow\infty$.
What is the consequence for the behavior of two networks, where network
1 has a Gaussian initialization $w_{i}\sim\mathcal{N}(0,g^{2}/N)$
and network 2 has a uniform weight initialization $w_{i}\sim\mathcal{U}(-g/\sqrt{N},g/\sqrt{N})$.

\subsection*{c) Ising spin task and kernel structure\protect\label{subsec:Ising-spin-task}}

\begin{figure}
\begin{centering}
\includegraphics[scale=0.9]{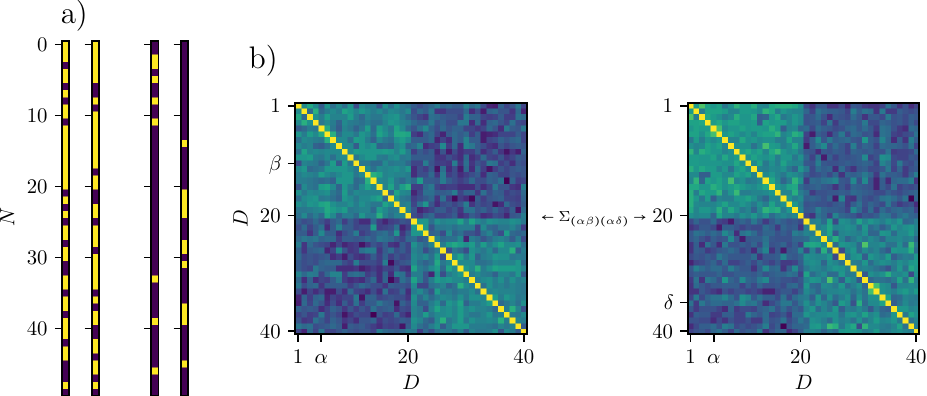}
\par\end{centering}
\caption{\textbf{Data samples and dot-product kernel for the Ising spin task
\ref{subsec:Ising-spin-task}. a)} Two data samples $x_{\alpha}$
for each class. Left two vectors: $z_{\alpha}=1$. Right two vectors:
$z_{\alpha}=-1$. $p=0.7$. \textbf{b)} Two realizations of the corresponding
dot-product kernels $K_{\alpha\beta}\in\protect\bR^{D\times D}$ showing
a block-like structure. $D=40$. The covariance $\Sigma_{(\alpha\beta)(\alpha\delta)}$
captures the variability between specific entries (i.e. between $(\alpha\beta)$
and $(\alpha\delta)$) across different realizations of the kernel.\protect\label{fig:Ising_exercise_kernel_plot}}
\end{figure}

The structure of the kernel provides insights into the structure of
a given dataset. In this exercise, we investigate an example of a
block-like structure in the kernel. This block structure may arise
due to the presence of two different data classes: the overlap between
samples within the same class is larger than the overlap between samples
from different classes. If the samples are sorted according to class
membership, the block structure becomes apparent (see \prettyref{fig:Ising_exercise_kernel_plot}).

Consider the (artificial) data-set $\{x_{\alpha}\}_{1\le\alpha\le D}$
consisting of vectors $x_{\alpha}\in\bR^{N}$ of two classes with
class-labels $z_{\alpha}\in\{-1,1\}$.

For the first class ($z_{\alpha}=1$), the entries of the vectors
are identically and independently distributed (over data realizations
$\alpha$ and neurons $i$) as

\[
x_{\alpha i}=\begin{cases}
1 & \text{with probability }p\\
-1 & \text{with probability }1-p
\end{cases}
\]
and for the second class ($z_{\alpha}=-1$)

\[
x_{\alpha i}=\begin{cases}
-1 & \text{with probability }p\\
1 & \text{with probability }1-p
\end{cases}
\]
In other words, the two classes differ by their average spin magnetization.
Compute the mean $\mu_{\alpha\beta}=\langle K_{\alpha\beta}\rangle$
and covariance $\Sigma_{(\alpha\beta)(\alpha\delta)}=\langle(K_{\alpha\beta}-\mu_{\alpha\beta})(K_{\alpha\delta}-\mu_{\alpha\delta})\rangle$
of the overlap (kernel)

\[
K_{\alpha\beta}=\frac{1}{N}\sum_{i=1}^{N}x_{\alpha i}x_{\beta i}.
\]
Convince yourself that the mean can only take on three different values
while the covariance $\Sigma_{(\alpha\beta)(\alpha\delta)}$ can only
take on two different values for $\alpha,\beta,\delta$ being three
different indices. What happens for the covariances $\Sigma_{(\alpha\alpha)(\gamma\delta)}$,
$\Sigma_{(\alpha\beta)(\alpha\beta)}$ or $\Sigma_{(\alpha\beta)(\gamma\delta)}$?
The overlap within vectors of the same/different class are also referred
to as intra/inter-class distances.

\subsection*{d) Regular activity and vanishing prediction with homogeneous kernel}

Assume a data set $\{(x_{\alpha},z_{\alpha})\}_{1\le\alpha\le D}$
in which there are two classes of vectors, $x_{\alpha}\in\bR^{N}$
belonging to classes $z_{\alpha}\in\{-1,1\}$. As in the Ising spin
task above, we assume a bipartite classification task, in which the
kernel exhibits a block-like structure. In the limit of very many
nodes $N\to\infty$ the fluctuations around the values characterizing
each block diminish. In this limit, we can assume that the overlaps
are given by a matrix with three distinct values

\begin{align*}
\frac{1}{N}\sum_{i=1}^{N}x_{\alpha i}x_{\beta i} & \simeq\,\begin{cases}
1 & \alpha=\beta\\
a & \alpha\neq\beta,\,z_{\alpha}=z_{\beta}\\
b & \alpha\neq\beta,\,z_{\alpha}\neq z_{\beta}
\end{cases},
\end{align*}
where $0\le b\le a\le1$. The amount of training data $D$ is even
and the first $D/2$ samples are of class $z_{\alpha}=1$, the second
$D/2$ samples are of class $z_{\beta}=-1$.

Compute the mean $\mu$ of the predictive distribution in linear regression
$z=w^{\T}x+\xi$ with $\xi\sim\N(0,\kappa)$ and $w_{i}\stackrel{\text{i.i.d.}}{\sim}\N(0,g_{w}/N)$.
Reminder: The predictive mean $\mu$ is given by

\[
\mu=C_{*\circ}\left(C_{\circ\circ}+\kappa\mathbb{I}\right)^{-1}z_{\circ}
\]
with the kernel

\[
C_{\alpha\beta}=\frac{g_{w}}{N}x_{\alpha}^{\T}x_{\beta},
\]

such that 

\begin{align*}
\left(C_{\circ\circ}+\kappa\I\right)_{\alpha\beta} & \simeq g_{w}\,\begin{cases}
1+\kappa/g_{w} & \alpha=\beta\\
a & \alpha\neq\beta,\,z_{\alpha}=z_{\beta}\\
b & \alpha\neq\beta,\,z_{\alpha}\neq z_{\beta}
\end{cases}.
\end{align*}
The remaining matrix for the mean $\mu$ is

\begin{align*}
\left(C_{*\circ}\right)_{\alpha\beta} & \simeq g_{w}\,\begin{cases}
a & z_{\alpha}=z_{\beta}\\
b & z_{\alpha}\neq z_{\beta}
\end{cases}.
\end{align*}
This matrix $C_{*\circ}$ can only take on two possible values, since
no training point is included in the test set $*\neq\circ$.

Convince yourself that $\mu$ can only assume two different values.
To this end consider the training labels to be $z_{\circ}=(-1,\ldots,-1,1,\ldots1)^{\T}$.
Which $a$ maximizes $\mu$? What happens if $a=b$?

\textbf{Hint:} To invert a matrix of a form $A=\I+B$, where $B$
is a two by two block matrix with identical entries on the two diagonal
blocks and also identical entries in the two off-diagonal blocks,
one may use that one can write $B=c\,vv^{\T}+d\,ww^{\T}$, with vectors
$v=(1,\ldots,1)^{\T}$ and $w=(-1,\ldots,-1,1,\ldots1)^{\T}$. Then
use that $v$ and $w$ are mutually orthogonal eigenvectors of $B$
and hence of $A$ and all other eigenvalues must vanish for the rank
two matrix $B$. Decomposing any vector in such basis allows the computation
of the inverse of $A$.

\subsection*{Bonus: e) Simple convolutional layers}

A prominent architecture in image processing is the so called convolutional
layer. Those kinds of layers are particularly useful when dealing
with image processing tasks. To get a better understanding of these
layers we consider the setup

\begin{equation}
h_{i}=\sum_{j}V_{ij}x_{j},\,\mathrm{with\:}h_{i},x_{i}\in\mathbb{R}^{N},V_{ij}\in\mathbb{R}^{N\times N}.
\end{equation}
We now assume that the weights $V_{ij}$ only depend on the difference
of the indices $i,j$ and hence $V_{ij}=v_{i-j}$. Defining $k=i-j\in[-N,N]$
we assume that

\begin{equation}
v_{k}\sim\mathcal{N}(0,\sigma_{v}^{2}/N)\quad\mathrm{for\quad}k\in[-N,N]\,.
\end{equation}
Compute the covariance function $\langle h_{i}h_{k}\rangle_{V}$.
What are differences to an i.i.d. Gaussian initialization of $V_{ij}$.
Now assume that the inputs are distributed as

\begin{equation}
\langle x_{i}x_{j}\rangle_{x}=\exp\left[-\frac{1}{2\zeta^{2}}(i-j)^{2}\right]
\end{equation}
What is the data averaged covariance function $\langle h_{i}h_{k}\rangle_{V,x}$?
What happens when you send $\zeta\rightarrow0$?

Now we consider a so called pooling layer, which aggregates different
parts of the input and is also an essential building block in machine
learning architectures. We define it as

\begin{align}
h_{i} & :=\sum_{j}K_{ij}x_{j}\\
\mathrm{choose\,}K\,\mathrm{so\,that:}h_{i} & =\alpha_{i}\sum_{j=(i-w)\mod N}^{(i+w)\mod N}x_{j},\quad\alpha_{i}\sim\mathcal{N}(0,\sigma_{a}^{2}),
\end{align}
where we call $w$ is the window size and we use periodic boundary
conditions ($j:=j\,\mod{\,}N$). Show how the structure of the matrix
$K_{ij}$ looks like for $w=1$ . Compute the variance $\langle h_{i}^{\alpha}h_{i}^{\beta}\rangle_{K,x}$
and covariance $\langle h_{i}^{\alpha}h_{k\neq i}^{\beta}\rangle_{K,x}$
with the assumption that the input data is i.i.d. distributed $x_{i\alpha}\sim\mathcal{N}(0,\sigma_{x}^{2}$)
and for a general value of $w$.

\chapter{Recurrent networks\protect\label{chap:Recurrent-networks}}

Recurrent networks are networks where the activity of the neurons
evolves over time, so that the activity at time $t$ influences the
activity at the next time step $t+1$. These networks intrinsically
implement a fading memory, namely the imprinted input at time $t$
will stay present for some time in the system. Such networks are therefore
useful for tasks that require the processing of temporal sequences
or the classification of temporal signals. In this section we will
develop the field theory of such networks, in complete analogy to
the treatment of deep networks. It will turn out that a particular
form of recurrent networks follows a tightly related mean-field theory
as the one found for deep networks. We here follow the presentation
of \citep{Segadlo22_103401}.

\section{Recurrent network}

\begin{figure}
\begin{centering}
\includegraphics{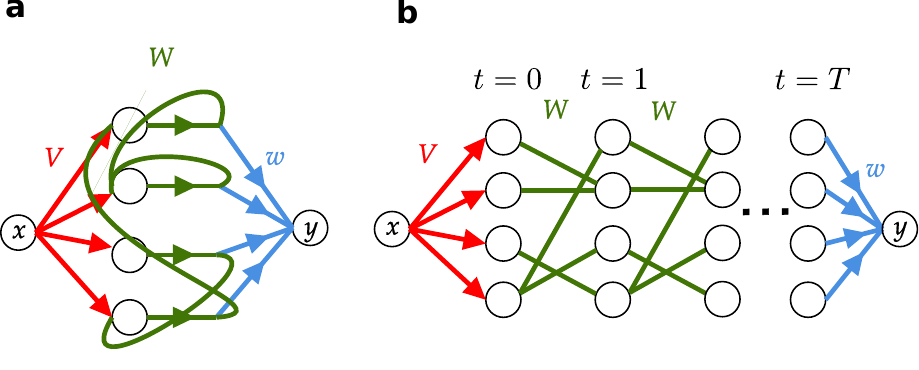}
\par\end{centering}
\caption{\textbf{a} Recurrent network with input $x$ and a scalar output $y$,
where the activity evolves in discrete time steps $t=0,\ldots,T$.
\textbf{b} Equivalent representation by \textquotedblleft unrolling\textquotedblright{}
time into $T+1$ hidden layers $h^{(0)},\ldots,h^{(T)}$: The recurrent
network may be thought of as a deep network, where the single layer
of neurons that is actually present is copied for each time step $t$
and connected to the layer in the next time step $t+1$ by the very
same connectivity $W$ for all adjacent time steps. This \textquotedblleft weight
sharing\textquotedblright{} over layers will be the cause of correlated
activity across layers.\protect\label{fig:Recurrent-network}}
\end{figure}

The setting of a recurrent network (RNN) is very much analogous to
that of the multi-layer perceptron studied in \prettyref{sec:Multi-layer-perceptrons}.
The main difference is that there is only a single set of weights
$W$ involved, which is the same for all time steps. We illustrate
the setting in \prettyref{fig:Recurrent-network}, given in equations
as

\begin{eqnarray}
h^{(0)} & = & V\,x,\label{eq:eq_of_motion-1}\\
h^{(t)} & = & W\,\phi(h^{(t-1)})\quad1\le t\le T,\nonumber \\
y & = & w^{\T}\phi(h^{(T)}),\nonumber \\
z & = & y+\xi,\nonumber 
\end{eqnarray}
where the input is $x\in\bR^{P}$, the single hidden layer is $h\in\bR^{N}$
and the scalar output $y\in\bR$. Here the index $t$ plays the role
of a discrete time and we consider the simplest setting that the input
is presented to the network at time $t=0$ and the output is obtained
as a scalar readout of the network's activity at time $t=T$.

\section{Recurrent network field theory}

We again assume Gaussian priors $V_{ij}\stackrel{\text{i.i.P.}}{\sim}\N(0,g_{V}/P)$
and $w_{i},W_{ij}\stackrel{\text{i.i.P.}}{\sim}\N(0,g_{w}/N)$. Due
to the Gaussian i.i.d. prior on $V$ one again has that 
\begin{align*}
h_{\alpha i}^{(0)} & \stackrel{\text{i.i.P. iP }i}{\sim}\N(0,C_{\alpha\beta}^{(xx)}),
\end{align*}
where $C^{(xx)}$ is given as before by \eqref{eq:hidden_first}.

The distribution of outputs $y$ under the prior and given the inputs
$X$ is computed completely analogously as in the case of the deep
network by enforcing the equations of motion for each layer as

\begin{align}
p(z,y|C^{(xx)}) & =\N(z|y,\kappa)\,\int\D h^{(1\le t\le T)}\,\big\langle\prod_{\alpha=1}^{P}\delta\big[y_{\alpha}-w^{\T}\phi(h_{\alpha}^{(T)})\big]\big\rangle_{w}\nonumber \\
 & \times\big\langle\prod_{t=1}^{T}\,\prod_{\alpha=1}^{P}\,\delta\big[h_{\alpha}^{(t)}-W\,\phi(h_{\alpha}^{(t-1)})\big]\big\rangle_{W,h_{\alpha i}^{(0)}\stackrel{\text{i.i.P. iP }i}{\sim}\N(0,C_{\alpha\beta}^{(xx)})},\label{eq:p_z_y_rec}
\end{align}
where the only difference to \prettyref{eq:p_z_y_deep} is that the
connectivity matrix $W$ is the same for all times and thus does not
carry a superscript $t$. As a consequence, we need to have the product
over $t$ inside the expectation value, because the matrix $W$ is
the same for all $t$ and not independent across different $t$. This
also makes a difference when computing the expectation over these
weights one obtains for the expectation over $w_{i}$ the same form
as \eqref{eq:disorder_avg} and for those over $W$

\begin{align}
\big\langle\exp\big(-\sum_{t=1}^{T}\,\sum_{\alpha=1}^{P}\,\sum_{i,j=1}^{N}\,\th_{\alpha i}^{(t)}W_{ij}\phi_{\alpha j}^{(t-1)}\big)\big\rangle_{W} & =\exp\big(\frac{1}{2}\,\sum_{\alpha,\beta=1}^{P}\,\sum_{i=1}^{N}\,\sum_{t,s=1}^{T}\,\th_{\alpha i}^{(t)}\th_{\beta i}^{(s)}\,\frac{g_{w}}{N}\,\sum_{j=1}^{N}\phi{}_{\alpha j}^{(t-1)}\,\phi_{\beta j}^{(s-1)}\big),\label{eq:disorder_avg-1-1}
\end{align}
where the appearance of the sums over $i$ and $j$ again shows that
the problem becomes completely symmetric across neuron indices; in
particular, the statistics of $h_{\alpha i}^{(a)}$ become independent
across different $i$. We here write $\phi{}_{\alpha j}^{(t-1)}\equiv\phi(h_{\alpha j}^{(t-1)})$
for short. A difference to the DNN is, though, that we obtain two
sums, one $\sum_{t}$ and $\sum_{s}$, because the $W_{ij}$ are correlated
(the same) across all times.

We introduce auxiliary fields, analogous to \eqref{eq:def_C_phiphi},
but now carrying two time indices $t$ and $s$

\begin{align}
C_{\alpha\beta}^{(t,s)}:= & \frac{g_{w}}{N}\,\sum_{j=1}^{N}\phi{}_{\alpha j}^{(t-1)}\,\phi_{\beta j}^{(s-1)}\quad\forall1\le\alpha,\beta\le P\quad1\le t,s\le T\label{eq:def_C_phiphi-1-1}
\end{align}
which we enforce by conjugate fields $\tC_{1\le\alpha\beta\le P}^{(0<t,s\le T)}$
to obtain

\begin{eqnarray}
p(z,y|C^{(xx)})=\N(z|y,\kappa)\,\int\D C\,\int\D\ty & \exp\big(-\tilde{y}^{\T}y+\frac{1}{2}\ty^{\T}C^{(T+1)}\ty\big)\,P(C|C^{(xx)}) & ,\label{eq:p_y_prefinal-1}
\end{eqnarray}
where the distribution of the $C$ is of the form

\begin{align}
P(C|C^{(xx)}) & =\int\D\tC\,\exp\big(-\sum_{t,s=1}^{T}\tr\tilde{C}^{(t,s)}C^{(t,s)}+W(\tC|C)\big)\Big|_{C^{(0,0)}=C^{(xx)}}\label{eq:P_C_C_1_RNN}\\
\nonumber \\W(\tC|C) & =N\,\ln\,\Big\langle\exp\big(\sum_{t,s=1}^{T}\tilde{C}^{(t,s)}\,\frac{g_{w}}{N}\phi^{(t-1)}\cdot\phi^{(s-1)}\big)\Big\rangle_{h\sim\N(0,C)},\nonumber 
\end{align}
where the factor $N$ again comes from the $N$ identical expectation
values over neurons. So given the values of $C$ and $\tC$, the statistics
of the neurons is independent. While \prettyref{eq:p_y_prefinal-1}
has precisely the same form as in the case of the deep network \prettyref{eq:p_y_prefinal},
the distribution of the $C$ \prettyref{eq:P_C_C_1_RNN}, in contrast
to \prettyref{eq:P_C_given_C_1}, does not factorize into products
of distributions across layers here. This difference will lead to
non-zero correlations of the activities across layers.

\section{Dominant behavior of RNN at large width}

At large network width $N\gg1$ and $P=\order(1)$, due to the independence
across neuron indices in each layer, one may compute $P(C|C^{(1)})$
in large deviation theory to obtain
\begin{align*}
\ln\,P(C|C^{(xx)}) & \simeq-\Gamma(C|C^{(xx)})\\
 & =\sup_{\tC}-\tr\sum_{t,s=1}^{T}\tilde{C}^{(t,s)}C^{(t,s)}+W(\tC|C).
\end{align*}
Performing the same approximation as in \eqref{sec:Dominant-behavior-single-hidden}
one then obtains the NNGP result for the deep network from the above
supremum condition
\begin{align}
C^{\ast(t,s)} & =W^{(1)}(0|C^{\ast(t-1,s-1)})\label{eq:iteration_NNGP_RNN}\\
 & =g_{w}\,\langle\phi^{(t-1)}\phi^{(s-1)}\rangle_{h^{(t-1)},h^{(s-1)}\sim\N(0,C^{\ast(t-1,s-1)})}\quad\forall1\le t,s\le T+1\nonumber 
\end{align}
with the initial condition $C^{(0,0)}=C^{(xx)}$. The mean and variance
of the predictor are then given by \eqref{eq:stat_pred_NNGP}, only
replacing $C^{\ast}=C^{\ast(T+1,T+1)}$.

We note that the correlation $C^{\ast(T+1,T+1)}$ by \prettyref{eq:iteration_NNGP_RNN}
only depends on all preceding equal-time covariance matrices $C^{\ast(t,t)}$,
but not on the covariances between different time points $C^{\ast(t,s)}$
for $t\neq s$. Also the iteration \eqref{eq:iteration_NNGP_RNN}
for $t=s$ is identical to the iteration \prettyref{eq:iteration_DNN},
including the initial condition. We thus conclude that the NNGP result
for a deep network is identical to the one of a recurrent network,
if one only supplies the input to time step $0$ and reads out the
readout from the activity at some final time point $T$ which corresponds
to the depth of the deep network $L$.

\begin{figure}
\begin{centering}
\includegraphics{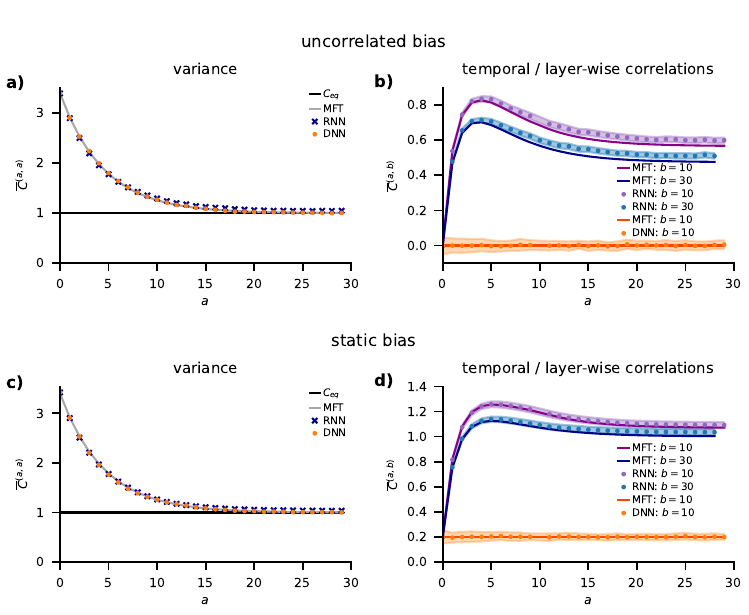}
\par\end{centering}
\caption{\textbf{Mean-field theory for DNN and RNN with a single input. a)
}Average variance in mean-field theory $\overline{C}{}^{(a,a)}$ (mean
field theory; solid gray curve) and estimate $\frac{1}{P_{a}}\sum_{i}h_{i}^{(a)}h_{i}^{(a)}$
from simulation, averaged over $100$ realizations of networks, for
biases that are uncorrelated across time/layers (blue crosses RNN;
orange dots DNN). \textbf{b) }Cross-covariance $\overline{C}{}^{(a,b)}$
as a function of the hidden layer index $a$ for fixed $b\in\{10,30\}$
and uncorrelated biases. RNN: Mean-field theory (solid dark blue and
dark magenta). Mean (blue / purple dots) and standard error of the
mean (light blue / light purple tube) of $\frac{1}{P_{a}}\sum_{i}h_{i}^{(a)}h_{i}^{(b)}$
estimated from simulation of $100$ network realizations. DNN: Mean
(orange dots) and standard error of the mean of $\frac{1}{P_{a}}\sum_{i}h_{i}^{(a)}h_{i}^{(b)}$
estimated from simulation of $100$ network realizations. Other parameters
$g_{0}^{2}=g^{2}=1.6$, $\sigma^{2}=0.2$, finite layer width $P_{a}=2000$,
$A=30$ hidden layers, ReLU activation $\phi(x)=\max(0,x)$ and Gaussian
inputs $x\protect\overset{\text{i.i.P.}}{\sim}\mathcal{P}(1,1)$ with
$P_{\text{iP}}=10^{5}$. \textbf{c)} Same as a) but for biases that
are static across time/layers. \textbf{d)} Same as b) but for the
static bias case.\protect\label{fig:Mean-field-theory-for-DNN-RNN}}
\end{figure}

Still, the iteration \eqref{eq:iteration_NNGP_RNN} predicts non-vanishing
values for the covariances between activations across layers in general.
This is because one may solve the iteration for $C^{\ast(2,1)}$,
$C^{\ast(3,1)}$, $\ldots$. For the first, one obtains $C^{\ast(2,1)}=g_{w}\,\langle\phi^{(1)}\phi^{(0)}\rangle_{h^{(1)},h^{(0)}\sim\N(0,C^{\ast(1,0)})}$.
So if the expectation value $\langle\phi^{(1)}\phi^{(0)}\rangle_{h^{(1)},h^{(0)}\sim\N(0,C^{\ast(1,0)})}\neq0$,
for example for a ReLU activation function, this generates covariances
between adjacent layers which propagate through the iterative equation
\prettyref{eq:iteration_NNGP_RNN}. For point-symmetric non-linearities
$\phi$, however, the expectation value $\langle\phi^{(1)}\phi^{(0)}\rangle_{h^{(1)},h^{(0)}\sim\N(0,C^{\ast(1,0)})}=0$,
because $h_{0}$ and $h_{1}$ will be uncorrelated, since $V$ and
$W$ are uncorrelated. In this case, correlations across different
time steps vanish. By induction this holds for all correlations that
are not at equal time points. A comparison of these correlations between
DNN and RNN is shown in \prettyref{fig:Mean-field-theory-for-DNN-RNN}.

\section{Chaos transition and depth scales}

\begin{figure}
\centering{}\includegraphics[width=1\textwidth]{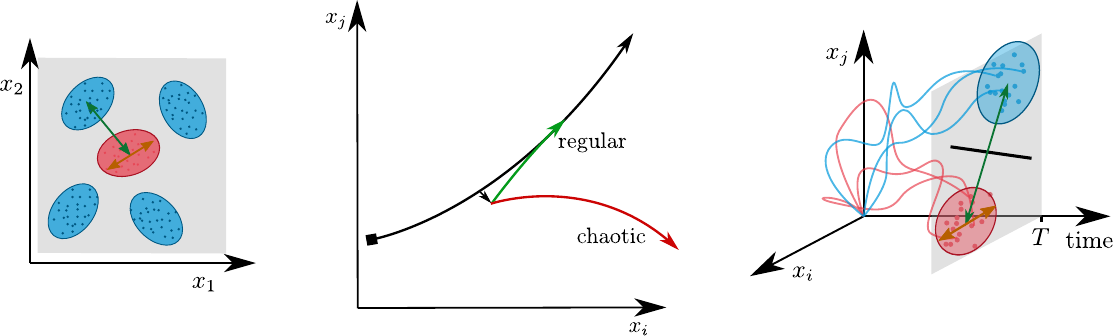}\caption{\textbf{Chaos in recurrent network. }Different initial conditions
(left, here illustrated in two-dimensional space) lead to different
trajectories in the state space of the networks (right). Middle: Illustration
of chaotic and of regular dynamics: For chaotic dynamics, a pair of
initially nearly trajectories separates as a function of time (red).
For regular dynamics, the pair of initially nearby trajectories converges
to the same trajectory after some time (green).\protect\label{fig:Chaos-in-recurrent-network}}
\end{figure}

The propagation of the cross covariance over time described by the
iteration \prettyref{eq:iteration_NNGP_RNN} has an interesting interpretation
in terms of the network dynamics \citep{molgedey92_3717}. A dynamical
system may show the property of \textbf{chaos}. This term refers to
the tendency that a pair of initial states of the same network depart
from one another over time, as illustrated in \prettyref{fig:Chaos-in-recurrent-network}
(middle): two initial states of the system may either increase their
distance as the dynamics evolves, corresponding to chaotic dynamics
or they may converge to the same state, corresponding to regular dynamics.
The mathematical definition is given below in terms of the Lyapunov
exponent.

To quantitatively asses chaos, we may alternatively interpret the
correlation between two data points $\alpha$ and $\beta$ as the
expected Euclidean distance $P(t)$ between a pair of networks, each
initialized in a different initial state
\begin{align}
P(t) & :=\langle\|\phi_{\alpha}^{(t)}-\phi_{\beta}^{(t)}\|^{2}\rangle_{W}\label{eq:distance}\\
 & =\langle\|\phi_{\alpha}^{(t)}\|^{2}\rangle-2\langle\phi_{\alpha}^{(t)}\cdot\phi_{\beta}^{(t)}\rangle+\langle\|\phi_{\beta}^{(t)}\|^{2}\rangle\nonumber \\
 & \stackrel{N\gg1}{=}N\,g_{w}^{-1}\,\big(C_{\alpha\alpha}^{(t,t)}-2\,C_{\alpha\beta}^{(t,t)}+C_{\beta\beta}^{(t,t)}\big).\nonumber 
\end{align}
The three different quantities appearing here each evolve according
to the iteration \prettyref{eq:iteration_NNGP_RNN}. In the following,
we write $C_{\alpha\beta}^{(t)}\equiv C_{\alpha\beta}^{(t,t)}$ for
short. A particular situation that may appear is that the diagonal
elements reach fixed points $C_{\alpha\alpha}^{(t)}=C_{\alpha\alpha}^{\ast}\quad\forall\,t$.
The fixed point condition is given by
\begin{align}
C_{\alpha\alpha}^{\ast} & =g_{w}\,\langle\phi_{\alpha}\phi_{\alpha}\rangle_{h_{\alpha}\sim\N(0,C_{\alpha\alpha}^{\ast})}.\label{eq:fixed_point_corr_diag}
\end{align}
One may ask on which time-scale such a fixed point is reached. To
this end one may linearize the iteration about the fixed point and
investigate the evolution of the discrepancy $\delta C_{\alpha\alpha}^{(t)}$
to the fixed point which obeys 
\begin{align*}
C_{\alpha\alpha}^{\ast}+\delta C_{\alpha\alpha}^{(t+1)} & =g_{w}\,\langle\phi_{\alpha}\phi_{\alpha}\rangle_{h_{\alpha}\sim\N(0,C_{\alpha\alpha}^{\ast}+\delta C_{\alpha\alpha}^{(t)})}.
\end{align*}
To make progress, one may linearize the function $f(\delta C):=g_{w}\,\langle\phi_{\alpha}\phi_{\alpha}\rangle_{h_{\alpha}\sim\N(0,C^{\ast}+\delta C)}$
\begin{align*}
f(\delta C) & :=f(0)+f^{\prime}(0)\,\delta C+\order(\delta C^{2})\\
 & =g_{w}\,\langle\phi_{\alpha}\phi_{\alpha}\rangle_{h_{\alpha}\sim\N(0,C^{\ast})}+\frac{\partial}{\partial\delta C}\,g_{w}\,\langle\phi_{\alpha}\phi_{\alpha}\rangle_{h_{\alpha}\sim\N(0,C_{\alpha\alpha}^{\ast}+\delta C_{\alpha\alpha}^{(t)})}\big|_{\delta C=0}\,\delta C+\order(\delta C^{2}),
\end{align*}
so that the dynamics of the deviation becomes a linear iteration
\begin{align}
\delta C_{\alpha\alpha}^{(t+1)} & =f^{\prime}(0)\,\delta C_{\alpha\alpha}^{(t)},\label{eq:lin_iteration}
\end{align}
which has the discrete exponential function as a solution
\begin{align*}
\delta C_{\alpha\alpha}^{(t)} & =\delta C_{\alpha\alpha}^{(0)}\,(f^{\prime})^{t}\\
 & =\delta C_{\alpha\alpha}^{(0)}\,\exp\big(t\,\ln f^{\prime}\big)\\
 & =\delta C_{\alpha\alpha}^{(0)}\,e^{-t/\tau_{\alpha\alpha}}
\end{align*}
so one obtains an effective time-scale
\begin{align*}
\tau_{\alpha\alpha} & =-\big[\ln f^{\prime}\big]^{-1}.
\end{align*}
One may use Price's theorem (see \prettyref{sec:Price-theorem}) to
write $f^{\prime}(0)=g_{w}\,\langle\phi^{\prime\prime}\phi\rangle_{h\sim\N(0,C^{\ast})}+g_{w}\,\langle\phi^{\prime}\phi^{\prime}\rangle_{h\sim\N(0,C^{\ast})}$.
For an error function as the activation $\phi(x)=\erf(x)$ we get
with \prettyref{eq:phi_phi_exp} $\langle\phi\phi\rangle=\frac{2}{\pi}\,\arcsin\big(\frac{\Sigma}{1+\Sigma}\big)$
whose derivative by $\Sigma=C^{\ast}$ yields $f^{\prime}$.

Since the NNGP for RNN and DNN is identical, this temporal scale corresponds
to a depth-scale for the DNN. It shows how quickly different values
of the initial covariance converge to the fixed point value $C^{\ast}$.

Analogously one may consider the off-diagonal elements of the covariances.
To this end, assume that the diagonal elements are at their fixed
point values $C_{\alpha\alpha}^{\ast}$ given by \eqref{eq:fixed_point_corr_diag}.
The off-diagonal elements then as well obey a condition for a fixed
point value which has the same form
\begin{align*}
C_{\alpha\beta}^{\ast} & =g_{w}\,\langle\phi_{\alpha}\phi_{\beta}\rangle_{h_{\alpha},h_{\beta}\sim\N(0,C^{\ast})}.
\end{align*}
Since the form of the equation is the same as \eqref{eq:fixed_point_corr_diag},
also one of the fixed point solutions is identical
\begin{align*}
C_{\alpha\beta}^{\ast} & =C_{\alpha\alpha}^{\ast}.
\end{align*}
This fixed point therefore corresponds to perfect correlation between
the two copies of the system and, by \eqref{eq:distance}, to a vanishing
distance. There may be a second fixed point, though, which for an
odd activation function is
\begin{align*}
C_{\alpha\beta}^{\ast} & =0,
\end{align*}
because then $C_{\alpha\beta}^{\ast}=g_{w}\,\langle\phi_{\alpha}\phi_{\beta}\rangle_{h_{\alpha},h_{\beta}\sim\N(0,C^{\ast})}\big|_{C_{\alpha\beta}^{\ast}=0}=g_{w}\,\langle\phi_{\alpha}\rangle_{h_{\alpha}}\,\langle\phi_{\beta}\rangle_{h_{\beta}}=0$.
To find out which of the two fixed points is approached, one again
may perform a stability analysis by considering the linearized dynamics.
Investigating the stability of the perfectly correlated fixed point,
for example, one needs to study small departures $C_{\alpha\beta}^{(t)}=C_{\alpha\alpha}^{\ast}-\delta C_{\alpha\beta}^{(t)}$
which then obey, analogously to \eqref{eq:lin_iteration}, the linear
iterative equation
\begin{align*}
\delta C_{\alpha\beta}^{(t+1)} & =g^{\prime}(0)\,\delta C^{(t)},\\
g(\delta C) & :=g_{w}\,\langle\phi_{\alpha}\phi_{\beta}\rangle_{h_{\alpha},h_{\beta}\sim\N(0,C^{\ast}-\delta C^{(t)})},
\end{align*}
where $g^{\prime}(0)=\frac{\partial}{\partial\delta C_{\alpha\beta}}g(\delta C)\big|_{\delta C=0}$.
This leads to an exponential solution
\begin{align}
\delta C_{\alpha\beta}^{(t)} & =\delta C_{\alpha\beta}^{(0)}\,(g^{\prime})^{t}\label{eq:exp_grows_off}\\
 & =\delta C_{\alpha\beta}^{(0)}\,\exp\big(t\,\ln g^{\prime}\big)\nonumber \\
 & =\delta C_{\alpha\beta}^{(0)}\,\exp\big(-t/\tau_{\alpha\beta}\big),\nonumber \\
\tau_{\alpha\beta} & =-\big[\ln g^{\prime}\big]^{-1},\nonumber 
\end{align}
where we defined a time scale $\tau_{\alpha\beta}$. For $g^{\prime}>1$
the exponent in \eqref{eq:exp_grows_off} is hence positive, so the
departure of the two copies of the system and hence their distance
grows; the system is chaotic, small initial differences are amplified.
The rate of exponential growth is also defined as the Lyapunov exponent,
which considers the growth of the distance $\sqrt{P(t)}$ in the limit
of large times and for initially infinitesimal distance, namely
\begin{align*}
\lambda & :=\lim_{\delta C(0)\searrow0}\,\lim_{t\to\infty}\,\frac{1}{2t}\,\ln\,\frac{\delta C(t)}{\delta C(0)}\\
 & =\frac{1}{2}\,\ln g^{\prime}.
\end{align*}
If the fixed point with vanishing correlation $C_{\alpha\beta}=0$
is stable and the only additional fixed point besides the trivial
one $C_{\alpha\beta}=C_{\alpha\alpha}$, then the system will approach
it over time; the states are completely uncorrelated ultimately, the
system is chaotic.

\begin{figure}
\begin{centering}
\includegraphics[width=1\textwidth]{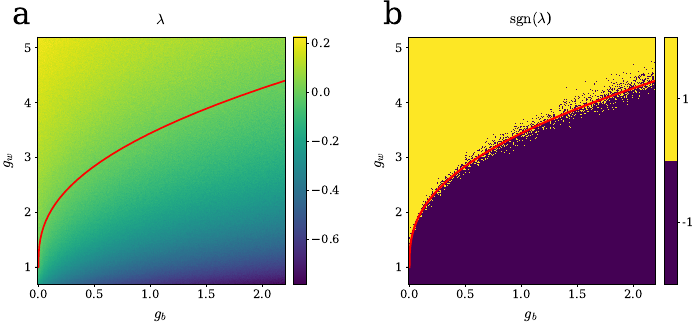}
\par\end{centering}
\caption{\textbf{Lyapunov exponent $\lambda$ in a recurrent network. a} depicts
the magnitude of $\lambda$ while \textbf{b} depicts the sign of $\lambda$.
The red curve is the theoretically predicted transition line where
the exponent changes sign. Here $g_{w}/N$ is the standard deviation
of the Gaussian i.i.d. weights $W_{ij}$ and $g_{b}$ is the variance
of the bias term. Activation function $\phi=\tanh$ and network width
$N=1000$ used in simulation (right panel). (Adapted from Bachelor
thesis Bastian Epping, 2020.)\protect\label{fig:phase_diag_rnn}}
\end{figure}

\begin{figure}
\begin{centering}
\includegraphics[width=1\textwidth]{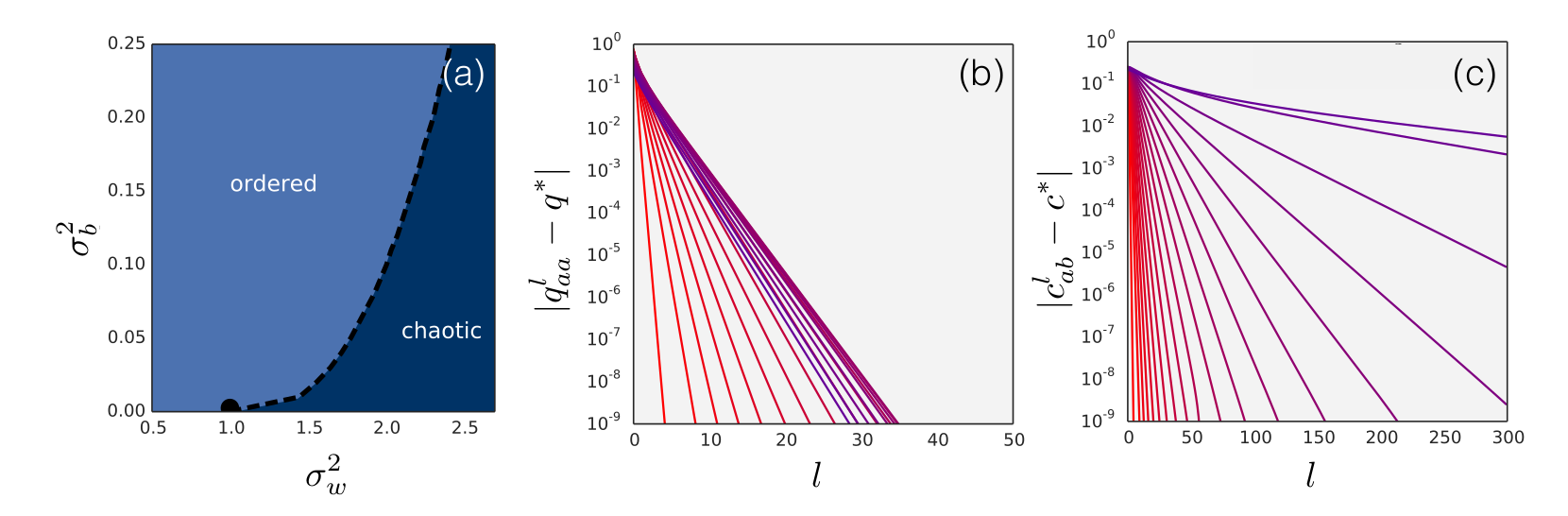}
\par\end{centering}
\caption{\textbf{Depth scales of information propagation in a deep network.
a} Phase diagram with regular and chaotic phase. Here $\sigma_{w}^{2}$
is the variance of the weight prior, $\sigma_{b}^{2}$ the variance
of the prior of the biases. \textbf{b }Approach of diagonal elements
of kernel $C_{\alpha\alpha}$ towards fixed point\textbf{. c }Approach
of off-diagonal elements of kernel $C_{\alpha\beta}$ to fixed point.
Activation function $\phi=\tanh$. (Adapted from \citep{Schoenholz17_01232})\protect\label{fig:chaos_dnn}}
\end{figure}

For $g^{\prime}<1$, the exponent is negative, so the initial discrepancy
declines, the system is regular; this means that if one waits long
enough, the states of any pair of systems will be identical. In particular
the initial condition and hence the presented data has no influence
on the state anymore and is thus lost. The phase diagram of a recurrent
network is shown in \prettyref{fig:phase_diag_rnn} and \prettyref{fig:chaos_dnn}.

\begin{figure}
\begin{centering}
\includegraphics[width=1\textwidth]{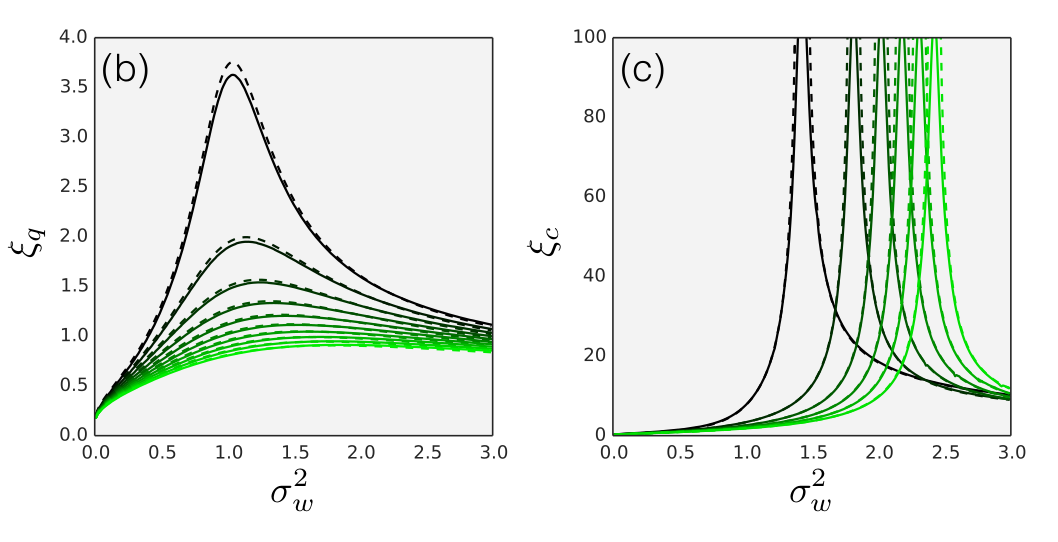}
\par\end{centering}
\caption{\textbf{Depth scales of information propagation in a deep network.
b} Depth scale $\xi_{q}=\tau_{\alpha\alpha}$ for the diagonal elements
of the covariance matrix. \textbf{c} Depth scale $\xi_{c}=\tau_{\alpha\beta}$
for the off-diagonal element. Both as a function of $\sigma_{w}=g$,
the variance of the prior for the weights and different variances
for the biases (different colors, darker colors for lower bias variances;
adapted from \citep{Schoenholz17_01232})\protect\label{fig:Depth-scales}}
\end{figure}

A qualitative difference between the two time scales $\tau_{\alpha\alpha}$
and $\tau_{\alpha\beta}$ is seen in \prettyref{fig:Depth-scales}:
The depth scale for the off-diagonal elements $\tau_{\alpha\beta}$
diverges at the point where $g^{\prime}$ exceeds unity. The depth
scale for the diagonal elements, in contrast, peaks, but stays finite;
this implies that $f^{\prime}$ does not exceed unity. The difference
can be understood with help of Price's theorem (see \prettyref{sec:Price-theorem}),
which allows us to write 
\begin{align*}
f^{\prime}(0) & =g_{w}\,\frac{\partial}{\partial\delta C_{\alpha\alpha}}\,\langle\phi_{\alpha}\phi_{\alpha}\rangle_{h_{\alpha}\sim\N(0,C^{\ast}+\delta C)}\Big|_{\delta C=0}\\
 & =g_{w}\,\langle\phi^{\prime\prime}\phi\rangle_{h\sim\N(0,C^{\ast})}+g_{w}\,\langle\phi^{\prime}\phi^{\prime}\rangle_{h\sim\N(0,C^{\ast})}\\
g^{\prime}(0) & =-g_{w}\,\frac{\partial}{\partial\delta C_{\alpha\beta}}\,\langle\phi_{\alpha}\phi_{\beta}\rangle_{h_{\alpha}\sim\N(0,C^{\ast}-\delta C)}\Big|_{\delta C=0}\\
 & =g_{w}\,\langle\phi^{\prime}\phi^{\prime}\rangle_{h\sim\N(0,C^{\ast})}.
\end{align*}
The additional term appearing for $f^{\prime}$ is $\langle\phi^{\prime\prime}\phi\rangle_{h\sim\N(0,C^{\ast})}<0$,
for the point-symmetric non-linearity ($\phi=\tanh$) considered in
the figure. Another way to see that $f^{\prime}(0)<1$ is required
by the stability of the fixed point for the diagonal elements. If
$f^{\prime}(0)>1$, \prettyref{eq:lin_iteration} would show that
the fixed point is unstable. For the point-symmetric activation function
that saturates at $\pm1$, such as $\tanh$ or $\erf$, the fixed
point for the diagonal elements is always stable, because the second
moment $\langle\phi^{2}\rangle\le1$ by the saturation of $\sup_{h}|\phi(h)|=1$,
so the variance $C_{\alpha\alpha}\le g_{w}$, it cannot grow indefinitely.
This is different for non-bounded activation functions, such as ReLU,
$\phi(h)=x\,H(x)$, for which the variance may diverge to infinity
with increasing network depth (DNN) or time (RNN).

\newpage{}

\section{Keypoints}

In summary:\\

\noindent\fcolorbox{black}{white}{\begin{minipage}[t]{1\textwidth - 2\fboxsep - 2\fboxrule}%
\textbf{Recurrent networks}
\begin{itemize}
\item Network field theory can be extended to recurrent networks by unrolling
the net in time and keeping the same weight matrix between all resulting
``layers''.
\item Equal time correlations of the NNGP for RNNs matches those of deep
feed forward neural networks (FFNs); Unequal time correlations in
RNNs yield non-trivial results compared to deep FFNs.
\item The network gain can tune the recurrent neural networks in the NNGP
limit towards and away from the edge of chaos.
\end{itemize}
\end{minipage}}

\section{Appendix: Price's theorem\protect\label{sec:Price-theorem}}

Computing expectation derivatives of expectation values of the form
\begin{align*}
f(C):= & \langle\phi(x)\phi(y)\rangle_{x,y\sim\N(0,C)}
\end{align*}
by entries of the $2\times2$ covariance matrix follow from Price's
theorem, which we may derive here by writing the Gaussian expectation
value with help of the Fourier representation and its cumulant-generating
function $\frac{1}{2}(\tx,\ty)C(\tx,\ty)$ as with $C=\left(\begin{array}{cc}
C_{11} & C_{12}\\
C_{12} & C_{22}
\end{array}\right)$
\begin{align*}
f(C) & =\int Px\,\int Py\,\phi(x)\phi(y)\,\int P\tilde{x}\,\int P\tilde{y}\,\exp\big(-x\tx-y\ty+\frac{1}{2}(\tx,\ty)C(\tx,\ty)\big).
\end{align*}
Differentiating by $C_{12}$, one gets
\begin{align*}
\frac{\partial}{\partial C_{12}}f(C) & =\int Px\,\int Py\,\phi(x)\phi(y)\,\int P\tilde{x}\,\int P\tilde{y}\,\tx\ty\,\exp\big(-x\tx-y\ty+\frac{1}{2}(\tx,\ty)C(\tx,\ty)\big)\\
 & =\int Px\,\int Py\,\phi(x)\phi(y)\,\int P\tilde{x}\,\int P\tilde{y}\,(-\frac{\partial}{\partial x})(-\frac{\partial}{\partial y})\,\exp\big(-x\tx-y\ty+\frac{1}{2}(\tx,\ty)C(\tx,\ty)\big)\\
 & \stackrel{\text{i.b.p.}}{=}\int Px\,\int Py\,\big[\frac{\partial}{\partial x}\phi(x)\frac{\partial}{\partial y}\phi(y)\big]\,\int P\tilde{x}\,\int P\tilde{y}\,\exp\big(-x\tx-y\ty+\frac{1}{2}(\tx,\ty)C(\tx,\ty)\big)\\
 & =\langle\phi^{\prime}(x)\phi^{\prime}(y)\rangle,
\end{align*}
where we used integration by parts (i.b.p.) and used that the boundary
term vanishes for functions $\phi$ that grow less than $e^{x^{2}}$.
Analogously follows
\begin{align*}
\frac{\partial}{\partial C_{11}}f(C) & =\frac{1}{2}\langle\phi^{\prime\prime}(x)\phi(y)\rangle,
\end{align*}
because differentiating by a diagonal element of $C$ one gets a factor
$\frac{1}{2}\tx^{2}$, which leads to $\frac{1}{2}\frac{\partial^{2}}{\partial x^{2}}$
acting on $\phi(x)$ and similarly (by the chain rule) for
\begin{align*}
g(C) & =\langle\phi^{2}(x)\rangle_{x\sim\N(0,C)}
\end{align*}
\begin{align*}
\frac{\partial}{\partial C}g(C) & =\langle(\phi^{\prime})^{2}+\phi^{\prime\prime}\phi\rangle_{x\sim\N(0,C)}.
\end{align*}

\section{Exercises}

\subsection*{a) Parameter average with and without weight sharing}

Here we want to recap some steps in detail that were left out in the
lecture. In the lecture we utilized the parameter average to infer
the distribution of the fields and ultimately the distribution of
the output $z$ in the limit of many neurons. The difference in the
calculations between deep neural networks (DNNs) and recurrent neural
networks (RNNs) lies in the former having different realizations of
the connectivity $W^{(l)}$ for each layer, whereas the latter has
the same realization $W^{(l)}\equiv W$ for each layer (Every layer
in an RNN corresponds to one timestep).

In both cases we are interested in the output statistics for the setup

\begin{eqnarray*}
h^{(0)} & = & V\,x,\\
h^{(l)} & = & W\,\phi(h^{(l-1)})\quad1\le l\le L,\\
y & = & w^{\T}\phi(h^{(L)}),\\
z & = & y+\xi,
\end{eqnarray*}
which are given by (see \eqref{eq:p_z_y_deep} and \eqref{eq:p_z_y_rec}) 

\begin{align}
p(z,y|C^{(xx)}) & =\N(z|y,\kappa)\,\int\D h^{(1\le t\le L)}\,\big\langle\prod_{\alpha=1}^{P}\delta\big[y_{\alpha}-w^{\T}\phi(h_{\alpha}^{(L)})\big]\big\rangle_{w}\nonumber \\
 & \times\big\langle\prod_{l=1}^{L}\,\prod_{\alpha=1}^{P}\,\delta\big[h_{\alpha}^{(l)}-W^{(l)}\,\phi(h_{\alpha}^{(l-1)})\big]\big\rangle_{W_{ij}^{(l)}\stackrel{\text{i.i.P. }}{\sim}\N(0,g_{w}/d),h_{\alpha i}^{(0)}\stackrel{\text{i.i.P. id }i}{\sim}\N(0,C_{\alpha\beta}^{(xx)})}.\label{eq:exc_disorder_average_recnets}
\end{align}
Evaluate the average over the connectivities in the second line of
\eqref{eq:exc_disorder_average_recnets} for DNNs and RNNs. Compare
the form of the expressions that show up in the exponent and the resulting
implications on the correlation between different layers $C^{(l,l^{\prime})}$.
Then introduce the auxiliary variables $C$ as in the lecture and
bring the result to the form \eqref{eq:p_y_prefinal-1} and \eqref{eq:P_C_C_1_RNN}
(for the RNN) and \eqref{eq:p_y_prefinal} and \eqref{eq:P_C_given_C_1}
(for the DNN). You may follow the analogous steps as in \prettyref{sec:Network-field-theory}.

\selectlanguage{english}%

\subsection*{b) \foreignlanguage{american}{Skip connections in ResNets }}

\selectlanguage{american}%
Consider the network with the following architecture
\begin{align*}
h^{(0)} & =V\,x,\\
h^{(1)} & =h^{(0)}+W\,\phi(h^{(0)}),\\
y & =w^{\T}\phi(h^{(1)}),\\
z & =y+\xi.
\end{align*}
with Gaussian priors $V_{ij}\stackrel{\text{i.i.d.}}{\sim}\N(0,g_{V}/d)$
and $w_{i},W_{ij}\stackrel{\text{i.i.d.}}{\sim}\N(0,g_{w}/N)$, $\xi_{\alpha}\sim\mathcal{d}(0,\kappa)$.
The first field has a direct connection to the readout, hence it is
referred to as a skip connection. Networks implementing these skip
connections are called residual networks (or ResNets in shortform).
These skip connections implement an identity mapping and thus facilitate
signal propagation to deeper layers. Resnet$50$ implements skip connections
and is one of the state of the art networks on CIFAR-10 \citep{He16_CVPR}.

Make use of the large width limit and compute the NNGP for this setup.
\textbf{Hint: }The sum of a pair of independent, Gaussian variables
is also distributed according to a Gaussian.

Can you see why the networks are called ResNets?

\subsection*{c) NNGP in recurrent neural networks for inputs at every timepoint}

Consider the Gaussian process of a recurrent neural network if a scalar
input is presented at each time step $x^{(t)}\in\mathbb{R}$ within
the interval $t\in[0,T]$
\begin{align}
h_{\alpha}^{(t+1)} & =W\,\phi(h_{\alpha}^{(t)})+v\,x_{\alpha}^{(t)},\\
y_{\alpha} & =u^{\top}\phi\left(h_{\alpha}^{(T)}\right),
\end{align}
with $\bR^{N}\ni h_{\alpha}^{(0)}=(0,\ldots,0)$ and $W_{ij}\sim\mathcal{\N}(0,g_{w}/N),\,v_{i}\sim\N(0,g_{v}),\,w_{i}\sim\N(0,g_{w}/N)$.
We want to understand how the covariances of the intermediate pre-activations
$h_{\alpha}^{(t)}$ and the network output $y_{\alpha}$ relate to
the covariances of the network input. For this purpose start from

\begin{align*}
p(y) & =\int\mathcal{P}h^{(1\leq t\leq T)}\mathcal{P}\tilde{h}^{(1\leq t\leq T)}\langle\prod_{\alpha=1}^{P}\delta\left[y_{\alpha}-w^{\top}\phi(h_{\alpha}^{(T)})\right]\rangle_{u}\\
 & \times\langle\prod_{t=1}^{T}\prod_{\alpha=1}^{P}\prod_{i,j=1}^{\acute{N}}\exp\left[\tilde{h}_{\alpha i}^{(t)}h_{\alpha i}^{(t)}-w_{ij}\tilde{h}_{\alpha i}^{(t)}\phi(h_{\alpha j}^{(t-1)})-v_{i}\tilde{h}_{\alpha i}^{(t)}x_{\alpha}^{(t-1)}+\tilde{h}_{\alpha i}^{(0)\top}h_{\alpha i}^{(0)}\right]\rangle_{w,v},
\end{align*}
and perform the expectation over $W,v$. It is useful to employ the
auxiliary fields

\begin{align}
C_{\alpha\beta}^{(t,s)} & =\frac{g_{w}}{N}\,\sum_{j}\phi(h_{\alpha j}^{(t-1)})\phi(h_{\beta j}^{(s-1)})\\
C_{x,\alpha\beta}^{(t,s)} & =g_{v}\,x_{\alpha}^{(t-1)}x_{\beta}^{(s-1)}
\end{align}
using auxiliary fields $\tilde{C}_{\alpha\beta}^{(t,s)}$ and $\tilde{C}_{x,\alpha\beta}^{(t,s)}$.
Compute the saddle point values for $C_{\alpha\beta}^{(t,s)},C_{x,\alpha\beta}^{(t,s)},\tilde{C}_{\alpha\beta}^{(t,s)},\tilde{C}_{x,\alpha\beta}^{(t,s)}$
(in the NNGP limit, $N\rightarrow\infty$). How do the results differ
from the derivations in the lecture notes? What is the result for
the network output covariance $\langle y_{\alpha}y_{\beta}\rangle$
in the NNGP limit? Assume now that we consider a linear activation
function. What is the result for the network output covariance $\langle y_{\alpha}y_{\beta}\rangle$
in the NNGP limit and how is it related to the input ? Assume stationary
statistics $\langle x_{\alpha}^{(t)}x_{\beta}^{(s)}\rangle:=\kappa_{2}^{(x)}(t-s)$.
What is the input averaged result for the output covariance $\left\langle \langle y_{\alpha}y_{\beta}\rangle_{y}\right\rangle _{x}$?
How does the choice of $g_{w}$ influence the behavior of $\left\langle \langle y_{\alpha}y_{\beta}\rangle_{y}\right\rangle _{x}$
for long times $T\gg1$ ?
\selectlanguage{english}%

\subsection*{d)\foreignlanguage{american}{ Fixed points for Neural Network Gaussian
Process Kernel in linear setting and non-linear setting}}

\selectlanguage{american}%
Suppose that we have a deep linear network with the network architecture

\begin{align}
h_{i\alpha}^{1} & =\sum_{j=1}^{d}V_{ij}x_{j\alpha}\,,V_{ij}\sim\mathcal{\N}(0,g_{v}/d)\\
h_{i\alpha}^{(l)} & =\sum_{j=1}^{N}W_{ij}^{(l)}h_{j\alpha}^{(l-1)}\quad l=2,\dots,L\,,W_{ij}^{(l)}\stackrel{\text{i.i.d.}}{\sim}\mathcal{\N}(0,g_{w}/N)\\
y_{\alpha} & =\sum_{i=1}^{N}w_{i}h_{i\alpha}^{(L)}\,,w_{i}\stackrel{\text{i.i.d.}}{\sim}\mathcal{\N}(0,g_{u}/N)\\
z_{\alpha} & =y_{\alpha}+\xi_{\alpha}\quad\xi_{\alpha}\stackrel{\text{i.i.d.}}{\sim}\N(0,\kappa)
\end{align}
We want to obtain the corresponding Neural Network Gaussian Process
kernel for this linear network. Instead of resorting to the field
theoretic approach we consider a more ``handwaving'' calculation,
which allows us to obtain the leading order results in $N$. First
show, that the mean for the network output and the hidden states yields
$0$ if we average over the network parameters $w,W^{(1)},\ldots,W^{(L)},V$.
Next show how the covariance $\langle h_{i\alpha}^{(l)}h_{j\beta}^{(l)}\rangle_{W^{(l)}}$
and $\langle h_{i\alpha}^{(l-1)}h_{j\beta}^{(l-1)}\rangle_{W^{(l-1)},\ldots,W^{(1)},V}$
are related to one another. Use that we consider the setting $N\rightarrow\infty$.
Where do you need to exploit this fact? What are the fixed points
for $C_{\alpha\beta}^{(z)}$ when we consider the case $L\rightarrow\infty$?
Hint: Consider the cases $g_{w}>1,g_{w}<1$ separately. What does
this mean for the inferred network output ? Now we consider the case
where the network layers get a bias terms $b$, which is also standard
practice in machine learning:

\begin{align}
h_{i\alpha}^{(1)} & =\sum_{j=1}^{d}V_{ij}x_{j\alpha}+b_{i}^{(1)}\,,V_{ij}\stackrel{\text{i.i.d.}}{\sim}\mathcal{\N}(0,g_{v}/d),\quad b_{i}^{(1)}\stackrel{\text{i.i.d.}}{\sim}\mathcal{\N}(0,g_{b})\\
h_{i\alpha}^{(l)} & =\sum_{j=1}^{\acute{N}}W_{ij}^{(l)}h_{j\alpha}^{(l-1)}+b_{i}^{(l)}\quad l=2,\ldots,L\,,W_{ij}^{(l)}\sim\N(0,g_{w}/N),\quad b_{i}^{(l)}\stackrel{\text{i.i.d.}}{\sim}\mathcal{\N}(0,g_{b}^{2})\\
y_{\alpha} & =\sum_{i=1}^{d_{h}}w_{i}h_{i\alpha}^{(L)}+b_{i}^{(\mathrm{out})}\,,w_{i}\stackrel{\text{i.i.d.}}{\sim}\N(0,g_{u}/N),\quad b_{i}^{(\mathrm{out})}\stackrel{\text{i.i.d.}}{\sim}\mathcal{\N}(0,g_{b}^{2})\\
z_{\alpha} & =y_{\alpha}+\xi_{\alpha}\quad\xi_{\alpha}\sim\mathcal{\N}(0,\kappa).
\end{align}
Considering this setting, how does the relation between $\langle h_{i\alpha}^{(l)}h_{j\beta}^{(l)}\rangle_{W^{(l)},\ldots,W^{(1)},V,b^{(1)}...b^{(l)}}$
and $\langle h_{i\alpha}^{(l-1)}h_{j\beta}^{(l-1)}\rangle_{W^{(l-1)},\ldots,W^{(1)},V,b^{(1)},\ldots,b^{(l-1)}}$
look like? What are the new fixed points for $C_{\alpha\beta}^{(z)}$
when we consider the case $L\rightarrow\infty$ ? Hint: Consider the
cases $g_{w}>1,g_{w}<1$ separately. What does this mean for the inferred
network output ? Having seen those results: What would be a reasonable
setting for $g_{w}$ if $\sigma_{b}=0$? 

As you can imagine, in the case of non-linear networks, the situation
might be a bit more complicated. Consider the non-linear network,
where the only difference is the activation function appearing in

\begin{align*}
h_{i\alpha}^{(l)} & =\sum_{j=1}^{N}W_{ij}^{(l)}\phi(h_{j\alpha}^{(l-1)})+b_{i}^{(l)}\quad l=2,\ldots,L,\\
y_{\alpha} & =\sum_{i=1}^{d_{h}}w_{i}\phi(h_{i\alpha}^{(L)})+b_{i}^{(\mathrm{out})}\,,
\end{align*}
where we choose the so called rectified linear unit (short ReLU) activation

\begin{equation}
\phi(x)=\mathrm{ReLU}(x)=\begin{cases}
0 & x\leq0\\
x & x>0
\end{cases}
\end{equation}
Compute the variance $C_{\alpha\alpha}^{y}$ of the NNGP for a network
with ReLu activation. Keep in mind that this means you need to compute
kernels of the form

\begin{equation}
\langle\phi(h_{\alpha})^{2}\rangle_{h\sim\mathcal{d}(0,C_{\alpha\alpha})}=\int_{-\infty}^{\infty}Ph_{\alpha}\phi^{2}(h_{\alpha})\frac{1}{\sqrt{2\pi C_{\alpha\alpha}}}\exp\left(-\frac{1}{2C_{\alpha\alpha}}h_{\alpha}^{2}\right)
\end{equation}
For $L\rightarrow\infty$, what is the value for $g_{w}$ which allows
for non-trivial fixpoints ?

\chapter{Fokker-Planck equation\protect\label{chap:Non-equilibrium-statistical-mech}}

To investigate the dynamics of learning in neuronal networks we need
to relate the time-dependent process of adapting the weights to the
distribution of weights that we have studied so far in the setting
of Bayesian inference. To this end we need the theoretical tool that
allows us to derive from an equation of motion, here the one for the
synaptic weights, the evolution of the probability distribution. This
tool is the Fokker-Planck equation.

\section{Stochastic differential equations}

For simplicity, we here consider the stochastic evolution of a scalar
variable $x(t)\in\mathbb{R}$, before we generalize this result to
the time-dependent stochastic evolution of multiple variables $x(t)\in\mathbb{R}^{N}$.
We therefore first consider the \textbf{stochastic differential equation}
(SDE)

\begin{eqnarray}
dx(t) & = & f(x)\,dt+g(x)\,dB(t)\label{eq:SDE-fokker}\\
x(0+) & = & a,\nonumber 
\end{eqnarray}

where $a$ is the initial value and $dB$ a stochastic increment.
Stochastic differential equations are defined as the limit $h\to0$
of a dynamics on a discrete time lattice of spacing $h$. For discrete
time $t_{l}=lh$, $l=0,\ldots,M$, the solution of the SDE consists
of the discrete set of points $x_{l}=x(t_{l})$. For the discretization
there are mainly two conventions used, the Ito and the Stratonovich
convention \citep{Gardiner09}. In case of additive noise ($g(x)=\mathrm{const.}$),
where the stochastic increment in \eqref{eq:SDE-fokker} does not
depend on the state $x$, the two conventions yield the same continuous-time
limit \citep{Gardiner09}. However, as we will see, different discretization
conventions of the drift term lead to different path integral representations,
as we will see later. The Ito convention defines the time discrete
notation of \eqref{eq:SDE-fokker} to be interpreted as 
\begin{eqnarray}
x_{i}-x_{i-1} & = & f(x_{i-1})\,h+g(x_{i-1})\,b_{i},\label{eq:SDE_Ito}
\end{eqnarray}
where $b_{i}$ is a stochastic increment that follows a probabilistic
law. A common choice for $b_{i}$ is a normal distribution $\rho(b_{i})=\N(b_{i}|0,hD)$,
called a \textbf{Wiener increment}. Here the parameter $D$ controls
the variance of the noise, as above. If the variance of the increment
is proportional to the time step $h$, this amounts to a $\delta$-distribution
in the autocorrelation of the noise $\xi=\frac{dB}{dt}$, because
$\langle b_{i}b_{j}\rangle=\delta_{ij}\,Dh$ and we see from \prettyref{eq:SDE_Ito}
that the correlation of the noise $\xi_{i}=b_{i}/h$ then is $\langle\xi_{i}/h\xi_{j}/h\rangle=\delta_{ij}\,\frac{D}{h}$,
which can be regarded as the discrete analogue of the Dirac distribution,
\begin{align*}
\langle\xi(t)\xi(s)\rangle & =D\,\delta(t-s).
\end{align*}

\subsection{Fokker-Planck equation}

We will now derive the Fokker-Planck equation, an equation describing
the time-evolution of the probability density function, closely following
the derivation in \citep{Risken96}. We note that the process has
the \textbf{Markov property}: the evolution of the system at time
point $t_{i}$ only depends on the current state $x_{i}$, but not
on the history $x_{j<i}$, the states by which state $x_{i}$ was
reached. This means that the knowledge of the state $y$ at some earlier
time point $s$ is sufficient to know the statistics in the future.
For such processes, one may write the evolution of the probability
distribution as the so-called \textbf{Chapman-Kolmogorov} \textbf{equation},

\begin{align}
p(x,t) & =\int dy\,p(x,t;y,s)\label{eq:Chapman_Kolmogorov}\\
 & =\int dy\,p(x,t|y,s)\,p(y,s),\nonumber 
\end{align}
where the first line marginalizes the joint probability of state $x$
at time $t$ and state $y$ at time $s$, $p(x,t;y,s)$, over the
state $y$ at time $s$. The second line is just the fundamental relation
between conditioned and unconditioned probabilities, $p(x,y)=p(x|y)\,p(y)$.
It is an expression of the conservation of probability: Every state
$y$ at time $s$ can move to any state $x$ at a later time $t>s$
by the transition probability $p(x,t|y,s)$. So the equation can be
regarded as a linear integral equation for the evolution of the joint
probability distribution for the state at the more advanced time point
$t$.

\begin{figure}
\centering{}\includegraphics[width=0.5\textwidth]{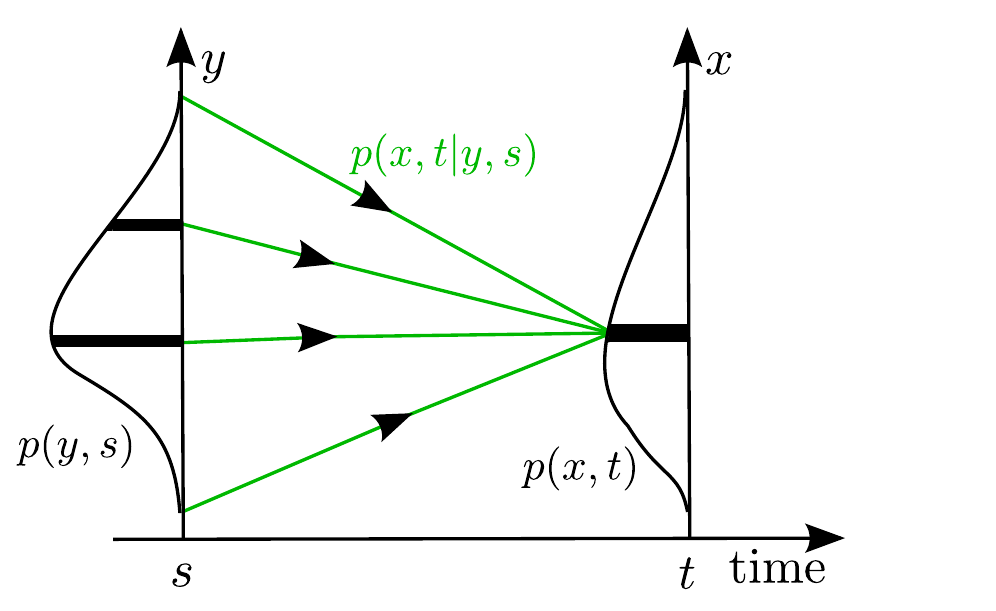}\caption{\textbf{Illustration of the Chapman-Kolmogorov equation. }The Chapman-Kolmogorov
equation describes the time evolution of the probability density.
The conditional probability (green) allows for the transition between
probabilities of different states and times.\protect\label{fig:Chapman-Kolmogorov-equation}}
\end{figure}
Our goal is now to derive, instead, a differential equation for this
evolution. Instead of providing the conditional probability $p(x,t|y,s)$
we may equivalently provide the moment-generating function for the
difference $x-y$
\begin{align*}
Z(j|y,s,\tau) & =\int\,dx\,e^{j(x-y)}\,p(x,s+\tau|y,s),
\end{align*}
where we set $t=s+\tau$. Here the triple $(y,s,\tau)$ plays the
role of parameters. Differentiating by $j$ obviously yields the moments
of $x-y$.

We may evaluate $Z(j|y,t,\tau=h)$ for the stochastic differential
equation \eqref{eq:SDE_Ito}. We read off that $x_{i}-x_{i-1}$ has
non-vanishing first and second cumulants only, which are
\begin{align*}
\llangle x_{i}-x_{i-1}\rrangle & =f(x_{i-1})\,h,\\
\llangle(x_{i}-x_{i-1})^{2}\rrangle & =g^{2}(x_{i-1})\,D\,h,\\
\llangle(x_{i}-x_{i-1})^{n>2}\rrangle & \equiv0,
\end{align*}
where we used that the right hand side is Gaussian distributed by
definition, cumulants of order higher than two vanish. Evaluating
these expressions at $x_{i-1}=y$ we may write $Z$ in terms of these
two cumulants
\begin{align}
Z(j|y,t,h) & =\exp\big(j\,f(y)\,h+\frac{j^{2}}{2}\,g^{2}(y)\,D\,h\big),\nonumber \\
 & =1+j\,f(y)\,h+\frac{j^{2}}{2}\,g^{2}(y)\,D\,h+\order(h^{2}),\label{eq:Kramers_Moyal}
\end{align}
where we may stop at order $h$ here, because we are after an evolution
equation for infinitesimal time steps $h\to0$, so this truncation
will not be an approximation.

We may hence express the conditional probability distribution as the
inverse Fourier transform

\begin{align*}
p(x,t+h|y,t) & =\frac{1}{2\pi i}\int_{-i\infty}^{i\infty}dj\,e^{-j(x-y)}\,Z(j|y,t,h)\\
 & =\frac{1}{2\pi i}\int_{-i\infty}^{i\infty}dj\,\Big(1+j\,f(y)\,h+\frac{j^{2}}{2}\,g^{2}(y)\,D\,h+\order(h^{2})\Big)\,e^{-j(x-y)}.
\end{align*}
The factors $j$ and $j^{2}$ appearing under the integral can also
be written as application of $-\partial_{x}$ and $\partial_{x}^{2}$
to $e^{-j(x-y)}$. This allows us to write the last line as
\begin{align}
 & \Big(1-\partial_{x}f(y)\,h+\frac{1}{2}\,\partial_{x}^{2}\,g^{2}(y)\,D\,h+\order(h^{2})\Big)\,\underbrace{\frac{1}{2\pi i}\int dj\,e^{-j(x-y)}}_{\delta(x-y)},\label{eq:intermedate_FPE}
\end{align}
so that we get from the Chapman-Kolmogorov relation \eqref{eq:Chapman_Kolmogorov},
by eliminating the integral over $y$ by using $\delta(x-y)$,
\begin{align*}
p(x,t+h)-p(x,t) & =\int dy\,p(x,t+h|y,t)\,p(y,t)-p(x,t)\\
 & =\Big(-\partial_{x}f(x)\,h+\frac{1}{2}\,\partial_{x}^{2}\,g^{2}(x)\,D\,h+\order(h^{2})\Big)\,p(x,t),
\end{align*}
where the term $1$ in \prettyref{eq:intermedate_FPE} has been canceled
versus the subtraction of the last term $p(x,t)$ in the penultimate
line. Note that the differential operators here also act on the functions
$f(x)$ and $g(x)$; this is because
\begin{align*}
\int dy\,\big[f(y)\,\partial_{x}\,\delta(x-y)\big]\,p(y,t) & =-\int dy\,f(y)\,p(y,t)\,\partial_{y}\,\delta(x-y)\\
 & \stackrel{\text{i.b.p.}}{=}\int dy\,\delta(x-y)\,\partial_{y}\big[f(y)\,p(y,t)\big]\\
 & =\partial_{x}\big[f(x)\,p(x,t)\big],
\end{align*}
where we used that the derivative of the Dirac distribution is defined
such that the derivative acts on the test function (here $f(y)\,p(y,t)$)
multiplying the distribution, so that integration by parts holds with
vanishing boundary terms. In the last step we used that the integral
together with the Dirac $\delta$ leads to the evaluation of the function
$y\mapsto\partial_{y}\big[f(y)\,p(y,t)\big]$ at the point $y=x$.
An analog argument holds for the diffusion term.

We write the obtained \textbf{Fokker-Planck equation} in infinitesimal
form (dividing by $h$ and taking the limit $h\to0$) as

\begin{align}
\partial_{t}p(x,t) & =-\partial_{x}\left(f(x)p(x,t)\right)+\frac{D}{2}\partial_{x}^{2}\left(g^{2}(x)p(x,t)\right)\label{eq:FPE_final}
\end{align}
We remark that:
\begin{itemize}
\item The Fokker-Planck equation is exact for the considered Langevin equation
with Gaussian white noise.
\item The first term is also called \textbf{drift term}, because it arises
from the deterministic drift; the second \textbf{diffusion term},
because it comes from the noise.
\item The expansion in terms or the moments of the transition probability
we performed in \prettyref{eq:Kramers_Moyal} is the so called \textbf{Kramers-Moyal
expansion}. It may also be performed for arbitrary Markov processes.
\item In general, for non-Gaussian increments, arbitrary many moments and
hence derivatives in \prettyref{eq:FPE_final} contribute. However,
\textbf{Pawula's theorem} states that the Kramers-Moyal expansion
terminates either after the first term (deterministic system), the
second term (as in the Gaussian case), or requires infinitely many
terms; this implies that for general statistics of the noise, the
Fokker-Planck equation becomes an approximation. Still, as frequently
employed SDE integrate many stochastic increments over time, employing
the Fokker-Planck equation may still be a very good approximation
even for non-Gaussian noise.
\end{itemize}
We may write the Fokker-Planck equation in a more suggestive form
as a \textbf{continuity equation} that expresses the conservation
of probability as
\begin{align*}
\partial_{t}p(x,t) & =-\partial_{x}\,J(x,t),\\
J(x,t) & =\Big(f(x)-\frac{1}{2}\,\partial_{x}\,g^{2}(x)\,D\Big)\,p(x,t),
\end{align*}
where the first line has the usual form of a continuity equation,
namely the temporal derivative equals minus the divergence (in one
dimension) of the probability current $J$. Note that here the derivative
in $\partial_{x}\,g^{2}(x)\,D$ also acts on $p(x,t)$. The second
equation expresses the current as the sum of the \textbf{drift term},
proportional to the product of the density of each state $x$ and
its velocity $f(x)$ and the second term is the \textbf{diffusion
term}, which describes the probability current due to the stochastic
jumps (it is proportional to $D$, the noise amplitude): it is directed
towards minus the gradient of the density. If we had an $N$-component
vector $x\in\bR^{N}$ instead of a scalar, the above formulation generalizes
to
\begin{align}
\partial_{t}p(x,t) & =-\sum_{i}\partial_{x_{i}}J_{i}(x,t)=-\nabla\cdot J(x,t)\label{eq:FP_N_dim}\\
J_{i}(x,t) & =\Big(f_{i}(x)-\frac{1}{2}\,\sum_{j}D_{ij}\partial_{x_{j}}g_{i}(x)\,g_{j}(x)\Big)\,p(x,t),\label{eq:J_current}
\end{align}
where $D_{ij}$ is the covariance matrix of the noise $\langle\xi_{i}(t)\xi_{j}(s)\rangle=D_{ij}\,\delta(t-s)$.
This can be seen by considering the Kramers-Moyal expansion \prettyref{eq:Kramers_Moyal},
in which $j_{i}f_{i}h$ will appear from the first cumulant and $\frac{1}{2}D_{ij}g_{i}(x)g_{j}(x)h$
from the second cumulant.

\subsection{Boltzmann distribution and detailed balance}

We are now equipped with the methods to make the link between the
Boltzmann distribution in equilibrium and the time-evolution described
by the Fokker-Planck equation.

For the stochastic differential equation \eqref{eq:SDE-fokker} with
uncorrelated \textbf{additive noise} with $D_{ij}=D\,\delta_{ij}$,
which is independent of the state, so $g=1$, and there the deterministic
force is conservative, which is
\begin{align*}
f(x) & =-\nabla V(x),
\end{align*}
the probability current of the Fokker-Planck equation in the continuum
formulation \prettyref{eq:J_current} takes the form

\begin{align}
J_{i}(x,t) & =\Big(-\frac{\partial}{\partial x_{i}}V(x)-\frac{D}{2}\frac{\partial}{\partial x_{i}}\Big)\,p(x,t).\label{eq:prob_current_continuum}
\end{align}
We see from the continuity equation that we get a particular stationary
distribution $\partial_{t}p(x,t)\equiv0$ if the probability current
is a constant as a function of $x$, because then the divergence in
\prettyref{eq:FP_N_dim} vanishes at each point in space. In such
a state hence all probability currents are identical. If we assume
that the amplitude of fluctuations of $x$ are limited, which is physically
certainly reasonable, there cannot be a probability current at arbitrary
large values of $x$ (the argument of the left hand side of \prettyref{eq:prob_current_continuum}).
Hence the only physical way how such a stationary distribution may
arise is if the current vanishes for all configurations $x$. This
condition is also called \textbf{detailed balance}, because it amount
to saying that the mutual probability fluxes between neighboring points
$x$ and $x+\delta x$ is vanishing at each configuration $x$, so
the fluxes between any such pair of configurations are of same magnitude
but have opposite direction – they are in balance.

We hence need to determine the solution of the differential equation
\begin{align*}
0= & \Big(-\nabla V-\frac{D}{2}\nabla\Big)\,p(x).
\end{align*}
We see that 
\begin{align}
p_{0}(x) & \propto\exp\big(-\frac{2}{D}\,V(x)\big)\label{eq:p_0_Boltzmann}
\end{align}
is such a stationary solution, which is of Boltzmann form, if we interpret
$V(x)$ as an energy and $D$ as the temperature $k_{B}T$.

\begin{figure}
\centering{}\includegraphics[width=0.5\textwidth]{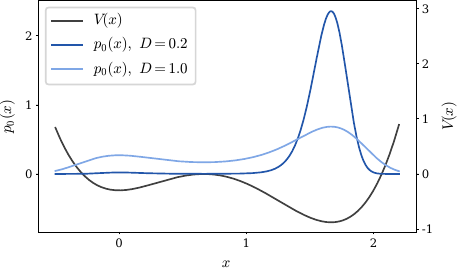}\caption{\textbf{Probability distribution in detailed balance. }Stationary
solution for the probability densities in a double well potential
with different strengths of the diffusion term. \protect\label{fig:Chapman-Kolmogorov-equation-1}}
\end{figure}

\section{Training by Langevin dynamics in relation to Bayesian inference\protect\label{sec:Training-by-Langevin}}

We are now in the position to relate the Bayesian view on training
neuronal networks to the stochastic dynamics of gradient descent,
following \citep{Naveh20_01190,Naveh21_064301}. Consider the quadratic
loss function \prettyref{eq:mse_loss}. Now consider changing the
weights proportional to the gradient with weight decay

\begin{align}
dW & =\big(-\gamma W-\nabla_{W}\mathcal{L}\big)\,dt+dB\label{eq:Langevin_training}\\
 & =-\nabla_{W}\,\big(\frac{\gamma}{2}\,\|W\|^{2}+\mathcal{L}\big)\,dt+dB,\nonumber 
\end{align}
where $dB$ is a Gaussian noise (Wiener increment) with variance $D$,
so $\langle dB_{i}(t)dB_{j}(s)\rangle=D\,\delta_{ij}\,\delta(t-s)\,dt$.
The equilibrium distribution of the weights then is of the Boltzmann
form, according to \prettyref{eq:p_0_Boltzmann}
\begin{align}
p_{0}(W) & \propto\exp\big(-\frac{\gamma}{D}\,\|W\|^{2}-\frac{2}{D}\,\mathcal{L}(z,y(W))\big).\label{eq:distribution_weights}
\end{align}
We thus obtain a distribution of the weights that is given by a Gaussian
term $e^{-\frac{1}{2}\frac{2\gamma}{D}\,\|W\|^{2}}$ and a term that
depends on the training data $e^{-\frac{2}{D}\mathcal{L}}$. The two
parameters $D$ and $\gamma$ allow us to separately control the variance
of the Gaussian and the importance of the data term. The data term
depends on the outputs $y_{\alpha}$ of the network. We wish to understand
the behavior of the network where the weights are drawn from the distribution
\eqref{eq:distribution_weights}, which implies a joint probability
distribution for $y$ and $z$, which is 
\begin{align}
p(z,y|x)\propto & \int dW\,\exp\big(-\frac{1}{2}\frac{2\gamma}{D}\,\|W\|^{2}-\frac{2}{D}\,\mathcal{L}(z,y)\big)\,\delta(y-y(W|x))\label{eq:p_z_y_Langevin}\\
= & \int dW\,\exp\big(-\frac{1}{2}\frac{2\gamma}{D}\,\|W\|^{2}-\frac{1}{D}\,\|y-z\|^{2}\big)\,\delta\big(y-y(W|x)\big)\nonumber \\
\propto & \Bigg\langle\exp\big(-\frac{1}{D}\,\|y-z\|^{2}\big)\,\delta\big(y-y(W|x)\big)\Bigg\rangle_{W_{ij}\stackrel{\text{i.i.d.}}{\sim}\N(0,\frac{D}{2\gamma})}\nonumber \\
\propto & \N(z|y,D/2)\,\langle p(y|W,X)\rangle_{W_{ij}\stackrel{\text{i.i.d.}}{\sim}\N(0,\frac{D}{2\gamma})},\nonumber 
\end{align}
which is the same as \prettyref{eq:pre_disorder_W} if one sets $g_{w}/M=D/(2\gamma)$
and $\kappa=D/2$. So the noise $D$ here plays the role of the regularization
and the weight decay term in \prettyref{eq:Langevin_training} controls
the width of the prior distribution of the weights in the Bayesian
approach.

On a conceptual level, the difference between the Bayesian approach
and the training of individual networks is that the posterior distribution
represents ensembles of trained networks. This means that one obtains
an uncertainty estimate for the trained network. One may practically
employ such ensembles by training multiple networks from different
initial conditions or by using the same initialization and stopping
the training process at different points in time after the equilibrium
distribution has been reached. Using multiple such networks one may
then obtain a practical estimate of the uncertainty of the network
output, for example by quantifying the variability of the outputs
of this ensemble.

\section{Keypoints}

We learned:\\

\noindent\fcolorbox{black}{white}{\begin{minipage}[t]{1\textwidth - 2\fboxsep - 2\fboxrule}%
\textbf{Fokker Planck equation}
\begin{itemize}
\item Assuming Markovian property we can utilize the Chapman-Kolmogorov
equation to compute path probability distribution $p(x,t)$ of SDEs.
\item We can approximate the time evolution of $p(x,t)$ yielding the Fokker
Planck equation; which is exact if the SDE noise is Gaussian white
noise.
\item The stationary distribution of the Fokker-Planck equation of an SDE
with Gaussian noise has a Boltzmann form if the deterministic force
is conservative.
\item Training networks using Langevin stochastic gradient descent (LSGD)
yields the same equilibrium distribution we describe analytically
using Bayesian inference. This links training dynamics with Bayesian
inference and allows one to test theoretical results numerically.
\end{itemize}
\end{minipage}}

\section{Exercises}

\selectlanguage{english}%

\subsection{Ornstein-Uhlenbeck process: time evolution of moments}

For a quadratic potential $V(\bx)=\frac{1}{2}\bx^{\T}m\bx$, the Langevin
dynamics (see \prettyref{sec:Training-by-Langevin}) become an Ornstein-Uhlenbeck
process

\begin{align}
\frac{\partial}{\partial t}\bx(t) & =-m\,\bx(t)+\boldsymbol{\xi}(t),\label{eq:exc_OUP}
\end{align}
with $m\in\bR^{N\times N}$, $\left\langle \xi_{i}(t)\right\rangle =0$
and $\left\langle \xi_{i}(t)\xi_{j}(s)\right\rangle =D\,\delta_{ij}\,\delta(t-s)$.

Show that

\begin{align}
\bx(t)= & \exp\left(-m\,t\right)\bx(0)\,+\,\int_{0}^{t}dt^{\prime}\,\exp\left(-m\,\left(t-t^{\prime}\right)\right)\boldsymbol{\xi}(t^{\prime})\label{eq:exc_OUP_formal_solution}
\end{align}
solves \prettyref{eq:exc_OUP}.

Using \prettyref{eq:exc_OUP_formal_solution}, derive expressions
for the mean $\left\langle \boldsymbol{x}(t)\right\rangle $ and the
covariance at two different points in time $C(t,s)=\left\langle \left(\boldsymbol{x}(t)-\left\langle \boldsymbol{x}(t)\right\rangle \right)\left(\boldsymbol{x}(s)-\left\langle \boldsymbol{x}(s)\right\rangle \right)^{\T}\right\rangle \in\bR^{N\times N}$.

Which differential equations do $C(t+\tau,t)$ $(\tau>0)$ and $C(t,t)$
obey? To this end take the derivatives $\frac{\partial}{\partial\tau}C(t+\tau,t)$
and $\frac{\partial}{\partial t}C(t,t)$. How would these time evolution
equations change if we had a noise that is non-diagonal in space $D\,\delta_{ij}\to D_{ij}$?

As you have seen in \prettyref{chap:Non-equilibrium-statistical-mech}
the Fokker-Planck equation describes the time evolution of the probability
density $p(\boldsymbol{x},t)$ exactly for Gaussian white noise. What
is the equilibrium distribution $p_{0}(\boldsymbol{x})$ the dynamics
\prettyref{eq:exc_OUP} relax to?

\selectlanguage{american}%

\subsection{Connecting Bayesian inference and Gradient descent for linear regression}

\global\long\def\Ntrain{N_{\mathrm{train}}}%
\global\long\def\av#1{\left\langle #1\right\rangle }%
In this exercise we want to explicitly see the connection between
the Bayesian approach and training by stochastic gradient descent
in the framework of linear regression. Just as one can calculate the
posterior of the outputs (see \prettyref{eq:posterior_y_star_inserted}),
we can also ask the question of how the weights are distributed after
conditioning on the training data (i.e. posterior of the weights,
see \prettyref{eq:weight_posterior} and \prettyref{fig:Prior-vs-posterior-Bastian-inspired}).
As for the outputs, we make use of the Bayesian formulation

\begin{equation}
p(w|X,z)=\frac{p(z|X,w)\,p(w)}{p(z|X)}.\label{eq:exc_lars_fokker_bayes}
\end{equation}
Here, $p(w)$ is the prior of the weights with

\[
w_{i}\stackrel{\text{i.i.d}}{\sim}\mathcal{N}(0,g_{w}/M)
\]
and the likelihood is given by

\begin{align*}
p(z|X,w) & =\int dy\,\N(z|y,\kappa)\,p(y|X,w)\\
 & =\int dy\,\N(z|y,\kappa)\,\delta\left(y-y(W|x)\right)\\
 & =\int dy\,\N(z|y,\kappa)\,\prod_{\alpha=1}^{D}\delta\left(y_{\alpha}-x_{\alpha}^{\T}w\right)\\
 & =\mathcal{N}(z|Xw,\kappa),
\end{align*}
with $X=\left(x_{1},\dots,x_{D}\right)^{\T}$, $X\in\mathbb{R}^{D\times M}$,
$w\in\mathbb{R}^{M}$ and the label noise $\kappa$. Show that $p(w|X,z)$
follows a Gaussian distribution. State the mean and covariance.

As you have have seen in the lecture, if one chooses $\kappa=\D/2$
and $g_{w}/M=\D/(2\gamma)$, this corresponds to the equilibrium distribution
of

\begin{equation}
dw=\big(-\gamma w-\nabla_{w}\mathcal{L}\big)\,dt+dB\label{eq:gradient_dyn}
\end{equation}
with $\langle dB_{i}(t)dB_{j}(s)\rangle=\D\,\delta_{ij}\,\delta(t-s)\,dt$
and $\mathcal{L}=\frac{1}{2}\sum_{\alpha=1}^{D}\left(z_{\alpha}-x_{\alpha}^{\T}w\right)^{2}$.
Derive the time evolution equation for the mean of the weights $\frac{d\langle w\rangle}{dt}$
and verify that the stationary point is given by the mean of \prettyref{eq:exc_lars_fokker_bayes}.
How do you solve such an equation in general? And how could you then
infer the effective time constant for $\av w$? To this end, decompose
the data overlap matrix $X^{\T}X=\sum_{i}\lambda_{i}v_{i}v_{i}^{\T}$
into its orthogonal eigenmodes.

Next we give you without proof for the linear stochastic differential
equation

\begin{align}
d\tilde{w}(t) & =-A\,\tilde{w}(t)\,dt+dB(t)\label{eq:linear_sde}
\end{align}
the time evolution for the second cumulant at equal times $C(t)=\langle\tilde{w}(t)\tilde{w}(t)^{\T}\rangle$

\begin{align}
\frac{d}{dt}C(t) & =-A\,C(t)-C(t)\,A^{\T}+\D\mathbb{I}.\label{eq:exc_lars_fokker_second_moment_evolution}
\end{align}
 Bring the stochastic gradient descent into the form \prettyref{eq:linear_sde}
and and verify that the variance of \prettyref{eq:exc_lars_fokker_bayes}
is a fixed point of \prettyref{eq:exc_lars_fokker_second_moment_evolution}.

\subsection{Lagrange multipliers and gradient descent}

\subsubsection{Recap: Method of Lagrange multipliers}

\begin{figure}
\begin{centering}
\includegraphics[scale=0.7]{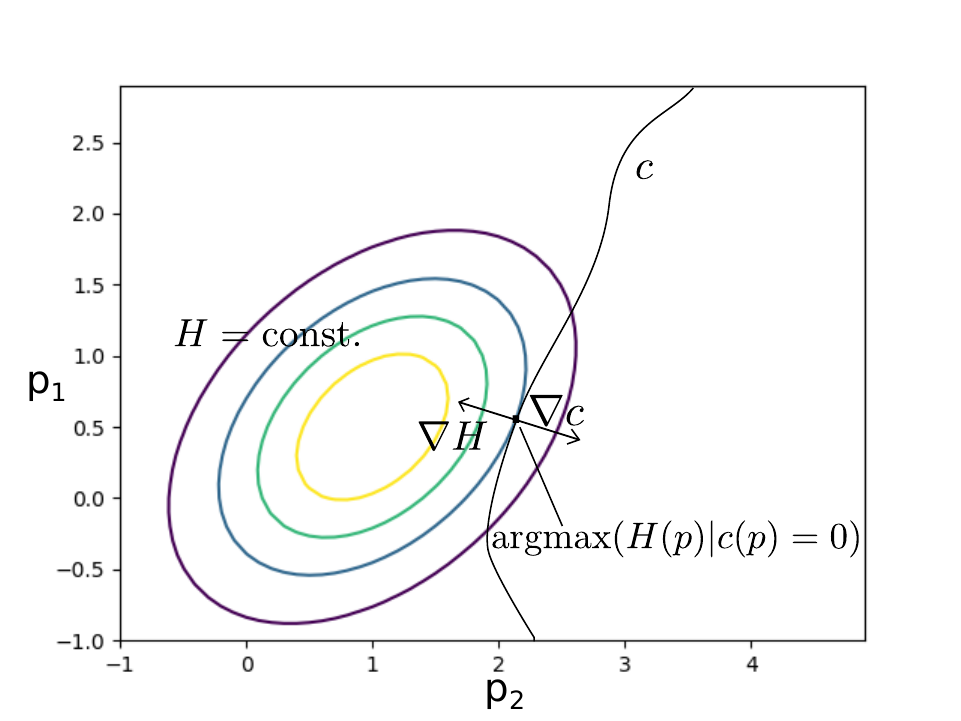}
\par\end{centering}
\caption{Maximization of a function $H(p_{1},p_{2})$ under the constraint
$c(p_{1},p_{2})=0$. The normal vector $\nabla_{p}H$ on the level
curves and the normal vector $\nabla_{p}c$ on the hypersurface defined
by the constraint $c$ must be parallel at the point of maximum, $\nabla_{p}H=\lambda\,\nabla_{p}c$.
This motivates the Lagrange multiplier $\lambda$ and leads to the
optimization of $\nabla_{p}(H-\lambda c)\stackrel{!}{=}0$.\protect\label{fig:Maximization-Lagrange-2}}
\end{figure}

Obtaining a neural network which solves a specific task, such as labeling
images or regressing time series, corresponds, in the end, to an optimization
task. One seeks to optimize a loss function under the constraint that
the output of the network obeys the defining equations of the network
that implements the mapping from data samples to outputs, $x\mapsto y$. 

In general maximizing a function $H$ in the variables $p_{1}...p_{N}$
with respect to some given constraints $c$ is solved by the method
of \textbf{Lagrange multipliers} (illustrated in \prettyref{fig:Maximization-Lagrange-2}).
To this end, one writes the constraints in the form $c(p_{1},\ldots,p_{N})=0$.
For any (not necessarily linear) constraint $c(p)=0$, the gradient
$\nabla_{p}c(p)$ is a vector that at each point is perpendicular
to the $N$-dimensional hypersurface formed by the constraint. Show
that this is indeed the case using the parametrization of a curve
for the constraint with $\varphi:\bR\to\bR^{N}$ so that $c(t)=c(\varphi(t))=0$.
Hint: Use that the multidimensional chain rule and the fact that $\varphi$
characterizes a path in the variables $p_{i=1...N}$. Later use the
relation between the velocity field $\partial_{t}\varphi(t)$ and
the curve $\varphi(t)$. 

Maximizing the function $H(p(x_{1}),\ldots,p(x_{N}))$ given the constraints,
we therefore have to consider all curves $\varphi$ within the plane
and search for the $t$ at which $H(\varphi(t))$ is stationary. Use
this statement to show, that the gradients $\nabla_{p}H$ and $\nabla_{p}c$
are parallel to each other. Using this we know, that the gradients
are related through a scalar $\lambda$, which we call the Lagrange
multiplier:
\begin{eqnarray}
\nabla_{p}H & = & \lambda\nabla_{p}c\\
\nabla_{p}(H(p)-\lambda c(p)) & = & 0.
\end{eqnarray}
Several constraints $c_{1},\ldots,c_{n}$ can be applied successively,
each reducing the space by one dimension, so that the final condition
is
\begin{eqnarray*}
\nabla_{p}(H(p)-\sum_{i}\lambda_{i}c_{i}(p)) & = & 0.
\end{eqnarray*}

\subsubsection{Application to derive backprogation through time}

We now want to use this general concept to understand the process
of training a discrete time recurrent neural network following \citep{Pearlmutter89}.
Consider the following setup where we assume scalar inputs, hidden
states, weights and labels $x_{\alpha}(t),\,h_{\alpha}(t)\,,w,\hat{y}_{\alpha}\in\mathbb{R}$
and $1\le\alpha\le D$ for $D$ training samples. The setup reads
for

\begin{align}
h(0) & =0,\label{eq:ContTime_NeuralNetwork}\\
h_{\alpha}(t+1) & =w\phi(h_{\alpha}(t))+x_{\alpha}(t)\quad t=0,\ldots,T-1,\nonumber \\
L & :=\frac{1}{2}\sum_{\alpha}(\hat{y}_{\alpha}-h_{\alpha}(T))^{2},\nonumber 
\end{align}
where we utilize a discrete time recurrent neural network in \prettyref{eq:ContTime_NeuralNetwork}
and $\hat{y}_{\alpha}$ denotes the true labels corresponding to the
input sequence $x_{\alpha}(t)$. Formulate the condition for the path
$h(t)$ for $t=1,\ldots,T$ that minimizes $L$ under the $DT$ constraints
that the equation of motion \prettyref{eq:ContTime_NeuralNetwork}
is fulfilled at each time point $t$ and for each training sample
$\alpha$, by introducing $DT$ Lagrange multipliers $\lambda_{\alpha}(t)$
and constraints $c$ to construct an action $S[h(t:1,\ldots T)]:=L[h]-\lambda\,c[h]$.
Compute the stationary point $\frac{\partial S}{\partial h}\stackrel{!}{=}0$
with regard to each point at the sequence $h(t:1,\ldots,T)$ and with
regard to each training sample $h_{\alpha}(t)$ (the $h_{\alpha}(t)$
here play the role of the parameters $p$ in the abstract formulation
above) to obtain an equation of motion for the Lagrange multiplier.
Think about the interpretation of the equation of motion for $\lambda$
in terms of error backpropagation.

Then determine the gradient with regard to $w$ of $S$ to derive
the learning rule. Here use that by $\lambda$ following the equation
of motion that ensures that $h$ is a minimum, inner derivatives $\frac{\partial S}{\partial h}\frac{\partial h}{\partial w}=0$,
because $\frac{\partial S}{\partial h}=0$ by construction. Also note
that $\frac{\partial S}{\partial\lambda}=c=0$ by construction, because
we choose the constraints such that $c=0$.

\chapter{Feature learning\protect\label{chap:Feature-learning}}

We here want to go beyond the NNGP as a theory of neuronal networks.
The latter has been obtained in the limit where the width $N\to\infty$
is taken to infinity, while the number of training points $P$ is
kept fixed. This led to the dominance of the rate function $\Gamma(C)\propto N$
in determining the most likely value for the kernels $C$. Because
$\Gamma$ is independent of the labels $z$, as a result the NNGP
kernels depend only on the input data point $x$, but not on the labels
$z$. This implies that the structures within the network described
by the kernels are independent of the joint statistics of $(x,z)$,
but rather only depend on the statistics of $x$ alone.

Learning the statistical relationship between $(x,z)$, on the other
hand, is often referred to as ``feature learning''; originally this
term refers to the fact that trained neuronal networks show the property
to learn ``features'' of the data, that is, neurons become selective
to certain patterns in the input data $x$ that are relevant to the
task. ``Relevance'' here means that these feature-selective neurons
help the network to represent the statistical relationship between
inputs $x$ and desired outputs $z$, for example to assign the right
label. The NNGP, being independent of $z$, hence cannot show such
behavior.

To study feature learning, we therefore need to consider a different
limit, namely sending the number of training points $P$ and the width
of the layer $N$ to infinity at the same rate, so $P=\alpha\,N\to\infty$,
where $\alpha=\order(1)$ stays of order unity.
In the first part, we will recover the works by Li et al. \citep{Li21_031059},
and by Ariosto et al. \citep{Ariosto2022}, exemplified for single
hidden layer networks. This theory shows that both these works compute
the maximum a posteriori estimate of the length of the readout weights
$\|w\|^{2}$. This section reformulates the original works such that
one does not need to introduce imaginary helping fields which do not
have physical meaning. We will then present the extension to deep
linear networks where the overlap between adjacent weight matrices
will appear as an order parameter. Subsequently we will look at another
formulation of feature learning that studies the adaptation of the
Gaussian process kernel itself to the joint statistics of $x$ and
$z$ \citep{seroussi23_908,Fischer24_10761}. We will here follow
the exposition of the latter work. 

\section{NNGP is label-agnostic}

We have obtained the NNGP approximation from \eqref{eq:p_z_y}, rewritten
with help of the auxiliary fields (in the case of a single hidden
layer network) as 
\begin{align*}
\int dy\,p(z,y|C^{(xx)}) & =\int\D C\,\exp\big(-\frac{1}{2}z^{\T}(C+\kappa\I)^{-1}z-\frac{1}{2}\ln\det(C+\kappa\I)-\Gamma(C)\big).
\end{align*}
The NNGP limit corresponds to computing the maximum of $\Gamma(C)\propto N$,
while neglecting the term $-\frac{1}{2}y^{\T}(C+\kappa\I)^{-1}y-\frac{1}{2}\ln\det(C+\kappa\I)\propto\order(P)$.
This may also be seen as computing the maximum of $C$ if one integrates
out $z$ in addition to $y$, because one is then left with
\begin{align*}
\int dz\,\int dy\,p(z,y|C^{(xx)}) & =\int\D C\,\exp\big(-\Gamma(C)\big).
\end{align*}
By comparing to \eqref{eq:p_z_y}, integrating out $z$ removes the
term that depends on the training labels. This shows that the posterior
distribution of the weights in this case is identical to the prior
distribution of the weights, because the ``data-term'' term $\propto\|z-y\|^{2}$,
which would change the prior distribution to the posterior, has dropped
out. This is also obvious from the fact that the resulting iteration
\eqref{eq:NNGP} or \eqref{eq:iteration_DNN} that yields the most
likely value for the $C$ is independent of $z$.

In this sense, the NNGP limit is data agnostic. It does not describe
how the distribution of the weights within the networks change due
to the presence of the training data. To capture this, we will need
to study the effect of ``data-term'' $\propto\|z-y\|^{2}$ on the
posterior distribution of the weights and hence on the posterior distribution
of the network outputs. So the distribution of the weights that influence
the statistics of $C$ is the same as their prior distribution. So
where is the learning then in this limit? Which parameters actually
change?

Since the mean and predictive distribution is still given by \eqref{eq:stat_pred_NNGP}
and since these expressions are the same as those of Bayesian linear
regression $y=w^{\T}x$, only with $C^{(L)}$ instead of $C^{(xx)}=\frac{g_{v}}{d}XX^{\T}$,
the only weights that change are these readout weights $w$. This
implies that the NNGP limit describes learning of the last layer of
the network, while all internal weights are essentially static, given
by their prior distribution. The reason is that due to the large $N$
and hence the dominance of $\Gamma(C)$, the inner part of the network
is very ``stiff''.

\section{Field theory of single hidden layer network}

We here again first consider the setup studied for the derivation
of the NNGP, a single hidden layer network, described by \eqref{eq:single_hidden_net}
with Gaussian i.i.d. priors for all weights and we consider $P$ tuples
of training data $\cD=\{(x_{\alpha},z_{\alpha})\}_{1\le\alpha\le P}$.
A difference will be that we consider $P\propto N$, the number of
data samples scaling linear with the number of hidden units.

The i.i.d. distribution of the input weights $V_{ij}\stackrel{\text{i.i.d.}}{\sim}\N(0,g_{V}/d)$
implies that

\begin{align}
h_{\alpha i} & \stackrel{\text{i.i.d. over }i}{\sim}\N(0,C^{(xx)}),\label{eq:hidden_first-1}\\
\bR^{P\times P}\ni C^{(xx)} & =\frac{g_{V}}{d}\,X\,X^{\T}.\nonumber 
\end{align}
Assuming an i.i.d. Gaussian regularization noise of variance $\kappa$,
the the central object of interest is the joint distribution of the
readouts $y=\{y_{\alpha}\}_{1\le\alpha\le P}$ and the network outputs
$z=\{z_{\alpha}\}_{1\le\alpha\le P}$, which follows from standard
manipulations as before (cf. \eqref{eq:pre_disorder_W}) as

\begin{align}
p(z,y|C^{(xx)}) & =\N(z|y,\kappa)\,\int\D\ty\,\big\langle\exp\big(-\ty^{\T}y+W(\ty|w)\,\big)\big\rangle_{w_{i}\stackrel{\text{i.i.d. }}{\sim}\N(0,\frac{g_{w}}{N})},\label{eq:p_z_y_feature}
\end{align}
where the cumulant-generating function $W$ appears that is a sum
of products of $w_{i}\phi(h_{\alpha i})$, which are independent across
$i$
\begin{align}
W(\ty|w):= & \ln\,\Big\langle\exp\big(\sum_{\alpha=1}^{P}\,\ty_{\alpha}\sum_{i=1}^{N}w_{i}\,\phi(h_{\alpha i})\big)\Big\rangle_{h_{\alpha i}\stackrel{\text{i.i.d. over }i}{\sim}\N(0,C^{(xx)})}.\label{eq:W_layer}
\end{align}

\section{Kernel scaling approach in linear single hidden networks}

Now consider the special case that the activation function $\phi(h)=h$
is the identity. Then \eqref{eq:W_layer} can be computed trivially
as
\begin{align}
W(\ty|w):= & \ln\,\Big\langle\exp\big(\sum_{\alpha}\ty_{\alpha}\sum_{i}w_{i}\,h_{\alpha i}\big)\Big\rangle_{h_{\alpha i}\stackrel{\text{i.i.d. over }i}{\sim}\N(0,C^{(xx)})}\label{eq:W_y_til_w_linear}\\
= & \frac{1}{2}\,\sum_{\alpha\beta}\ty_{\alpha}C_{\alpha\beta}^{(xx)}\ty_{\beta}\,\sum_{i=1}^{N}w_{i}^{2}\,,\nonumber 
\end{align}
where we used the definition of the cumulant-generating function of
a Gaussian. The sum over $i$ appears because of the independence
of the $h_{i}$ across $i$. This shows that the readout weights only
appear in the form of the squared norm $\|w\|^{2}\equiv\sum_{i=1}^{N}w_{i}^{2}$.
The distribution of the output of the network \eqref{eq:p_z_y_feature}
is hence
\begin{align*}
p(z,y|C^{(xx)}) & =\N(z|y,\kappa)\,\int\D\ty\,\big\langle\exp\big(-\ty^{\T}y+\frac{1}{2}\,\ty^{\T}C^{(xx)}\ty\,\|w\|^{2}\big)\big\rangle_{w_{i}\stackrel{\text{i.i.d. }}{\sim}\N(0,\frac{g_{w}}{N})}.
\end{align*}
In the limit of large $N$ we may assume the norm $\|w\|$ to concentrate,
namely to obey a large deviation principle, so we introduce this quantity
as an auxiliary variable, which we name $Q:=\|w\|^{2}=\sum_{i=1}^{N}w_{i}^{2}$;
also note that, given $\|w\|$, the integral over $\ty$ in \eqref{eq:p_z_y_feature}
simply yields $y|_{\|w\|^{2}}\sim\N(0,\|w\|^{2}C^{(xx)})$, so
\begin{align}
p(z,y|C^{(xx)}) & =\N(z|y,\kappa)\,\int dQ\,\N(y|0,Q\,C^{(xx)})\,p(Q).\label{eq:p_z_y_Q}
\end{align}
Here the distribution of the squared norm is 
\begin{align}
p(Q) & =\big\langle\delta[-Q+\|w\|^{2}]\big\rangle_{w_{i}\stackrel{\text{i.i.d.}}{\sim}\N(0,\frac{g_{w}}{N})}\label{eq:p_Q}\\
 & =\int_{-i\infty}^{i\infty}\frac{d\tilde{Q}}{2\pi i}\,\big\langle\exp\big(\tilde{Q}\big[-Q+\sum_{i=1}^{N}w_{i}^{2}\big]\big)\big\rangle_{w_{i}\stackrel{\text{i.i.d.}}{\sim}\N(0,\frac{g_{w}}{N})}\nonumber \\
 & =\int_{-i\infty}^{i\infty}\frac{d\tilde{Q}}{2\pi i}\,\exp\big(-\tilde{Q}Q-\frac{N}{2}\,\ln\big[1-2\frac{g_{w}}{N}\tilde{Q}\big]\big),\nonumber 
\end{align}
where we performed the $d$ Gaussian integrals over the $N$ mutually
independent $w_{i}$ in the last step; in detail
\begin{align*}
\big\langle\exp\big(\tilde{Q}w_{i}^{2}\big]\big\rangle_{w_{i}\sim\N(0,\frac{g_{W}}{N})} & =\frac{1}{\sqrt{2\pi\,g_{w}/N}}\,\int\,dw\,\exp\big(-\frac{1}{2}\,\Big[\frac{N}{g_{w}}-2\tilde{Q}\Big]\,w^{2}\big)\\
 & =\frac{1}{\sqrt{2\pi g_{w}/N}}\,\sqrt{2\pi\,\Big[\frac{N}{g_{w}}-2\tilde{Q}\Big]^{-1}}\\
 & =\sqrt{\frac{1}{\frac{g_{w}}{N}\,\Big[\frac{N}{g_{w}}-2\tilde{Q}\Big]}}=\Big[1-2\,\frac{g_{w}}{N}\,\tilde{Q}\Big]^{-\frac{1}{2}}.
\end{align*}
So far all steps are exact.

Eq. \eqref{eq:p_z_y_Q} shows that the auxiliary variable $Q$ being
a scalar may only carry fluctuations in the direction of the overall
scaling of the kernel. As we will be seeking a mean-field approximations
within this variable, the only result can be that the scale of the
covariance of the readout $y$ will be $QC^{(xx)}$. We will see in
the following that this procedure can be iterated across layers. The
reason is that within the mean-field approximation, namely neglecting
fluctuations of $Q$, preactivations remain independent Gaussian across
neuron indices $i$, which is required for the step \eqref{eq:W_y_til_w_linear}
to be applied iteratively.

\section{Approximation for large $N$}

One expects that $Q$ concentrates, because it is the distribution
of $Q=\|w\|^{2}=\sum_{i=1}^{N}w_{i}^{2}$ for large $N$ and for i.i.d.
$w_{i}\sim\N$. Formally, this is shown by the scaled cumulant generating
function of the form $N\,\lambda_{Q}\,(\frac{\tilde{Q}}{N}\,)=-\frac{N}{2}\,\ln\big[1-2g_{W}\,\frac{\tilde{Q}}{N}\big]$
appearing in \eqref{eq:p_Q}, which implies a mean of order $\langle Q\rangle=\order(1)$
and all higher order cumulants of $Q$ being suppressed by at least
$\order(N^{-1})$. So on exponential scales, one obtains the distribution
of $Q$ from the rate function, which is the Legendre transform of
the cumulant-generating function, as we have seen in \prettyref{chap:The-law-of-large-number}
\begin{align}
p(Q) & \simeq e^{-\Gamma(Q)}\,,\nonumber \\
\Gamma(Q) & =\sup_{\tilde{Q}}\tilde{Q}Q-N\,\lambda_{Q}(\frac{\tilde{Q}}{N})\label{eq:GammaQ}\\
 & =\sup_{\tilde{Q}}\tilde{Q}Q+\frac{N}{2}\,\ln\big[1-2g_{w}\,\frac{\tilde{Q}}{N}\big]\,,\nonumber 
\end{align}
which in field theory terms is a saddle point approximation in $\tilde{Q}$.
The supremum condition can be computed exactly which yields the condition
$0\stackrel{!}{=}Q-g_{w}\,\big[1-2g_{w}\frac{\tilde{Q}}{N}\big]^{-1}$,
solved for $1-\frac{2g_{w}}{N}\tilde{Q}=\frac{g_{w}}{Q}$ and 
\begin{align}
\tilde{Q} & =\frac{N}{2g_{w}}\,\big[1-\frac{g_{w}}{Q}\big]\nonumber \\
 & \equiv\Gamma^{\prime}(Q)\label{eq:eq_state_Q}
\end{align}
 and inserted into \eqref{eq:GammaQ} yields the rate function for
$Q$
\begin{align}
\Gamma(Q) & =\frac{N}{2g_{w}}\,\big[Q-g_{w}\big]-\frac{N}{2}\,\ln\,\frac{Q}{g_{w}}.\label{eq:Gamma_final}
\end{align}
So the final expression for the joint probability of $z$ and $y$
is
\begin{align}
p(z,y|C^{(xx)}) & \simeq\N(z|y,\kappa)\,\int dQ\,\N(y|0,Q\,C^{(xx)})\,e^{-\Gamma(Q)}.\label{eq:p_z_y_pre_saddle}
\end{align}

\subsection{Recovering the NNGP result}

For the previously obtained NNGP result \eqref{eq:GP_final} we neglected
the data term or, in other words, were only interested in the maximum
of the rate function. Thus we can recover this result by searching
for the most probable $Q$ according to the rate function.

\begin{align*}
0 & \overset{!}{=}\Gamma^{\prime}\left(Q\right)\\
 & =\frac{N}{2g_{w}}-\frac{N}{2Q}
\end{align*}

\[
\Rightarrow\,Q=g_{w}\,.
\]
This rescaling by $g_{w}$ is exactly what we also would have obtained
with the iteration derived in \prettyref{sec:Intuitive-approach-to-NNGP}

\begin{align*}
C_{\alpha\beta} & =g_{w}\langle\phi_{\alpha}\phi_{\beta}\rangle_{h\sim\N(0,C^{(xx)})}\\
 & =g_{w}\langle h_{\alpha}h_{\beta}\rangle_{h\sim\N(0,C^{(xx)})}\\
 & =g_{w}\,C^{(xx)}
\end{align*}
such that

\begin{align*}
p(z,y|C^{(xx)}) & \overset{\mathrm{NNGP}}{\simeq}\N(z|y,\kappa)\,\N(y|0,g_{w}\,C^{(xx)})\,.
\end{align*}

\section{Posterior}

We will now compute the saddle point in $Q$ of \eqref{eq:p_z_y_pre_saddle}.
One may wonder what is the meaning of this step: By \eqref{eq:p_z_y_Q},
this has the meaning of computing the maximum of the posterior for
$Q=\|w\|^{2}$. To see this, first note that we may obtain an action
which only contains $z$ if one marginalizes \eqref{eq:p_z_y_pre_saddle}
over the network outputs $y$, because this yields
\begin{align}
p(z|C^{(xx)}) & \equiv\int dy\,p(z,y|C^{(xx)})\label{eq:p_z_Q}\\
 & =\int dQ\,\exp\big(S(Q|z)\big),\nonumber 
\end{align}
with the action

\begin{align}
S(Q|z) & =-\frac{1}{2}z^{\T}\,\big(QC^{(xx)}+\kappa\I\big)^{-1}\,z-\frac{1}{2}\ln\det\big(QC^{(xx)}+\kappa\I\big)-\Gamma(Q)\,,\label{eq:S_Q_z}
\end{align}
where in the rate function $\Gamma(Q)=\frac{N}{2g_{w}}\,Q-\frac{N}{2}\,\ln\,Q+\const$
given by \eqref{eq:Gamma_final} we may drop all terms that are independent
of $Q$ and we here used that the integral over $y$ can be considered
as the convolution of the probability distribution $\N(z|y,\kappa)$
with the distribution of the readout $y$, giving rise to the addition
of the corresponding variances.

This form agrees to the original work \citep{Li21_031059}, their
Eq. A11 after we have inserted the rate function \eqref{eq:Gamma_final}.
This means that in the saddle point equation for $Q$ one ignores
the effect of $y$. Taking into account that the form of \eqref{eq:p_z_Q}
is $p(z)=\int PQ\,p(z|Q)\,p(Q)$, computing the $Q$-integral in saddle
point approximation has the meaning that one determines the maximum
a posteriori for $Q$, because

\begin{align*}
p(Q|z) & =\frac{p(z|Q)\,p(Q)}{p(z)}\,,
\end{align*}
whose maximum $Q^{\ast}$ only depends on the numerator and is therefore
given by the stationary point of \eqref{eq:S_Q_z}, because $\ln\,p(z|Q)\,p(Q)=S(Q|z)+\const$
has the same stationary point as $p(z|Q)\,p(Q)$. So the result is
the most likely value for the length of the readout vector $Q^{\ast}=\|w\|^{2}$.
In the following we will see how one can compute from here other useful
properties of the predictor.

The saddle point approximation of $Q$ in \prettyref{eq:p_z_Q}, obtained
as the maximum of \eqref{eq:S_Q_z}, is given by
\begin{align}
0 & =\frac{\partial S}{\partial Q}=\frac{1}{2}z^{\T}\big(QC^{(xx)}+\kappa\I\big)^{-1}C^{(xx)}\,\big(QC^{(xx)}+\kappa\I\big)^{-1}z-\tr\,C^{(xx)}\big(QC^{(xx)}+\kappa\I\big)^{-1}-\frac{N}{2}\big(\frac{1}{g_{w}}-\frac{1}{Q}\big).\label{eq:stat_Q}
\end{align}
To obtain the derivative of $\big(QC^{(xx)}+\kappa\I\big)^{-1}=:N^{-1}$
by $Q$, we first considered $[N^{-1}\,N]_{\gamma\delta}=\delta_{\gamma\delta}$,
differentiated by $N_{\alpha\beta}$ yields $[N^{-1}]_{\gamma\alpha}\,\delta_{\beta\delta}+\sum_{\epsilon}\,\frac{\partial[N^{-1}]_{\gamma\epsilon}}{\partial N_{\alpha\beta}}\,N_{\epsilon\delta}=0$,
multiplied from right with $[N^{-1}]_{\delta\iota}$ and summed over
$\delta$ one has $\frac{\partial[N^{-1}]_{\gamma\iota}}{\partial N_{\alpha\beta}}=-[N^{-1}]_{\gamma\alpha}[N^{-1}]_{\beta\iota}$.
The using the chain rule $\frac{\partial N_{\alpha\beta}}{\partial Q}=C_{\alpha\beta}^{(xx)}$.
Also we used that $\frac{\partial}{\partial N}\,\ln\det(N)=N^{-1}$,
which follows from writing $-\frac{1}{2}\ln\det(N)=\ln\,\int Px\,\exp\big(-\frac{1}{2}x^{\T}Nx\big)+\const$.
Differentiating the right hand side by $N_{\alpha\beta}$ yields the
second moment $-\frac{1}{2}\langle x_{\alpha}x_{\beta}\rangle\equiv-\frac{1}{2}\,[N^{-1}]_{\alpha\beta}$
(also shown in the earlier exercises).

\section{Predictor statistics}

To obtain predictions beyond the length of the readout $\|w\|^{2}$
we may start from \eqref{eq:p_y_z_given_X-1}, but generalized such
that instead of variance $\kappa$ we insert a general covariance
matrix $K$ and we perform an integration over $y$
\begin{align}
p(z|K,C^{(xx)}) & =\int dy\,\N(z|y,K)\,\big\langle\,\prod_{\alpha=1}^{P}\delta\,\big[y_{\alpha}-\sum_{i=1}^{N}w_{i}\,\phi(h_{\alpha i})\big]\big\rangle_{w_{i}\stackrel{\text{i.i.P.}}{\sim}\N(0,\frac{g_{w}}{N}),\quad h_{\alpha i}\stackrel{\text{i.i.d. over }i}{\sim}\N(0,C^{(xx)})}.\label{eq:p_y_z_given_X-1}
\end{align}
If one sets the matrix $K=\kappa\I$, we again arrive at the same
expression \eqref{eq:p_y_z_given_X-1} as before. For this case, note
that one may use $z$ as a source variable to differentiate by to
obtain the mean discrepancy $\langle\Delta\rangle:=z-\langle y\rangle$
between target and network output, because of $\N(z|y,\kappa\I)\propto\exp\big(-\|z-y\|^{2}/(2\kappa)\big)$
\begin{align}
\frac{\partial}{\partial z_{\alpha}}\,\ln p(z|\kappa\I,C^{(xx)}) & =-\frac{1}{\kappa}\,\big\langle z_{\alpha}-y_{\alpha}\big\rangle.\label{eq:mean_pred_Delta}
\end{align}
We obtain an expression in the mean-field approximation for this observable
as
\begin{align*}
\frac{\partial}{\partial z_{\alpha}}\,\ln p(z|\kappa\I,C^{(xx)}) & \stackrel{\text{mean-field}}{\simeq}\frac{d}{dz_{\alpha}}\,\sup_{Q}S(Q|z,\kappa\I)\\
 & =\frac{\partial}{\partial z_{\alpha}}\,S(Q^{\ast}|z,\kappa\I)+\underbrace{\frac{\partial S}{\partial Q}}_{=0}\,\frac{\partial Q}{\partial z_{\alpha}}\big|_{Q=Q^{\ast}}\,,
\end{align*}
where the derivative by $Q$ vanishes because $Q^{\ast}$ has been
determined by the supremum condition as the stationary point of the
action. The partial derivative by $z_{\alpha}$ only acts on $-z^{\T}\,\big(QC^{(xx)}+\kappa\I\big)^{-1}\,z/2$
in the expression for \eqref{eq:S_Q_z}, which hence yields
\begin{align}
\langle\Delta_{\alpha}\rangle & =\kappa\,\big(QC^{(xx)}+\kappa\I\big)^{-1}\,z\,.\label{eq:mean_discrepancies}
\end{align}
Likewise, the presence of the general matrix $K$ allows us to measure
the statistics of the discrepancies $\Delta_{\alpha}:=z_{\alpha}-y_{\alpha}$,
because, writing the Gaussian $\N(z|y,K)=\exp\big(-\frac{1}{2}(z-y)^{\T}K^{-1}(z-y)+\frac{1}{2}\,\ln\,\det\,(K^{-1})\big)$
explicitly we observe that derivatives by $\big[K^{-1}\big]_{\alpha\beta}$
yield
\begin{align}
\frac{\partial}{\partial[K]_{\alpha\beta}^{-1}}\ln p(z|K,C^{(xx)})\Big|_{K=\kappa\I} & =-\frac{1}{2}\langle(z-y)_{\alpha}(z-y)_{\beta}\rangle+\frac{1}{2}\,\kappa\,\delta_{\alpha\beta}\,,\label{eq:gen_disc}
\end{align}
where we used that $\partial\,\ln\,\det\,(K^{-1})/\partial K^{-1}=K$,
as before. With the same manipulations that led to \eqref{eq:p_z_Q}
one has with $C(Q)=Q\,C^{(xx)}$ the action

\begin{align*}
S(Q|z,K) & =-\frac{1}{2}z^{\T}\,\big(C(Q)+K\big)^{-1}\,z-\frac{1}{2}\ln\det\big(C(Q)+K\big)-\Gamma(Q)\,.
\end{align*}
So in mean-field approximation for $Q$ we get
\begin{align*}
\frac{\partial}{\partial[K]_{\alpha\beta}^{-1}}\ln p(z|K,C^{(xx)})\Big|_{K=\kappa\I} & \stackrel{\text{mean-field}}{\simeq}\frac{d}{d[K]_{\alpha\beta}^{-1}}\,\sup_{Q}S(Q|z,K)\Big|_{K=\kappa\I}\\
 & =\frac{\partial}{\partial[K]_{\alpha\beta}^{-1}}S(Q^{\ast}|z,K)\Big|_{K=\kappa\I}+\underbrace{\frac{\partial S}{\partial Q}\Big|_{K=\kappa\I}}_{=0}\,\frac{\partial Q}{\partial[K]_{\alpha\beta}^{-1}}\,,
\end{align*}
where the inner derivative by $\partial S/\partial Q$ again drops
out due to stationarity of the action at $Q^{\ast}$, which is given
by the solution of \eqref{eq:stat_Q}. The latter partial derivative
evaluates to
\begin{align}
\frac{\partial}{\partial[K]_{\alpha\beta}^{-1}}S(Q|z,K)\Big|_{K=\kappa\I} & =\Big[-\frac{1}{2}z^{\T}\,\big[C+K\big]^{-1}\,KK\,\big[C+K\big]^{-1}\,z+\frac{1}{2}\,K\,(C+K)^{-1}\,K\Big]_{\alpha\beta}\Big|_{K=\kappa\I}\label{eq:var_as_first_deriv}\\
 & =\kappa^{2}\,\Big[-\frac{1}{2}\,\big[C+\kappa\I\big]^{-1}zz^{\T}\,\big[C+\kappa\I\big]^{-1}+\frac{1}{2}\,(C+\kappa\I)^{-1}\Big]_{\alpha\beta}\Big|_{K=\kappa\I}\,,\nonumber 
\end{align}
where we used that $\partial K_{\gamma\delta}/\partial[K]_{\alpha\beta}^{-1}=-K_{\gamma\alpha}\,K_{\beta\delta}$,
which follows by symmetry from $\partial K_{\gamma\delta}^{-1}/\partial K_{\alpha\beta}=-K_{\gamma\alpha}^{-1}\,K_{\beta\delta}^{-1}$.

So the second moment of the discrepancies with \eqref{eq:gen_disc}
is
\begin{align}
\langle\Delta_{\alpha}\Delta_{\beta}\rangle & =\kappa^{2}\,\Big[\,\big[C^{\ast}+\kappa\I\big]^{-1}zz^{\T}\,\big[C^{\ast}+\kappa\I\big]^{-1}\label{eq:exp_discrepancies_matrix}\\
 & +\I/\kappa-(C^{\ast}+\kappa\I)^{-1}\,\Big]_{\alpha\beta}\nonumber 
\end{align}
and the average training loss is
\begin{align}
\langle\cL\rangle= & \frac{1}{2}\,\sum_{\alpha=1}^{P}\langle(z_{\alpha}-y_{\alpha})^{2}\rangle\label{eq:avg_training_loss}\\
= & \frac{1}{2}\,\tr\,\langle\Delta\Delta^{\T}\rangle.\nonumber 
\end{align}
Note that the computation of the variance by a first derivative in
\eqref{eq:var_as_first_deriv} ignores fluctuations of $C$; this
limits the accuracy of this result. Indeed, one finds that in certain
settings, such fluctuation corrections produce sizable corrections
to the predictor statistics (see, e.g. \citep{Rubin24_iclr,Rubin25}).

The expression for the expected discrepancies can be seen in the light
of the bias-variance decomposition, because the first line in \eqref{eq:exp_discrepancies_matrix}
is, by \eqref{eq:mean_discrepancies}, the outer product of the mean
discrepancies $\langle\Delta_{\alpha}\rangle\langle\Delta_{\beta}\rangle$.
To this end, note that the mean of the network output for a training
point, according to \eqref{eq:stat_pred_NNGP} is $\langle y\rangle=C\,[C+\kappa\I]^{-1}z$.
The mean discrepancy therefore becomes 
\begin{align}
\langle\Delta\rangle & =z-\langle y\rangle\label{eq:mean_mismatch-1}\\
 & =\big[\I-C\,[C+\kappa\I]^{-1}\big]\,z\nonumber \\
 & =\big[C+\kappa\I-C\big]\,[C+\kappa\I]^{-1}\,z\nonumber \\
 & =\kappa\,[C+\kappa\I]^{-1}\,z.\nonumber 
\end{align}
Likewise, the variance of the predictor for a training point is with
\eqref{eq:stat_pred_NNGP} $\llangle yy^{\T}\rrangle=C-C[C+\kappa\I]^{-1}C$,
which can also be written as
\begin{align}
\llangle yy^{\T}\rrangle & =C-C\,[C+\kappa\I]^{-1}\,C\label{eq:cov_output_NNGP}\\
 & =\overbrace{C-\underbrace{[C+\kappa\I]\,[C+\kappa\I]^{-1}}_{\I}\,C}^{=0}+\kappa\I\,[C+\kappa\I]^{-1}\,C\nonumber \\
 & =\kappa\I\,\underbrace{[C+\kappa\I]^{-1}\,[C+\kappa\I]}_{\I}-\kappa^{2}\,[C+\kappa\I]^{-1}\nonumber \\
 & =\kappa\I-\kappa^{2}\,[C+\kappa\I]^{-1}.\nonumber 
\end{align}
So comparing \eqref{eq:mean_mismatch-1} and \eqref{eq:cov_output_NNGP}
with \eqref{eq:exp_discrepancies_matrix} we observe that we get the
bias-variance decomposition \eqref{eq:bias_variance_decomp}
\begin{align*}
\langle\Delta_{\alpha}\Delta_{\beta}\rangle & =\langle\Delta_{\alpha}\rangle\,\langle\Delta_{\beta}\rangle+\llangle y_{\alpha}y_{\beta}\rrangle.
\end{align*}

\section{Comparison to numerics}

\begin{figure}[H]
\centering{}%
\begin{tabular}{ll}
\textbf{a} & \textbf{b}\tabularnewline
\includegraphics[width=0.5\textwidth]{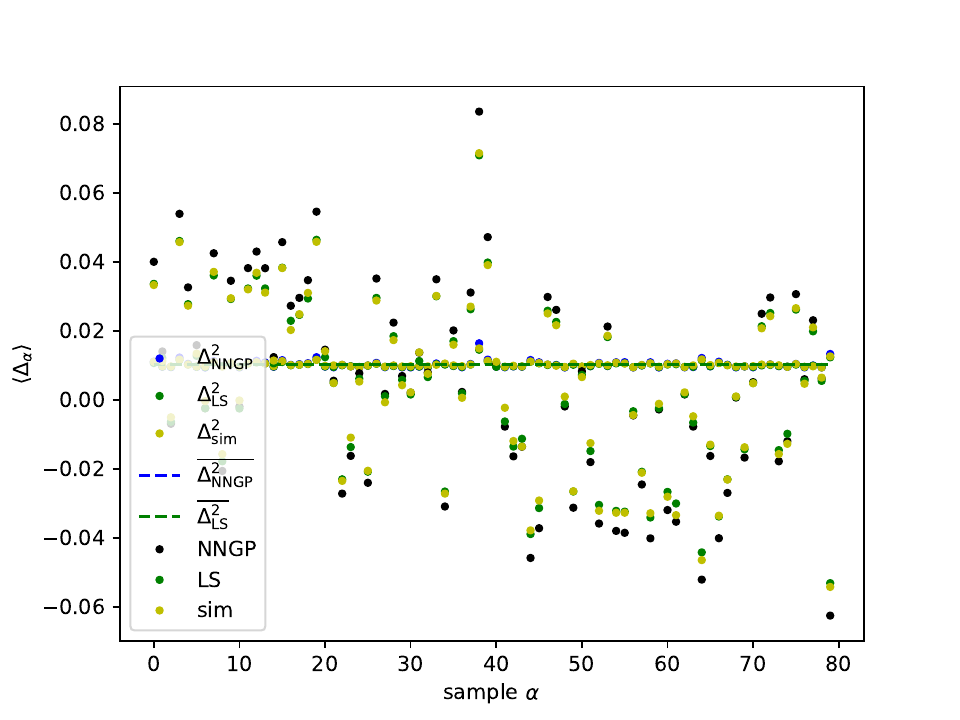} & \includegraphics[width=0.5\textwidth]{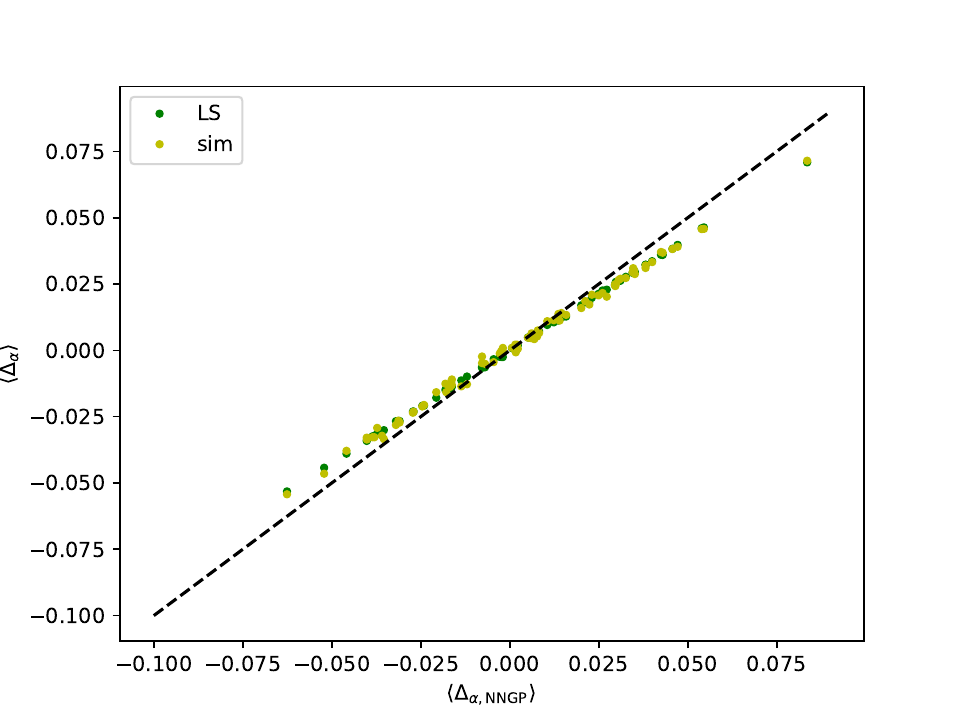}\tabularnewline
\end{tabular}\caption{\textbf{Feature learning in a linear single hidden layer network.}
Comparison of NNGP (black), Li \& Sompolinsky (green) and simulation
(yellow) for mean discrepancies $\langle\Delta_{\alpha}\rangle$.
$P=80$ samples, $N=100$, $d=200$, Ising task with $p=0.1$, regulator
$\kappa=0.01$. Numerical sampling by Langevin dynamics with $1,000,000$
training steps, $20,000$ samples taken.\protect\label{fig:Linear-feature-learning-1-1}}
\end{figure}

\section{Deep linear networks}

\textbf{}We here reformulate the scaling approach by \citep{Li21_031059}
in a manner so that one obtains a physical meaning for the introduced
order parameters and that one may formulate the theory by help of
a large deviation approach.

\section{Setup deep linear network}

We here consider is a network with a $L$ hidden layers whose activations
are called $h^{(a)}\in\bR^{N}$, where $0\le a\le L$.

\begin{eqnarray}
h^{(0)} & = & V\,x,\label{eq:single_hidden_net-2-1}\\
h^{(a)} & = & W^{(a)}\,h^{(a-1)}\quad1\le a\le L,\nonumber \\
y & = & w^{\T}h^{(L)},\nonumber \\
z & = & y+\xi,\nonumber 
\end{eqnarray}
with Gaussian i.i.d. priors $V_{ij}\stackrel{\text{i.i.P.}}{\sim}\N(0,g_{v}/d)$
and $W_{ij}^{(a)}\stackrel{\text{i.i.P.}}{\sim}\N(0,g_{w}/N)$ and
readout noise $\xi_{\alpha}\stackrel{\text{i.i.P.}}{\sim}\N(0,\kappa)$.

\subsection{Backward approach}

Start with the joint probability of the output and the readout

\begin{align}
p(z,y|C^{(xx)}) & =\N(z|y,\kappa)\,\int\big\{\prod_{a=1}^{L}dh^{(a)}\big\}\,p(y|h^{(L)})\,p(h^{(L)}|h^{(L-1)})\cdots p(h^{(2)}|h^{(1)})p(h^{(1)}|X).\label{eq:p_z_y_lin_net_initial_backward}
\end{align}
To follow the backward approach, we first consider the readout conditioned
on the second-last hidden-layer's activations.

\subsubsection{Idea of the approach}

We want to know the joint distribution of the $y_{\alpha}$. Consider
the case that we condition on the last hidden activations $h^{(L-1)}$:
The randomness of $y_{\alpha}$ has two sources, the readout $w$
and the projection from the previous layer $W^{(L)}$. The matrix
$h_{\alpha i}^{(L-1)}$ is given and fixed. The only appearance of
the index $\alpha$ is in $h_{\alpha i}^{(L-1)}$, so the output is
written as
\begin{align}
y_{\alpha} & =\sum_{ij}\,h_{\alpha j}^{(L-1)}W_{ji}^{(L)\T}w_{i}.\label{eq:starting_point}
\end{align}
The first factor plays the role of a projection from the neuron space
to the sample space and is fixed. The two other factors fluctuate.
Since the weights are all drawn i.i.d., we expect that also the product
$\sum_{i}W_{ji}^{(L)\T}w_{i}$ will be i.i.d. and can be approximated
as Gaussian $\eta_{j}\stackrel{\text{i.i.P.}}{\sim}\N(0,\frac{g_{w}}{N})$
but with a variance $Q_{1}$ to be determined
\begin{align}
y_{\alpha} & =\sum_{j}\,h_{\alpha j}^{(L-1)}\,\sqrt{Q_{1}}\,\eta_{j},\label{eq:replaced_by_Gauss}
\end{align}
so that the kernel of the network readout would be
\begin{align*}
\langle y_{\alpha}y_{\beta}\rangle & =\frac{g_{w}}{N}Q_{1}\,\sum_{j=1}^{N}h_{\alpha j}^{(L-1)}h_{\beta j}^{(L-1)}.
\end{align*}
Once replaced by a Gaussian, the approach can be iterated to conditioning
on the layer before, because $\eta_{j}$ then plays the role of $w_{i}$
in the first step. So conditining on the pre-activations of the layer
before, one obtains from \eqref{eq:replaced_by_Gauss}
\begin{align}
y_{\alpha}=\sqrt{Q_{1}}\,\sum_{ij}\,h_{\alpha j}^{(L-2)}W_{ji}^{(L-1)\T}\eta_{i},\label{eq:intuition_iteration}
\end{align}
which is of the same form as \eqref{eq:starting_point}, only that
$w_{i}\to\eta_{i}$ and $L\to L-1$. So the same procedure can be
iterated. The need for the formal approach arises, because we need
to know the most likely value for $Q$. So in a way this corresponds
to a variational Gaussian approximation for the quantity $W_{ji}^{(L)\T}w_{i}$
in the first step and of $W_{ji}^{(L-1)\T}\eta_{i}$ in the following
step, etc.

\subsubsection{Formal approach}

To compute the distribution of the readout \eqref{eq:starting_point}
conditioned on $h^{(L-1)}$
\begin{align*}
p(y|h^{(L-1)}) & =\int\D\ty\,\exp\big(-\ty^{\T}y+W_{1}(\ty|h^{(L-1)})\big),
\end{align*}
we write down its cumulant-generating function
\begin{align}
W_{1}(\ty|h^{(L-1)}) & =\ln\big\langle\exp\big(\sum_{\alpha}\ty_{\alpha}\,\sum_{ij}\,h_{\alpha j}^{(L-1)}W_{ji}^{(L)\T}w_{i}\big)\big\rangle_{W^{(L)},w}.\label{eq:step_one}
\end{align}
Taking the expectation over $W_{ij}^{(L)}\stackrel{\text{i.i.P.}}{\sim}\N(0,g_{w}/N)$
yields
\begin{align*}
W_{1}(\ty|h^{(L-1)})=\ln\big\langle\exp\big(\frac{1}{2}\frac{g_{w}}{N}\,\sum_{\alpha\beta}\ty_{\alpha}\ty_{\beta}\,\sum_{j}\,h_{\alpha j}^{(L-1)}h_{\beta j}^{(L-1)}\,\|w\|^{2}\big)\big\rangle_{w}.
\end{align*}
Taking the disorder average over $w$ is the same problem as in the
single hidden layer network \eqref{eq:W_y_til_w_linear}. So one introduces
$Q:=\|w\|^{2}$ as an auxiliary variable, the length of the readout,
and in the limit of large $N$, one hence obtains

\begin{align}
W_{1}(\ty|h^{(L-1)}) & \simeq\ln\,\int dQ_{1}\,\exp\big(\frac{1}{2}\frac{g_{w}}{N}\,Q_{1}\,\sum_{\alpha\beta}\ty_{\alpha}\ty_{\beta}\,\sum_{j}\,h_{\alpha j}^{(L-1)}h_{\beta j}^{(L-1)}\big)\,e^{-\Gamma(Q_{1})},\label{eq:W_1_ldp}
\end{align}
where the rate function is the same as before, given by \eqref{eq:GammaQ}.

The latter result may be rewritten with help of a Gaussian $\eta_{i}\stackrel{\text{i.i.P.}}{\sim}\N(0,\frac{g_{w}}{N})$
as
\begin{align*}
W_{1}(\ty|h^{(L-1)})\simeq\ln\,\int dQ_{1}\,\big\langle\exp\big(\sqrt{Q_{1}}\,\sum_{\alpha}\ty_{\alpha}\sum_{j}\,h_{\alpha j}^{(L-1)}\eta_{j}\big)\big\rangle_{\eta_{j}\stackrel{\text{i.i.d.}}{\sim}\N(0,\frac{g_{w}}{N})}\,e^{-\Gamma(Q_{1})}.
\end{align*}
This step shows that a saddle point equation in $Q_{1}$, fixing the
variance of $\sqrt{Q_{1}}\eta_{j}$, amounts to replacing 
\begin{align}
\sum_{i}W_{ji}^{(L)\T}w_{i} & =\sqrt{Q_{1}}\,\eta_{j}.\label{eq:eta_overlap}
\end{align}
In particular, the non-Gaussian product of two Gaussian variables
has been replaced by one effective Gaussian variable, as in the conceptual
step \eqref{eq:intuition_iteration}.

Now condition on the pre-activations of one layer before, on $h^{(L-2)}$
\begin{align*}
W_{2}(\ty|h^{(L-2)}) & =\ln\,\int dh^{(L-1)}\,\big\langle\exp\big(W_{1}(\ty|h^{(L-1)}\big)\,\delta\big[h^{(L-1)}-W^{(L-1)\T}h^{(L-2)}\big]\big\rangle_{W^{(L-1)}}\\
 & =\ln\,\int dQ_{1}\,\big\langle\exp\big(\sqrt{Q_{1}}\,\sum_{\alpha}\ty_{\alpha}\sum_{ij}\,h_{\alpha j}^{(L-2)}W_{ji}^{(L-1)\T}\eta_{i}\big)\big\rangle_{\eta_{i}\stackrel{\text{i.i.d.}}{\sim}\N(0,\frac{g_{w}}{N}),W^{(L-1)}}\,e^{-\Gamma(Q_{1})},
\end{align*}
where we computed the trivial (due to the Dirac $\delta$) integral
over $h^{(L-1)}$.

Comparing the expectation value $\big\langle\exp\big(\ldots\sum_{\alpha}\ty_{\alpha}\sum_{ij}\,h_{\alpha j}^{(L-2)}W_{ji}^{(L-1)}\eta_{i}\big)\rangle$
to the expectation value computed in \eqref{eq:step_one}, one realizes
that, apart from the factor $\sqrt{Q_{1}}$, the two are identical
if one renames $\eta_{i}\to w_{i}$ and $L-1\to L$. The expectation
value over $W^{(L-1)}$ hence yields by the same steps as before,
effectively replacing 
\begin{align}
\sum_{i}W_{ji}^{(L-1)\T}\eta_{i} & =\sqrt{Q_{2}}\,\xi_{j}\label{eq:replace_Q2}
\end{align}
in the cumulant-generating function

\begin{align*}
W_{2}(\ty|h^{(L-2)}) & =\ln\,\int dQ_{1}\,\int dQ_{2}\,\big\langle\exp\big(\sqrt{Q_{1}Q_{2}}\,\sum_{\alpha}\ty_{\alpha}\,\sum_{j}\,h_{\alpha j}^{(L-2)}\xi_{j}\big)\big\rangle_{\xi_{j}\stackrel{\text{i.i.d.}}{\sim}\N(0,\frac{g_{w}}{N})}\,e^{-\Gamma(Q_{1})-\Gamma(Q_{2})}.
\end{align*}
We here notice that the meaning of $Q_{2}$ is different from the
meaning in the first step: in contrast to the readout layer, where
the order parameter $Q_{1}=\|w\|^{2}$ has the physical meaning of
the length of the readout vector, here it has the meaning of the length
$\|\eta\|^{2}$, so is also contains the overlap of $\sum_{i}\,W_{ji}^{(L)\T}w_{i}$
which is part of the definition of $\eta_{i}$ by \eqref{eq:eta_overlap},
so its meaning is
\begin{align}
Q_{2} & =\|\eta\|^{2}=\frac{1}{Q_{1}}\,\|W^{(L)\T}w\|^{2},\label{eq:meaning_order_paramQ2}
\end{align}
which contains the projection of the output directions of $W^{(L)}$
on the readout vector $w$. In line with \eqref{eq:meaning_order_paramQ2},
the meaning of the order parameter $Q_{3}$ is with \eqref{eq:replace_Q2}
\begin{align*}
Q_{3} & =\|\xi\|^{2}=\frac{1}{Q_{2}}\,\|W^{(L-1)\T}\eta\|^{2}\\
 & =\frac{1}{Q_{1}Q_{2}}\,\|W^{(L-1)\T}W^{(L)\T}\,w\|^{2}.
\end{align*}
Iterating this approach until the first preactivation one has
\begin{align*}
W_{L}(\ty|X) & =\ln\,\int d^{L}Q\,\big\langle\exp\big(\sqrt{\prod_{l=1}^{L}Q_{l}}\,\sum_{\alpha}\ty_{\alpha}\,\sum_{j}\,x_{\alpha j}\,\eta_{j}\big)\big\rangle_{\eta_{j}\stackrel{\text{i.i.d.}}{\sim}\N(0,\frac{g_{v}}{d})}\,e^{-\sum_{l=1}^{L}\Gamma(Q_{l})}\\
 & =\ln\,\int d^{L}Q\,\exp\big(\frac{1}{2}\prod_{l=1}^{L}Q_{l}\,\sum_{\alpha\beta}\ty_{\alpha}\ty_{\beta}\,C_{\alpha\beta}^{(xx)}\big)\,e^{-\sum_{l=1}^{L}\Gamma(Q_{l})},
\end{align*}
where the only difference in the last step is that we take the disorder
average over $V_{ij}\stackrel{\text{i.i.P.}}{\sim}\N(0,g_{v}/d)$\@.
By induction we also find the meaning of the order parameters as
\begin{align}
\prod_{a=1}^{l}Q_{l} & =\|\Big\{\prod_{a=L}^{L-l+2}W^{(a)\T}\Big\}\,w\|^{2}.\label{eq:prod_Q}
\end{align}
We notice that the result for all $Q_{l}$ fixed is a Gaussian distribution
in $y$, so we obtain for \eqref{eq:p_z_y_pre_saddle}

\begin{align*}
p(z,y|C^{(xx)}) & =\N(z|y,\kappa)\,\int d^{L}Q\,\N\big(y|0,\prod_{l=1}^{L}Q_{l}\,C^{(xx)}\big)\,e^{-\sum_{l=1}^{L}\Gamma(Q_{l})},
\end{align*}
\begin{align}
\int P^{L}Q\,\exp\big(S(Q_{1},\ldots,Q_{L}|z)\big) & :=\int dy\,p(z,y|C^{(xx)})\label{eq:S_Q}\\
S(Q_{1},\ldots,Q_{L}|z) & =-\frac{1}{2}z^{\T}\big(C+\kappa\I\big)^{-1}z-\frac{1}{2}\ln\det\big(C+\kappa\I\big)-\sum_{l=1}^{L}\Gamma(Q_{l})\Big|_{C=\prod_{l=1}^{L}Q_{l}\,C^{(xx)}}.\nonumber 
\end{align}

\subsection{Saddle point solution}

\begin{figure}
\begin{centering}
\begin{tabular}{ll}
\textbf{a} & \textbf{b}\tabularnewline
\includegraphics[width=0.5\textwidth]{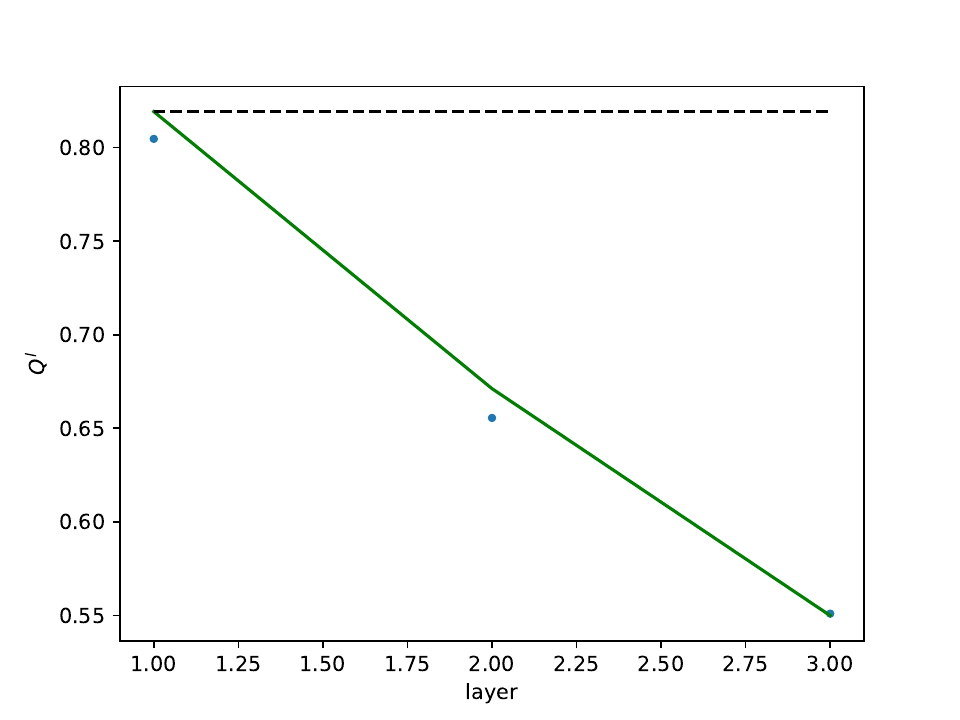} & \includegraphics[width=0.5\textwidth]{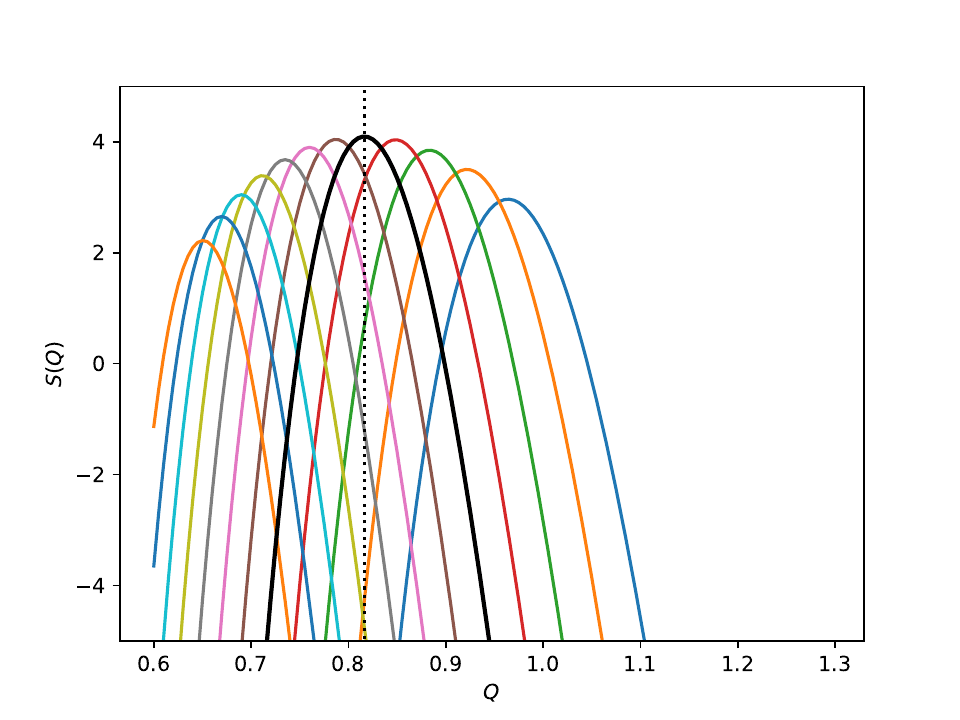}\tabularnewline
\end{tabular}
\par\end{centering}
\caption{\textbf{Feature learning in linear $L=4$ hidden layer network.} \textbf{a}
Order parameters $\prod_{a=1}^{l}Q_{l}$ \eqref{eq:prod_Q} compared
to simulation $\|\prod_{a=L}^{L-l+2}W^{(a)\protect\T}\,w\|^{2}$ for
$T=50,000$ steps. Dashed line is self-consistent solution of \eqref{eq:sc_Q}
for $Q^{\ast}$. \textbf{b} Comparison of action $S(Q)$ for symmetric
solution (black fat), and symmetry broken states (colors, different
values of the respective other $Q$). Dashed fat: Assumption of $Q=g_{w}$
for all but the final layer. Other parameters: $d=200$, $N=100$,
$g_{w}=0.7$, $\kappa=0.01$.\protect\label{fig:Feature-learning-kernel-scaling}}
\end{figure}

Assuming a symmetric solution is compatible with the result by \citep{Li21_031059}
(their Eqs. 45 and 46 in the main text and Appendix A statement after
Eq. (A19)). This symmetry is not clear a priori; one might, of course,
in principle find solutions that break this symmetry. One can check
that the symmetric solution is indeed the one with the highest probability,
as shown in \eqref{fig:Feature-learning-kernel-scaling}. Assuming
this symmetry
\begin{align}
Q^{(a)} & \stackrel{\text{symmetry}}{=}Q\quad\forall1\le a\le L,\label{eq:symmetry_Q}
\end{align}
one obtains $L$ identical stationarity conditions (compare text after
their Eq. (A19)) from the saddle point condition from \eqref{eq:S_Q}
\begin{align}
0 & \stackrel{!}{=}\frac{\partial S}{\partial Q^{(a)}}\big|_{Q^{(a)}=Q}\label{eq:stationarity_Q}\\
 & =\frac{\partial S}{\partial C}\,\frac{\partial C}{\partial Q^{(a)}}\big|_{Q^{(a)}=Q}-\Gamma^{\prime}(Q)\nonumber \\
 & =Q^{L-1}\Big[\frac{1}{2}\,z^{\T}\big[C+\kappa\I\big]^{-1}C^{(xx)}\,\big[C+\kappa\I\big]^{-1}z-\frac{1}{2}\,\tr\,C^{(xx)}\,(C+\kappa\I)^{-1}\Big]\Big|_{C=Q^{L}\,C^{(xx)}}-\frac{N}{2g_{w}}\,\big[1-\frac{g_{w}}{Q}\big],\nonumber 
\end{align}
where we used the equation of state \eqref{eq:eq_state_Q} for $\Gamma^{\prime}$.
Introducing the short hand 
\begin{align*}
r(Q) & :=-\frac{2Q}{N}\,\frac{\partial S}{\partial C}\,\frac{\partial C}{\partial Q^{(a)}}\big|_{Q^{(a)}=Q}\\
 & =\frac{1}{N}\,\Big[-z^{\T}\,\big[C+\kappa\I\big]^{-1}C\,\big[C+\kappa\I\big]^{-1}z+\tr\,C\,(C+\kappa\I)^{-1}\Big]\Big|_{C=Q^{L}\,C^{(xx)}}
\end{align*}
\prettyref{eq:stationarity_Q} is a self-consistency equation
\begin{align*}
0 & =-\frac{N}{2Q}\,r(Q)-\frac{N}{2g_{w}}+\frac{N}{2Q}.
\end{align*}
Multiplied by $2Q/N$ and rearranged

\begin{align}
1-\frac{Q}{g_{w}} & =r(Q),\label{eq:sc_Q}
\end{align}
which is identical to Eq. (A46) in \citep{Li21_031059} and can be
solved by means of bisectioning.

\section{Extension to the non-linear case}

Instead of a linear network, we here want to consider an odd non-linear
activation function $\phi(h)$ for the case of a non-linear network
\citep{Ariosto2022}, here applied to the single hidden layer case.
To this end, start from the setup \eqref{eq:single_hidden_net} and
use \eqref{eq:p_z_y_feature} and \eqref{eq:W_layer}. From step \eqref{eq:W_y_til_w_linear},
we perform an expansion of the cumulant-generating function

\begin{align}
W(\ty|w) & =\frac{1}{2}\,\ty_{\alpha}\,C_{\alpha\beta}^{(\phi\phi)}\,\ty_{\beta}\,\sum_{i}w_{i}^{2}\,+\order\big(\ty^{4}\big),\label{eq:W_phi}
\end{align}
where $C_{\alpha\beta}^{(\phi\phi)}:=\langle\phi(h_{\alpha})\phi(h_{\beta})\rangle_{h_{\alpha}\sim\N(0,C^{(xx)})}$
and where we used the pairwise independence of the $\phi(h_{\alpha i})$
across $i$, so that only diagonal terms appear. This approximation
therefore corresponds to stating that the $\phi_{\alpha}$ be jointly
Gaussian distributed; it is not quite the same as stating that $\phi$
be the identity. The latter would correspond to taking $C^{(\phi\phi)}=C^{(xx)}$
in addition, which is the result for the linear network. So we obtain
the result by \citep{Ariosto2022} from the action \eqref{eq:S_Q_z}
by replacing $C^{(xx)}$ by $C^{(\phi\phi)}$, which corresponds to
their action Eq. (33) in \citep{Ariosto2022}.

\subsection{Recovering the NNGP result}

By neglecting the first two terms in \eqref{eq:S_Q} when searching
for the most probable value of $Q_{l}$, we recover the NNGP limit.

\begin{align*}
0 & \overset{!}{=}\Gamma^{\prime}\left(Q_{l}\right)\\
 & \Rightarrow\,Q_{l}=g_{w}
\end{align*}

\begin{align*}
\Rightarrow\,p(z,y|C^{(xx)}) & =\N(z|y,\kappa)\,\N\big(y|0,g_{w}^{L}\,C^{(xx)}\big)
\end{align*}
This we could have also obtained by using the iterative equation \eqref{eq:iteration_DNN}

\begin{align*}
\bar{C}^{(a)} & =g_{w}\langle\phi^{(a-1)}\phi^{(a-1)}\rangle_{h^{(a-1)}\sim\N(0,\bar{C}^{(a-1)})}\\
 & =g_{w}\langle h^{(a-1)}h^{(a-1)}\rangle_{h^{(a-1)}\sim\N(0,\bar{C}^{(a-1)})}\\
 & =g_{w}\,\bar{C}^{(a-1)}\quad\forall\,1\le a\le L+1
\end{align*}

\[
\Rightarrow\,\bar{C}^{(L)}=g_{w}^{L}\,C^{(xx)}.
\]

\section{Kernel adaptation approach}

So far we have followed the feature learning approach that results
in a rescaling of the data kernel $C^{(xx)}$ by some constant. We
here would like to investigate a complimentary approach, similar to
the one followed by \citep{seroussi23_908} and by \citep{Fischer24_10761}.
The setup here is again the single hidden layer network as in \eqref{eq:single_hidden_net}.
We use \eqref{eq:p_y_given_Cxx_as_Gauss-1} as the starting point
\begin{align}
p(z,y|C^{(xx)}) & \simeq\N(z|y,\kappa)\,\int\D C\,\N(y|0,C)\,e^{-\Gamma(C)},\label{eq:p_z_y_feature_kernel}\\
\Gamma(C) & =\sup_{\tC}\,\tr\,\tC^{\T}C-W\,\big(\tC|C^{(xx)}\big),\nonumber \\
W(\tC|C^{(xx)}) & =N\,\ln\,\Big\langle\exp\,\big(\frac{g_{w}}{N}\,\phi^{\T}\tC\,\phi\big)\Big\rangle_{h_{\alpha}\sim\N(0,C^{(xx)})}.\nonumber 
\end{align}
If we are interested in the maximum posterior of $C_{\alpha\beta}$
\begin{align}
p(C|z) & =\frac{p(z|C)\,p(C)}{p(z)},\label{eq:p_C_given_y}
\end{align}
we need to compute the stationary point of the numerator. First note
that we may obtain the statistics of the predictor by using $z$ or
$\kappa\to K$ as sources to differentiate by; one may therefore integrate
over $y$ without losing information about its statistics; this implies
that the variance of $y$ and the variance $\kappa$ of the regularization
simply add up and one has

\begin{align*}
p(z|C^{(xx)}) & :=\int p(z,y|C^{(xx)})\,Py\\
 & =\int\D C\,\exp\big(S(C|z)\big),
\end{align*}
where the action is

\begin{align}
S(C|z) & =-\frac{1}{2}z^{\T}\big[C+\kappa\I\big]^{-1}z-\frac{1}{2}\ln\det(C+\kappa\I)-\Gamma(C)+\const.,\label{eq:actionC}
\end{align}
whose stationary point $C^{\ast}$ obeys $\partial S/\partial C\stackrel{!}{=}0$
so it fulfills
\begin{align}
0\stackrel{!}{=}\frac{\partial S}{\partial C_{\alpha\beta}}=\frac{1}{2}\big(\big[C+\kappa\I\big]^{-1}\,zz^{\T}\,\big[C+\kappa\I\big]^{-1}\big)_{\alpha\beta}-\frac{1}{2}\big[C+\kappa\I\big]_{\alpha\beta}^{-1}- & \frac{\partial\Gamma(C)}{\partial C_{\alpha\beta}}\quad\forall\alpha,\beta.\label{eq:map_C}
\end{align}
For the rate function holds the equation of state \eqref{eq:eq_of_state_GammaC},
so
\begin{align}
\frac{\partial\Gamma(C)}{\partial C_{\alpha\beta}} & =\tC_{\alpha\beta}.\label{eq:eq_of_state}
\end{align}
So \eqref{eq:map_C} shows that this fixes the $\arg\sup=\tilde{C}$
of the supremum condition in \eqref{eq:gamma_C}. By this latter equation
we hence see how the stationary value of $C$ not only depends on
$p(C)$, but is also shifted due to the presence of the first two
terms (if the first two terms would be absent, we would have $\partial\Gamma(C)/\partial C_{\alpha\beta}=0=\tilde{C}_{\alpha\beta}$,
which is given by \eqref{eq:NNGP}; this would be the uncorrected
NNGP kernel. We will find that $\tilde{C}\neq0$ produces corrections
to the NNGP kernel.

\section{Predictor statistics}

The statistics of the predictor follows completely analogous to the
approach of the linear network. Again, the training label $z$ may
be used as a source to differentiate by, due to the term $\N(z|y,\kappa\I)\propto\exp\big(-\|z-y\|^{2}/(2\kappa)\big)$
in \eqref{eq:p_z_y_feature} as in \eqref{eq:mean_pred_Delta}, to
obtain the mean of the discrepancy
\begin{align*}
\frac{\partial}{\partial z_{\alpha}}\,\ln p(z|C^{(xx)}) & =-\frac{1}{\kappa}\,\big\langle z_{\alpha}-y_{\alpha}\big\rangle.
\end{align*}
Since we compute the integrals over $C$ in \eqref{eq:map_C} in saddle
point approximation, the action is stationary with regard to $C$,
so
\begin{align*}
\frac{\partial}{\partial z_{\alpha}}\,\ln p(z|\kappa\I,C^{(xx)}) & =\frac{d}{dz_{\alpha}}\sup_{C}\,S(C|z)\\
 & =\frac{\partial S}{\partial z_{\alpha}}+\underbrace{\frac{\partial S}{\partial C}}_{=0}\,\frac{\partial C}{\partial z_{\alpha}}\Big|_{C=C^{\ast}}\\
 & =-\big[C+\kappa\I\big]^{-1}z,
\end{align*}
which together yields the same expression \eqref{eq:mean_discrepancies}
as before
\begin{align}
\langle\Delta\rangle & \equiv z-\langle y\rangle\label{eq:mean_discrepancy}\\
 & =\kappa\,\big[C^{\ast}+\kappa\I\big]^{-1}z.\nonumber 
\end{align}
By the same argument, again using that $C^{\ast}$ is a stationary
point of the action, we get, completely analogous as \eqref{eq:exp_discrepancies_matrix},
by differentiating by $[K]_{\alpha\beta}^{-1}$ the matrix of second
moments of the discrepancies
\begin{align}
\langle\Delta_{\alpha}\Delta_{\beta}\rangle & =\kappa^{2}\,\Big[\,\big[C^{\ast}+\kappa\I\big]^{-1}zz^{\T}\,\big[C^{\ast}+\kappa\I\big]^{-1}+\I/\kappa-\big[C^{\ast}+\kappa\I\big]^{-1}\,\Big]_{\alpha\beta}\label{eq:second_moment_discrepancy}\\
 & =\langle\Delta_{\alpha}\rangle\,\langle\Delta_{\beta}\rangle+\llangle y_{\alpha}y_{\beta}\rrangle.\nonumber 
\end{align}
From this result, one obtains an insight for the meaning of the field
$\tC$ given by \eqref{eq:map_C}: We observe that
\begin{align*}
2\kappa^{2}\,\tC & =\langle\Delta\rangle\langle\Delta\rangle^{\T}+\llangle yy^{\T}\rrangle-\kappa\I.
\end{align*}
So the auxiliary field $2\kappa^{2}\tC$ on the off-diagonal elements
equals to the second moment of the discrepancies; on the diagonals
one needs to subtract the variance of the regulator.

\section{Linear single hidden layer network}

To get an idea of the meaning of the mean-field equations, we again
consider the case of $\phi(h)=h$ the identity function. The rate
function $\Gamma(C)$ can then be computed explicitly, because then
\eqref{eq:W_C_tilde} has the closed form
\begin{align*}
W(\tC|C^{(xx)}) & =N\,\ln\,\Big\langle\exp\,\big(\frac{g_{w}}{N}\,h^{\T}\tC\,h\big)\Big\rangle_{h\sim\N(0,C^{(xx)})}.
\end{align*}
Writing the Gaussian expectation value over $h$ explicitly as
\begin{align*}
 & \big\langle\exp\big(\frac{g_{w}}{N}\,h^{\T}\tC h\big)\big\rangle_{h\sim\N(0,C^{(xx)})}\\
= & \frac{1}{(2\pi)^{\frac{P}{2}}\,\det\big(C^{(xx)}\big)^{\frac{1}{2}}}\,\int\,\prod_{\alpha}dh_{\alpha}\,\exp\big(-\frac{1}{2}h^{\T}\,\big([C^{(xx)}]^{-1}-2\frac{g_{w}}{N}\,\tC\big)\,h\big)\\
= & \frac{1}{\det\big(C^{(xx)}\big)^{\frac{1}{2}}\,\det\big([C^{(xx)}]^{-1}-2\frac{g_{w}}{N}\tC\big)^{\frac{1}{2}}}\\
= & \frac{1}{\det\big(\I-2\frac{g_{w}}{N}\,C^{(xx)}\,\tC\big)^{\frac{1}{2}}}.
\end{align*}
Because the latter expression is the normalization condition of a
Gaussian, the argument of the determinant must be symmetric; so are
the matrices $C$ and $\tilde{C}$, which implies that they also commute,
because
\begin{align}
C\,\tC\stackrel{!}{=}(C\,\tC)^{\T} & =\tC^{\T}C^{\T}=\tC\,C.\label{eq:C_commute}
\end{align}
So for $W$ we have
\begin{align}
W(\tC|C^{(xx)}) & =-\frac{N}{2}\,\ln\det\,\big(\I-2\frac{g_{w}}{N}\,C^{(xx)}\,\tC\big).\label{eq:W_closed_form}
\end{align}
The supremum condition in \eqref{eq:gamma_C} then reads as $0\stackrel{!}{=}C-g_{w}\,\big(\I-2\frac{g_{w}}{N}\,C^{(xx)}\tC\big)^{-1}\,C^{(xx)}$
(Note: here the order of the factors must be compatible with differentiating
$-\partial/\partial\tC_{\alpha\beta}\,\frac{N}{2}\ln\,\det\big([C^{(xx)}]^{-1}-2\frac{g_{w}}{N}\tC\big)=g_{w}\,\big([C^{(xx)}]^{-1}-2\frac{g_{w}}{N}\tC\big)^{-1}=g_{w}\,\big[[C^{(xx)}]^{-1}\,\big(\I-2\frac{g_{w}}{N}C^{(xx)}\tC\big)\big]^{-1}=g_{w}\,\big(\I-2\frac{g_{w}}{N}C^{(xx)}\tC\big){}^{-1}C^{(xx)}$).
Rearranging this as $g_{w}\,C^{(xx)}[C]^{-1}=\I-2\frac{g_{w}}{N}\,C^{(xx)}\tC$
and solving for $C^{(xx)}\tC=\frac{N}{2g_{w}}\,\big(\I-g_{w}\,C^{(xx)}[C]^{-1}\big)$.
Multiplying from left with $[C^{(xx)}]^{-1}$ we get
\begin{align}
\tC & =\frac{N}{2}\,\big([g_{w}\,C^{(xx)}]^{-1}-C^{-1}\big).\label{eq:C_tilde_explicit_linear}
\end{align}
So together with \eqref{eq:map_C} one has
\begin{align}
0\stackrel{!}{=}2\frac{\partial S(C)}{\partial C}= & [C+\kappa\I]^{-1}\,zz^{\T}\,[C+\kappa\I]^{-1}-[C+\kappa\I]^{-1}-N\,\big([g_{w}\,C^{(xx)}]^{-1}-C^{-1}\big),\label{eq:lin_MAP_C}
\end{align}
which is a quadratic matrix equation in $C^{-1}$. For large $N$
and $\kappa=0$ this simplifies slightly to
\begin{align*}
0 & \stackrel{N\gg1}{\simeq}C^{-1}\,zz^{\T}\,C^{-1}+N\,\big(C^{-1}-[g_{w}\,C^{(xx)}]^{-1}\big).
\end{align*}

\begin{figure}
\begin{centering}
\begin{tabular}{ll}
\textbf{a} & \textbf{b}\tabularnewline
\includegraphics[width=0.5\textwidth]{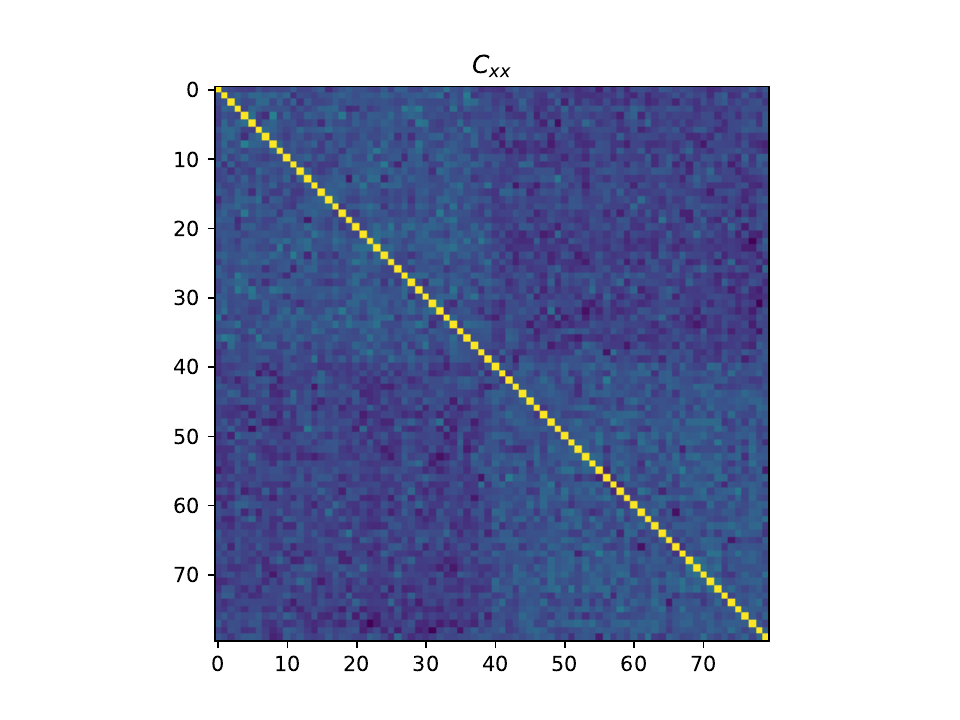} & \includegraphics[width=0.5\textwidth]{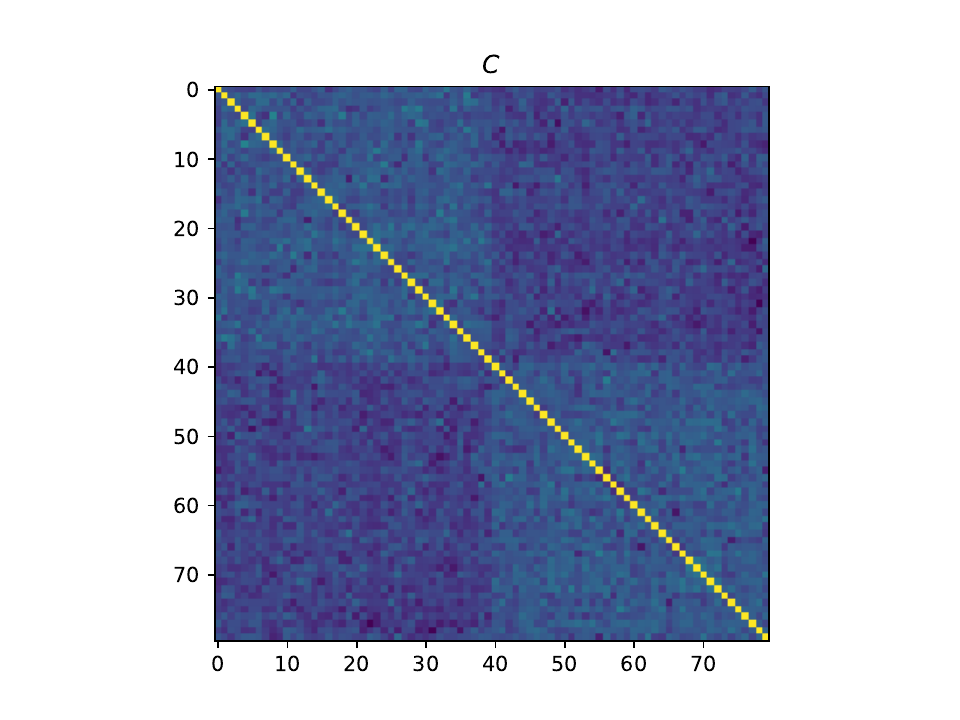}\tabularnewline
\textbf{c} & \textbf{d}\tabularnewline
\includegraphics[width=0.5\textwidth]{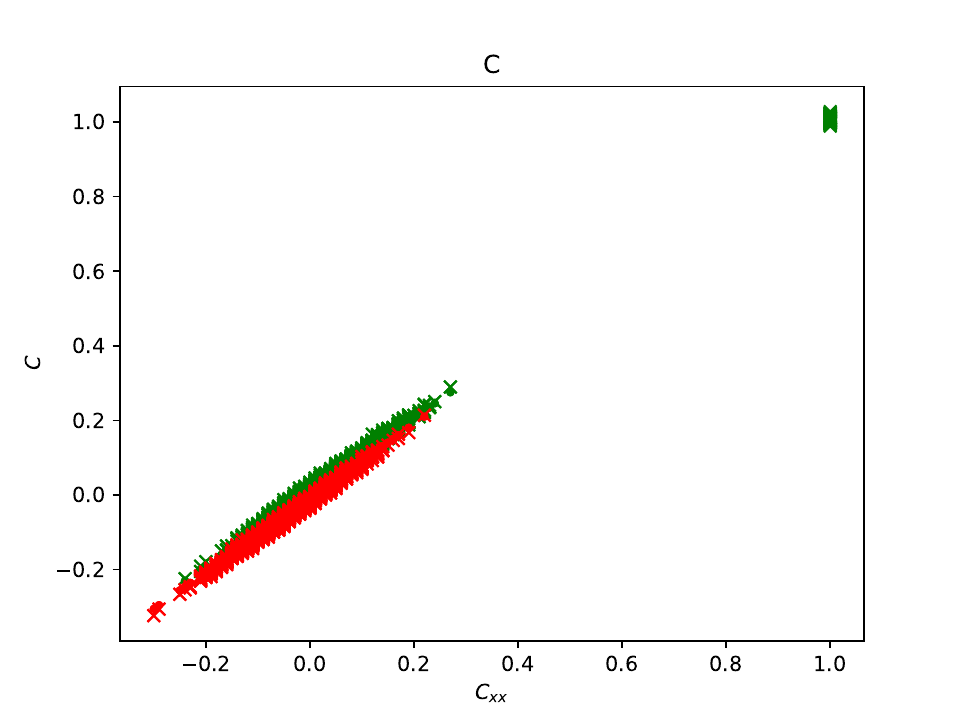} & \includegraphics[width=0.5\textwidth]{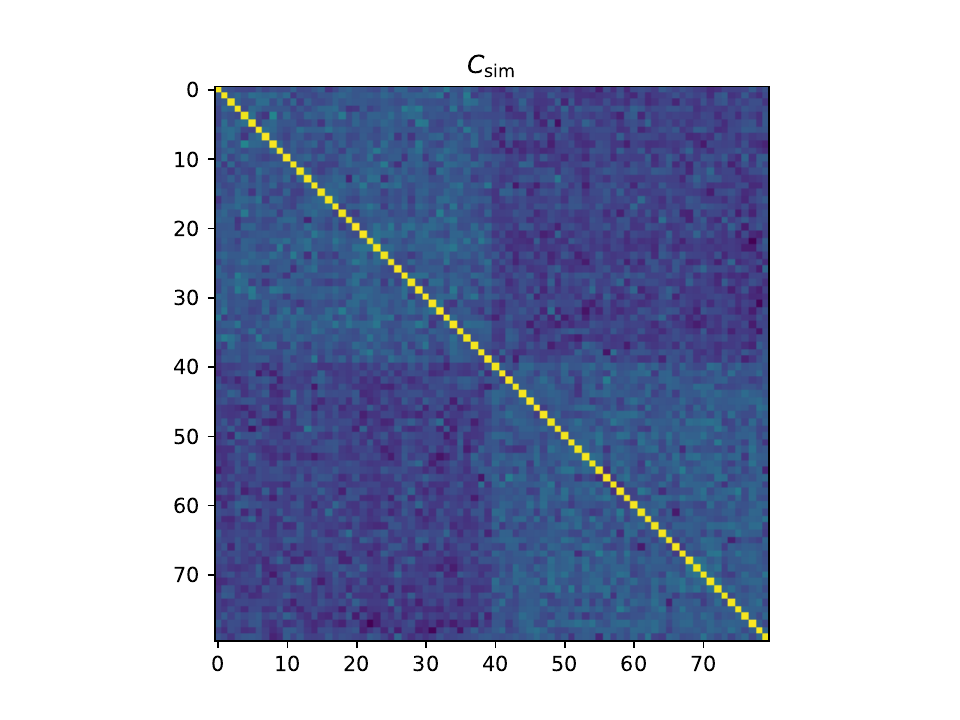}\tabularnewline
\end{tabular}
\par\end{centering}
\caption{\textbf{Comparison of kernels between theory and simulation. a} Input
kernel $C^{(xx)}$. \textbf{b} Kernel $C^{\ast}$ of hidden layer
from theory. \textbf{c} Scatter plot of input kernel versus hidden
layer kernel (crosses: simulation; dots: theory). \textbf{d} Measured
kernel $\frac{g_{w}}{N}\sum_{i=1}^{N}h_{\alpha i}h_{\beta i}$ of
hidden layer from simulation. Other parameters: Single hidden layer
linear network. Regulator \textbf{$\kappa=0.01$}. Ising task with
$p=0.1$, $d=200$, $g_{v}=g_{w}=0.5$, number of hidden units $N=100$;
$P=80$ patterns.\protect\label{fig:Linear-feature-learning-2}}
\end{figure}

The term $zz^{\T}$ for the case of binary classification is a matrix
with a pair of blocks: diagonal blocks with positive sign, off-diagonal
blocks with negative sign, in the case the the labels of the two classes
are $\pm1$. The term $[C+\kappa\I]^{-1}\,zz^{\T}\,[C+\kappa\I]^{-1}=\langle\Delta\rangle\langle\Delta\rangle^{\T}/\kappa^{2}$
in \eqref{eq:lin_MAP_C} can be understood as the outer product of
the vectors of mean discrepancies (by comparing to \eqref{eq:mean_discrepancy}).
It shows that the inverse kernel $C^{-1}$ gets a correction precisely
into that direction compared to the inverse NNGP kernel $[g_{w}\,C^{(xx)}]^{-1}$,
also also visible in \prettyref{fig:Linear-feature-learning-2}b,
c.

Even though the corrections in the level of the kernel elements are
small (order $1/N$), a change of $1/N\,zz^{\T}$ corresponds to a
change of the action of the kernel on $z$ as $1/N\,z\,\|z\|^{2}$,
where the length $\|z\|^{2}=P$, so in the proportional limit $P\propto N$,
this causes an outlier eigenvalue of the kernel matrix of $\order(1)$.
More striking effects of this kernel approach to feature learning
are described in \citep{Rubin25}, demonstrating that the sample complexity
can be changed by these corrections.

A more stable way of solving this equation is by a gradient ascent
with regard to $C$. To this end we use that $C_{xx}$ and $C$ as
covariance matrices are both positive semidefinite, so that we compute
instead of the gradient $PS/PC$
\begin{align*}
2C\,\frac{\partial S(C)}{\partial C}C^{(xx)} & =C\,\Big([C+\kappa\I]^{-1}\,zz^{\T}\,[C+\kappa\I]^{-1}-\frac{1}{2}[C+\kappa\I]^{-1}\Big)\,C^{(xx)}\\
 & -N\,\big([g_{w}^{-1}C-C^{(xx)}\big),
\end{align*}
which we use as a gradient ascent to obtain a solution shown in \prettyref{fig:Linear-feature-learning-2}.
The latter equation avoids computing the inverses $C^{-1}$ and $C^{(xx)-1}$,
which may not exist; the inverses including the regulator term $\kappa\I$
for $\kappa>0$ are uncritical.

\section{Nonlinear single hidden layer network}

To treat the nonlinear case, one may expand the cumulant-generating
function $W$ in \eqref{eq:W_C_tilde} into the first two leading
cumulants to compute $\Gamma(C)$ given by \eqref{eq:gamma_C}. Since
the cumulants are the coefficients in the Taylor expansion of the
cumulant-generating function, we may approximate the latter by a truncation
of the former. For the specific scaling form of \eqref{eq:W_C_tilde}
$W(\tC|C^{(xx)})=N\,\lambda\big(\tQ/N)$ a truncation at second order
amounts to neglecting terms of order $N^{-2}$. The cumulant expansion
of $W(\tC|C^{(xx)})$ reads (in Einstein's summation convention, summing
over repeated indices on the right)
\begin{align*}
W(\tC|C^{(xx)}) & =\tC_{\alpha\beta}\,g_{w}\,C_{\alpha\beta}^{(\phi\phi)}+\frac{g_{w}^{2}}{2N}\,\tC_{\alpha\beta}\,C_{\alpha\beta,\gamma\delta}^{(\phi\phi,\phi\phi)}\,\tC_{\gamma\delta}+\order(N^{-2})\\
\\C_{\alpha\beta}^{(\phi\phi)} & =\big\langle\phi_{\alpha}\phi_{\beta}\big\rangle_{h\sim\N(0,C^{(xx)})}\,,\\
C_{\alpha\beta,\gamma\delta}^{(\phi\phi,\phi\phi)} & =\llangle\phi_{\alpha}\phi_{\beta},\phi_{\gamma}\phi_{\delta}\rrangle_{h\sim\N(0,C^{(xx)})}\,,
\end{align*}
where $\llangle\phi_{\alpha}\phi_{\beta},\phi_{\gamma}\phi_{\delta}\rrangle=\langle\phi_{\alpha}\phi_{\beta}\,\phi_{\gamma}\phi_{\delta}\rangle-\langle\phi_{\alpha}\phi_{\beta}\rangle\,\langle\phi_{\gamma}\phi_{\delta}\rangle$
is the second order cumulant of $\phi_{\alpha}\phi_{\beta}$. The
supremum condition for $\Gamma$ then yields a linear equation in
$\tC$
\begin{align}
0 & \stackrel{!}{=}C_{\alpha\beta}-g_{w}\,C_{\alpha\beta}^{(\phi\phi)}-\frac{g_{w}^{2}}{N}\,C_{\alpha\beta,\gamma\delta}^{(\phi\phi,\phi\phi)}\,\tC_{\gamma\delta}\,.\label{eq:stat_condition_for_tilC}\\
\tC_{\alpha\beta} & =\frac{N}{g_{w}^{2}}\,[C^{(\phi\phi,\phi\phi)}]_{\alpha\beta,\gamma\delta}^{-1}\,\big(C_{\gamma\delta}-g_{w}\,C_{\gamma\delta}^{(\phi\phi)}\big)\nonumber 
\end{align}
Since the cumulant expansion to second order keeps Gaussian fluctuations
of $\phi_{\alpha}\phi_{\beta}$, at this order the replacement of
$p(C)$ by $e^{\Gamma(C)}$ is an exact operation; this is so because
integrating over $\tC$ in \eqref{eq:p_C} and taking the supremum
yields the same expression for a Gaussian, namely in both cases one
obtains (apart from algebraic prefactors)

\begin{align}
\Gamma(C) & =\sup_{\tC}\,\tC_{\alpha\beta}C_{\alpha\beta}-W\big(\tC|C^{(xx)}\big).\label{eq:Gamma_explicit}\\
 & =\frac{N}{2}\,\big[C_{\alpha\beta}-g_{w}\,C_{\alpha\beta}^{(\phi\phi)}\big]\,[C^{(\phi\phi,\phi\phi)}]_{\alpha\beta,\gamma\delta}^{-1}\,\big[C_{\gamma\delta}-g_{w}\,C_{\gamma\delta}^{(\phi\phi)}\big],\nonumber 
\end{align}
where $[C^{(\phi\phi,\phi\phi)}]^{-1}$ is the inverse appearing from
solving \eqref{eq:stat_condition_for_tilC}. This shows that $C$
here follows a quadratic potential which may be thought of as a superposition
of Gaussian kernels, where each kernel appears with the Gaussian measure
$e^{-\Gamma(C)}$.

The maximum a posteriori estimate for $C$ thus obeys with \eqref{eq:map_C}
and with \eqref{eq:stat_condition_for_tilC}, where $\tC_{\alpha\beta}$
is the solution of \eqref{eq:stat_condition_for_tilC}, the derivative
of \eqref{eq:Gamma_explicit}
\begin{align*}
\frac{\partial\Gamma(C)}{\partial C_{\alpha\beta}}= & \tC_{\alpha\beta}\stackrel{(\ref{eq:stat_condition_for_tilC})}{=}\frac{N}{g_{w}^{2}}\,[C^{(\phi\phi,\phi\phi)}]_{\alpha\beta,\gamma\delta}^{-1}\,\big[C_{\gamma\delta}-g_{w}\,C_{\gamma\delta}^{(\phi\phi)}\big].
\end{align*}

\section{Keypoints}

In summary:

\noindent\fcolorbox{black}{white}{\begin{minipage}[t]{1\textwidth - 2\fboxsep - 2\fboxrule}%
\textbf{Feature learning}
\begin{itemize}
\item The NNGP kernel only depends on the statistics of the inputs, but
not on the joint statistics of inputs and training labels.
\item One needs to go beyond the NNGP limit and consider a setting where
$P/N\neq0$ to uncover feature learning.
\item There are different approximations to go beyond the NNGP limit in
linear DNNs. Two of them include the kernel scaling approach and the
kernel adaptation approach for large networks.
\item For linear networks, the kernel gets rescaled by a factor $Q$ when
moving across layers, much like the NNGP limit where this factor $Q$
is the variance of the weight prior $g_{w}$; the difference is that
$Q$ is different from $g_{w}$ due to the training process.
\item One can obtain the predictor statistics, such as the mean of the deviations
from the true training label $z_{\alpha}-y_{\alpha}$ and the MSE
training error, from the conjugate fields $\tilde{C}$ in the presented
field theoretic formalism.
\end{itemize}
\end{minipage}}

\section{Appendix: Large deviation principle for Wishart matrices}

The presented kernel adaptation approach relies on the approximation
of the distribution $p(C)$ for the matrix $C\in\bR^{P\times P}$.
We here employed a large deviation approach. In general one needs
to be careful when relying on large deviation results for high-dimensional
random variables. This can be easiest appreciated on the current example
of a single hidden later linear network. The prior distribution of
the matrix 
\begin{align}
C_{\alpha\beta} & :=\frac{1}{N}\sum_{i=1}^{N}\,h_{\alpha i}h_{\beta i}\label{eq:def_Wishart}
\end{align}
defined in \eqref{eq:def_C_phiphi}, for $\phi(h)=h$ is known as
the Wishart distribution \citep{Wishart1928_32}, where $h_{\alpha i}\stackrel{\text{i.i.P. in }i}{\sim}\N(0,C^{(0)})$ 
.

The probability of the matrix is known exactly
\begin{align}
-\ln p(C)/N & =\frac{1}{2}\tr\,[C^{(0)}]^{-1}C-\frac{1-\frac{P-1}{N}}{2}\,\ln\det\big(C\big)+\const.\label{eq:Wishart_exact}
\end{align}
We may compare this expression to the result obtained from the rate
function, which is obtained from \eqref{eq:gamma_C} with \eqref{eq:W_closed_form},
and the supremum condition on $\tC$ which yields \eqref{eq:C_tilde_explicit_linear}
\begin{align*}
\Gamma(C) & :=\sup_{\tC}\,\tr\,\tC^{\T}C-W\,\big(\tC|C^{(0)}\big)\,,\\
W(\tC|C^{(0)})= & -\frac{N}{2}\,\ln\det\,\big(\I-\frac{2}{N}\,C^{(0)}\,\tC\big)\,,\\
\tC= & \frac{N}{2}\,\big([C^{(0)}]^{-1}-C^{-1}\big)\,,
\end{align*}
which yields the final expression
\begin{align}
\Gamma(C)/N & =\frac{1}{2}\tr\,[C^{(0)}]^{-1}\,C-\frac{1}{2}\,\ln\det\,\big(C\big)+\const.\label{eq:Wishart_ldp}
\end{align}
Comparing the expressions \eqref{eq:Wishart_exact} and \eqref{eq:Wishart_ldp},
we notice that the former differs by the factor $\frac{1-\frac{P-1}{N}}{2}$
from the latter, where it is $\frac{1}{2}$. So the two distributions
are close if $P\ll N$.

In the proportional limit $P\propto N$ that we investigate for feature
learning, the assumption that $P\ll N$ is not necessarily justified.
In a typical setting of a neuronal network, the matrix $C^{(0)}=XX^{\T}$
with $X\in\bR^{P\times d}$ has a rank that is $\mathrm{rk}(C^{(0)})=\min(P,d)$.
This implies for its spectrum that only $\min(P,d)$ of its eigenvalues
are non-zero. If we are interested in the limit $P\to\infty$, we
get $\min(P,d)=d$, so we can expand $C^{(0)}$ into the set of the
$d$ eigenvectors $u_{1\le\mu\le d}$ 
\begin{align*}
C^{(0)} & =\sum_{\mu=1}^{d}\,\lambda_{\mu}\,u_{\mu}u_{\mu}^{\T}\,,
\end{align*}
where $\lambda_{\mu}$ are the corresponding eigenvalues. We may write
the vectors $h$ in this basis $\{u_{\alpha}\}$, too,
\begin{align}
h_{\alpha i} & =\sum_{\mu=1}^{d}\sqrt{\lambda_{\mu}}\,z_{\mu,i}\,u_{\alpha\mu},\label{eq:h_decomp_lin}\\
\bR^{d\times N}\ni z_{\mu,i} & \stackrel{\text{i.i.d.}}{\sim}\N(0,1)\,,\nonumber 
\end{align}
so that $C_{\alpha\beta}^{(0)}\,\delta_{ij}=\langle h_{\alpha i}h_{\beta j}\rangle$,
as it should and the $h$ are obviously Gaussian.

The Wishart matrix is defined as \eqref{eq:def_Wishart}. Expressing
the $h_{\alpha i}$ as linear combinations \eqref{eq:h_decomp_lin},
we notice that only a finite number $d\times d$ of projectors $\{u_{\mu}u_{\nu}^{\T}\}_{1\le\mu,\nu\le d}$
appears in its expansion
\begin{align}
C_{\alpha\beta} & =\sum_{\mu,\nu=1}^{d}\sqrt{\lambda_{\mu}}\sqrt{\lambda_{\nu}}\,\sum_{i=1}^{N}z_{\mu,i}z_{\nu,i}\,u_{\alpha\mu}u_{\beta\nu}\nonumber \\
 & =:\sum_{\mu,\nu=1}^{d}\sqrt{\lambda_{\mu}}\sqrt{\lambda_{\nu}}\,\bar{K}_{\mu\nu}\,u_{\alpha\mu}u_{\beta\nu}\,.\label{eq:K_K_bar_transform}
\end{align}
The $\bar{K}_{\mu\nu}:=\sum_{i=1}^{N}z_{\mu,i}z_{\nu,i}$ are random
and follow a Wishart distribution themselves, namely (cf. \eqref{eq:Wishart_exact})
\begin{align*}
-\ln\,p(\bar{K})/N & =\frac{1}{2}\,\tr\,\bar{K}-\frac{1-\frac{d-1}{N}}{2}\,\ln\det\,\big(\bar{K}\big)+\const\,.
\end{align*}
In the limit $P,N\to\infty$, while $d$ stays constant, we may replace
the density with

\begin{align*}
-\ln\,p(\bar{K})/N & \stackrel{d\ll N}{\simeq}\frac{1}{2}\,\tr\,\bar{K}-\frac{1}{2}\,\ln\det\,\big(\bar{K}\big)+\order(P/N)+\const
\end{align*}
The density for the matrix given by \eqref{eq:K_K_bar_transform}
written as $C=U\sqrt{\Lambda}\,\bar{K}\,\sqrt{\Lambda}\,U^{\T}$ with
diagonal matrix $\sqrt{\Lambda}=\mathrm{diag}(\sqrt{\lambda_{1}},\ldots,\sqrt{\lambda_{P}})$
and unitary matrix $U=(u_{1},\ldots,u_{P})$, $U^{\T}U=\bI$ is then
obtained with $\bar{K}=\sqrt{\Lambda}^{-1}U^{\T}\,C\,U\,\sqrt{\Lambda}^{-1}$
and $\tr\,\bar{K}=\tr\,U\,\Lambda^{-1}\,U^{\T}\,C=\tr\,[C^{(0)}]^{-1}\,C$
(exploiting the cyclic invariance of the trace) as well as $\det\big(\bar{K}\big)=\det\big(\sqrt{\Lambda}^{-1}U^{\T}\,C\,U\,\sqrt{\Lambda}^{-1}\big)=\det\big([C^{(0)}]^{-1}\big)\,\det\big(C\big)$
(exploiting that $\det(AB)=\det(A)\det(B)$ as well as $\det(U)=1$
and $\det(\Lambda)=\det(C^{(0)}$)) as
\begin{align*}
-\ln p(C)/N & \stackrel{d\ll N}{\simeq}\frac{1}{2}\,\tr\,[C^{(0)}]^{-1}\,C-\frac{1}{2}\,\ln\det\,\big(C\big)+\const.\,,
\end{align*}
which is the same result as the rate function \eqref{eq:Wishart_ldp}.
In the last step we suppressed additional constant terms that come
from the substitution of the random variable $\bar{K}\to C$.

This result shows that, in the proportional limit where $P\propto N\to\infty$,
but the dimension of the data vectors $d$ stays constant, the large
deviation approach is justified.

\section{Appendix: Expectation values of error functions}

One often considers as an activation function a sigmoidal function,
for example the error function $\phi(x)=\erf(x)$. This can also be
written with
\begin{align*}
g(x) & :=\frac{1}{\sqrt{2\pi}}\,e^{-\frac{x^{2}}{2}},\\
G(x) & :=\int_{-\infty}^{x}g(t)\,dt
\end{align*}
as
\begin{align*}
\phi(x) & =\erf(x)\\
 & =\frac{2}{\sqrt{\pi}}\,\int_{0}^{x}\,e^{-t^{2}}\,Pt\\
 & \stackrel{s=\sqrt{2}t}{=}\sqrt{\frac{2}{\pi}}\,\int_{0}^{\sqrt{2}\,x}\,e^{-\frac{s^{2}}{2}}\,ds\\
 & =2\,\frac{1}{\sqrt{2\pi}}\,\int_{0}^{\sqrt{2}\,x}\,e^{-\frac{s^{2}}{2}}\,ds\\
 & =2\,\big(G(\sqrt{2}\,x)-G(0)\big)\\
 & =2\,G(\sqrt{2}\,x)-1.
\end{align*}
We need to evaluate expectation values of $\phi$ and $\phi^{2}$
with regard to Gaussian distributed arguments, such as
\begin{align}
\langle\phi(x)\rangle_{x\sim\N(\mu,\sigma^{2})} & =2\,\langle G(y)\rangle_{\N(M,\Sigma)}-1\label{eq:gauss_exp_single_G}\\
 & =2\,G(\frac{M}{\sqrt{1+\Sigma}})-1\nonumber \\
 & =\phi(\frac{M}{\sqrt{1+\Sigma}}/\sqrt{2})\nonumber \\
 & =\phi(\frac{\mu}{\sqrt{1+2\sigma^{2}}}),\nonumber 
\end{align}
where $M=\sqrt{2}\,\mu$ and $\Sigma=2\sigma^{2}$. The last result
follows from \citep{vanMeegen21_043077}; one may also see this by
writing $\erf$ as the expectation of the Heaviside function under
a unit variance Gauss, so that variances $1$ and $\Sigma$ add up.

Likewise, we need
\begin{align}
\langle\phi(x)\phi(y)\rangle_{x,y\sim\N\Bigg[\left(\begin{array}{c}
\mu_{1}\\
\mu_{2}
\end{array}\right),\left(\begin{array}{cc}
\sigma_{11}^{2} & \sigma_{12}^{2}\\
\sigma_{21}^{2} & \sigma_{22}^{2}
\end{array}\right)\Bigg]} & =\langle(2G(\sqrt{2}\,x)-1)(2G(\sqrt{2}\,y)-1)\rangle\label{eq:phi_phi_ad_G_G}\\
 & =4\,\langle G(x^{\prime})G(y^{\prime})\rangle-2\langle G(x^{\prime})\rangle-2\langle G(y^{\prime})\rangle+1\nonumber \\
 & =4\,\langle G(x^{\prime})G(y^{\prime})\rangle-2G\big(\frac{M_{1}}{\sqrt{1+\Sigma_{11}}}\big)-2G\big(\frac{M_{2}}{\sqrt{1+\Sigma_{22}}}\big)+1,\nonumber \\
 & =4\,\langle G(x^{\prime})G(y^{\prime})\rangle-\phi(\frac{\mu_{1}}{\sqrt{1+2\sigma_{11}^{2}}})-\phi(\frac{\mu_{2}}{\sqrt{1+2\sigma_{22}^{2}}})-1,\nonumber 
\end{align}
where $x^{\prime},y^{\prime}\sim\N\Bigg[\left(\begin{array}{c}
M_{1}\\
M_{2}
\end{array}\right),\left(\begin{array}{cc}
\Sigma_{11} & \Sigma_{12}\\
\Sigma_{21} & \Sigma_{22}
\end{array}\right)\Bigg]$. We decompose the bivariate Gauss into two uncorrelated parts driven
by a pair of uncorrelated unit variance Gaussian variables $z_{1}$
and $z_{2}$, respectively, and a joint Gaussian variable $z$, writing
them as
\begin{align*}
x^{\prime} & =M_{1}+A\,z_{1}+\sqrt{|\Sigma_{12}|}\,z\\
y^{\prime} & =M_{2}+B\,z_{2}+\mathrm{sgn(\Sigma_{12})\,}\sqrt{|\Sigma_{12}|}\,z.
\end{align*}
This ensures that $\llangle x^{\prime}y^{\prime}\rrangle=\Sigma_{12}$
and $\llangle x^{\prime2}\rrangle=A^{2}+|\Sigma_{12}|$ as well as
$\llangle y^{\prime2}\rrangle=B^{2}+|\Sigma_{12}|$, so
\begin{align*}
A & =\sqrt{\Sigma_{11}-|\Sigma_{12}|},\\
B & =\sqrt{\Sigma_{22}-|\Sigma_{12}|}.
\end{align*}
Then we may write with \eqref{eq:gauss_exp_single_G}
\begin{align*}
\langle G(M_{1}+A\,z_{1}+\sqrt{|\Sigma_{12}|}\,z)\rangle_{z_{1}\sim\N(0,1)} & =G(\frac{M_{1}+\sqrt{|\Sigma_{12}|}\,z}{\sqrt{1+A^{2}}})\\
\langle G(M_{2}+B\,z_{2}+\mathrm{sgn}(\Sigma_{12})\,\sqrt{|\Sigma_{12}|}\,z)\rangle_{z_{2}\sim\N(0,1)} & =G(\frac{M_{2}+\mathrm{sgn}(\Sigma_{12})\,\sqrt{|\Sigma_{12}|}\,z}{\sqrt{1+B^{2}}})
\end{align*}
\begin{align*}
4\,\langle G(x^{\prime})G(y^{\prime})\rangle & =4\,\int g(z)\,G(\frac{M_{1}+\sqrt{|\Sigma_{12}|}\,z}{\sqrt{1+A^{2}}})\,G(\frac{M_{2}+\mathrm{sgn}(\Sigma_{12})\,\sqrt{|\Sigma_{12}|}\,z}{\sqrt{1+B^{2}}})\,dz,
\end{align*}
so that we may use the idea by \citep{vanMeegen21_043077}, which
employs the result of \citep[20,010.4]{Owen80_389}
\begin{align}
4\,\int_{-\infty}^{\infty}G(a+b\,z)\,G(c+d\,z)g(z)\,dz & =2\,G(\frac{a}{\sqrt{1+b^{2}}})+2\,G(\frac{c}{\sqrt{1+d^{2}}})\label{eq:general_Owen-1}\\
 & -4\,T(\frac{a}{\sqrt{1+b^{2}}},\frac{c+cb^{2}-abd}{a\sqrt{1+b^{2}+d^{2}}})\nonumber \\
 & -4\,T(\frac{c}{\sqrt{1+P^{2}}},\frac{a+ad^{2}-bcd}{c\sqrt{1+b^{2}+d^{2}}})\nonumber \\
 & -\begin{cases}
0 & \text{if }ac>0\text{ or if }ac=0\text{ and }a\text{ or }c>0\\
2 & \text{if }ac<0\text{ or if }ac=0\text{ and }a\text{ or }c<0
\end{cases}\nonumber 
\end{align}
where $a=\frac{M_{1}}{\sqrt{1+A^{2}}}$, $c=\frac{N_{2}}{\sqrt{1+B^{2}}}$
, $b=\frac{\sqrt{|\Sigma_{12}|}}{\sqrt{1+A^{2}}}$, $P=\frac{sgn(\Sigma_{12})\,\sqrt{|\Sigma_{12}|}}{\sqrt{1+B^{2}}}$, 
 and 
\begin{align*}
T(h,y) & =\frac{1}{2\pi}\,\int_{0}^{y}dx\,\frac{e^{-\frac{1}{2}h^{2}(1+x^{2})}}{1+x^{2}}\\
 & \stackrel{h=0}{=}\frac{1}{2\pi}\,\arctan\big(y\big).
\end{align*}
The case $N_{1}=0\Rightarrow a=0$ and $c\neq0$ therefore leads to
\begin{align*}
4\,T(\frac{a}{\sqrt{1+b^{2}}},\frac{c+cb^{2}-abd}{a\sqrt{1+b^{2}+d^{2}}}\big) & \stackrel{a\searrow0}{=}4\,\frac{1}{2\pi}\,\arctan\big(y\big)\big|_{y\to\frac{c+cb^{2}}{a\sqrt{1+b^{2}+d^{2}}}}\\
 & =4\,\frac{1}{2\pi}\,\begin{cases}
\frac{\pi}{2} & a\,c>0\\
-\frac{\pi}{2} & a\,c<0
\end{cases},
\end{align*}
so 

\begin{align}
4\,\int_{-\infty}^{\infty}G(b\,z)\,G(c+d\,z)g(z)\,dz & =2\,G(\frac{c}{\sqrt{1+d^{2}}})\label{eq:general_Owen-1-1}\\
 & -4\,T(\frac{c}{\sqrt{1+d^{2}}},\frac{-bd}{\sqrt{1+b^{2}+d^{2}}}).\nonumber 
\end{align}
and likewise for the other term if $M_{2}=0\Rightarrow c=0$ and $M_{1}\neq0$
\begin{align*}
4\,\int_{-\infty}^{\infty}G(a+b\,z)\,G(d\,z)g(z)\,dz & =2\,G(\frac{a}{\sqrt{1+b^{2}}})\\
 & -4\,T(\frac{a}{\sqrt{1+b^{2}}},\frac{-bd}{\sqrt{1+b^{2}+d^{2}}}).
\end{align*}
Finally, if $N_{1}=N_{2}=0\Rightarrow a=c=0$ 
 one has \citep{Segadlo21_arxiv}
\begin{align}
\langle\phi\phi\rangle & =\frac{2}{\pi}\,\arcsin\big(\frac{\Sigma_{12}}{\sqrt{(1+\Sigma_{11})(1+\Sigma_{22}})}\big).\label{eq:phi_phi_exp}
\end{align}
The terms in the first line are $2G(\frac{a}{\sqrt{1+b^{2}}})=\phi(\frac{a/\sqrt{2}}{\sqrt{1+b^{2}}})+1$.

\global\long\def\tY{\tilde{Y}}%

\section{Exercises}

\subsection*{a) Feature learning in linear networks}

\subsubsection*{a.1) Stochastic gradient descent}

In this exercise we want to compare the results from the feature learning
theory with the numerical results obtained with network training.
We hence need to implement a version of network training which corresponds
to the Bayesian approach we have been studying so far. Ultimately
we have a network

\begin{align}
h_{\alpha} & =Vx_{\alpha}\quad V_{ij}\stackrel{\text{i.i.d.}}{\sim}\mathcal{N}(0,g_{v}/d)\\
y_{\alpha} & =w^{\T}h_{\alpha}\quad w_{i}\stackrel{\text{i.i.d.}}{\sim}\mathcal{N}(0,g_{w}/N)\\
z_{\alpha} & =y_{\alpha}+\xi_{\alpha}\quad\xi_{\alpha}\stackrel{\text{i.i.d.}}{\sim}\mathcal{N}(0,\kappa)\\
L & =\frac{1}{2}\sum_{\alpha}(y_{\alpha}-z_{\alpha})^{2}
\end{align}
with $x_{\alpha}\in\mathbb{R}^{d},h_{\alpha}\in\mathbb{R}^{N},y_{\alpha}\in\mathbb{R},V\in\mathbb{R}^{N\times d},W\in\mathbb{R}^{1\times N}$
and squared error loss terms and the training data set $\mathcal{P}=(x_{\alpha},y_{\alpha})$,
$\alpha=1,\ldots,P$. We want to sample from the distribution

\begin{equation}
p(W,V\vert\mathcal{P})\propto\exp\left(-L(\mathcal{P},W,V)-\frac{N}{2g_{w}}\|w\|^{2}-\frac{d}{2g_{V}}\tr V^{\T}V\right)\quad.
\end{equation}
As we saw during the lecture we can sample from this distribution
by Langevin gradient descent, a specific implementation of stochastic
gradient descent. First consider the simplified example with $w(t)\in\bR$

\begin{equation}
dw(t)=-\gamma w(t)dt+dB(t)
\end{equation}
with the Gaussian noise $\langle dB(t)dB(t^{\prime})\rangle=P\delta_{ij}\delta\left(t-t^{\prime}\right)dt$.
Calculate the equilibrium distribution of $w$. Calculate the time
discrete version of the equation using the Ito convention: see \eqref{eq:SDE_Ito}
in the script. Similarly we want to use the following stochastic
differential equation to sample from $p(W,V\vert\mathcal{P})$ via

\begin{align}
d\Theta & =(-\gamma\Theta-\nabla_{\Theta}L)\,dt+dB\quad,\quad\Theta=\{W,V\}\\
\langle dB_{i}(t)dB_{j}(s)\rangle & =P\delta_{ij}\delta(t-s)dt\quad.
\end{align}
How do we need to choose $\gamma$ and $P$ in our case above? Choose
$P$ in such a way that we use the same noise distribution for $W,V$.
. Use your results and fill the gaps denoted with TODO in the corresponding
Python program.

\subsubsection*{a.2) Theoretical results}

We also want to compare the numerics to the theory. In particular
we want to compare the length of the trained readout weights $Q=\|w\|^{2}$
to the value predicted by the feature learning theory for $Q$. For
this we need to solve 

\begin{equation}
0=\frac{\partial S}{\partial Q}=\frac{1}{2}z^{\T}\big(QC^{(xx)}+\kappa\I\big)^{-1}C^{(xx)}\,\big(QC^{(xx)}+\kappa\I\big)^{-1}z-\tr\,C^{(xx)}\big(QC^{(xx)}+\kappa\I\big)^{-1}-\frac{N}{2}\big(\frac{1}{g_{w}}-\frac{1}{Q}\big)
\end{equation}
for $Q$. We find the root of this equation using the so called bisection
method. Follow the instructions in the program in the method ``return\_Q\_Sompolinsky\_biscection''.

\subsubsection*{a.3) Comparison of numerics and theory}

We want you to plot the theoretical value of $Q$ and numerical values
for $\|w\|^{2}$. In order to do this we need to make sure that the
stochastic differential equations above actually sample from the distribution
$p(W,V\vert\mathcal{P})$. We do this by training the network. We
set a so called burn-in-time (T0) which is the amount of gradient
descent steps we wait, before we start sampling. How would you set
this burn-in-time and how would you see whether the stochastic differential
equations converged to $p(W,V\vert\mathcal{P})$? After the burn-in-time
we sample a certain number (sample\_amount) of weight configurations
from $p(W,V\vert\mathcal{P})$. However, we wait some amount of time
(delta\_T) between the samples. Why do we do this and how would you
set this?\\
\\
Using the program produce a plot where $Q$ is plotted against $N\in[10^{1},\ldots,10^{4}]$
. First for $P=10$ data points: How do the empirical results match
the theory ? Where do you see deviations from the theory and why?
Produce plots for $P=4,10,20$. If there are deviations, do they depend
on $P$? Produce plots for different values of $g_{w}$. Does this
change the match between theory and empirical results? If so why?

\subsubsection*{a.3) Notes regarding the program}

The program utilizes JAX in order to be efficient and fast. If you
want to debug and e.g. check the numerical values of the parameters
you can turn jax arrays to numpy arrays by numpy.array(...). Some
hints
\begin{itemize}
\item checks parts of the program successively
\item save your theory and data traces and do pretty plots afterwards using
the saved traces
\end{itemize}

\subsubsection*{b) Most likely network output}

Consider is a network with a single hidden layer defined by \prettyref{eq:single_hidden_net},
but with a different scaling (also called ``mean-field scaling'')
in the readout weights $w_{i}\stackrel{\text{i.i.d.}}{\sim}\N(0,g_{w}/N^{2})$.
Follow the derivation in \prettyref{sec:Network-field-theory} to
obtain

\begin{align*}
p(z,y|C^{(xx)}) & =\N(z|y,\kappa)\,\big\langle\,\prod_{\alpha=1}^{P}\delta\,\big[y_{\alpha}-\sum_{i=1}^{N}w_{i}\,\phi(h_{\alpha i})\big]\big\rangle_{w_{i},h_{\alpha i}}\\
 & =\N(z|y,\kappa)\,\int\D\ty\,\big\langle\exp\big(-\ty^{\T}y+W(\ty|C^{(xx)})\big)\big\rangle,
\end{align*}
where the cumulant-generating function is
\begin{align*}
W(\ty|C^{(xx)}) & =N\,\ln\,\Big\langle\exp\big(\frac{1}{2}\sum_{\alpha\beta}\frac{\ty_{\alpha}}{N}\frac{\ty_{\beta}}{N}g_{w}\,\phi(h_{\alpha})\phi(h_{\beta})\big)\Big\rangle_{h_{\alpha}\sim\N(0,C^{(xx)})}.
\end{align*}
Compute $p(z,y|C^{(xx)})$ for large $N$ by introducing the rate
function 
\begin{align}
\Gamma(y) & :=\sup_{\ty}\,\ty^{\T}y-W(\ty|C^{(xx)}).\label{eq:Gamma_y}
\end{align}
Use this approximation to obtain a set of equations for the most likely
network output $y^{\ast}$ given the training labels by considering
$0\stackrel{!}{=}\frac{\partial}{\partial y_{\alpha}}\,\ln\,p(z,y|C^{(xx)})$.
Show that this condition relates the auxiliary fields $\ty$ and the
discrepancies as
\begin{align*}
\ty_{\alpha} & =\frac{\Delta_{\alpha}}{\kappa}\equiv\frac{z_{\alpha}-y_{\alpha}}{\kappa}.
\end{align*}
Obtain a second set of equations for $\ty$ from the supremum condition
in \eqref{eq:Gamma_y}; show that the result is
\begin{align}
0\stackrel{!}{=} & y_{\alpha}-g_{w}\,\sum_{\beta}\big[\phi(h_{\alpha})\phi(h_{\beta})\big]_{h}\frac{\ty_{\beta}}{N},\label{eq:sup_condition_general}
\end{align}
where the measure $[\ldots]_{h}$ is given by

\begin{align}
\big[\ldots\big]_{h}:= & \frac{\int dh\,\ldots\,\exp\big(\frac{1}{2}\frac{g_{w}}{N^{2}}\,\ty^{\T}\phi(h)\phi(h)\ty-\frac{1}{2}h^{\T}[C^{(xx)}]^{-1}h\big)}{\int dh\,\exp\big(\frac{1}{2}\frac{g_{w}}{N^{2}}\,\ty^{\T}\phi(h)\phi(h)\ty-\frac{1}{2}h^{\T}[C^{(xx)}]^{-1}h\big)}.\label{eq:measure_h}
\end{align}
Show that the most likely value of the discrepancy is
\begin{align*}
\Delta_{\alpha} & =\kappa\,\big(\frac{g_{w}}{N}\,\big[\phi(h)\phi(h)^{\T}\big]_{h}+\kappa\I\big)^{-1}\,z.
\end{align*}
Show that for the case of a linear activation function $\phi(h)=h$
one has
\begin{align*}
W(\ty):= & -\frac{N}{2}\,\ln\det\big(\big[[C^{(xx)}]^{-1}-g_{w}\frac{\ty}{N}\frac{\ty^{\T}}{N}\big]\big)-\frac{N}{2}\ln\det(C^{(xx)})
\end{align*}
and derive the expression corresponding to \eqref{eq:sup_condition_general}
from it. Show that for the most likely value of the discrepancies
holds

\begin{align*}
\Big[\,\big\{[g_{w}C^{(xx)}]^{-1}-\frac{\Delta}{\kappa N}\frac{\Delta}{\kappa N}^{\T}\big\}^{-1}+\kappa\,\I\Big]\,\frac{\Delta}{\kappa N} & =z.
\end{align*}
Compare this to the result in NNGP (in standard scaling $w_{i}\sim\N(0,g_{w}/N)$)
$\Delta=\kappa\,(g_{w}C^{(xx)}+\kappa\,\I)^{-1}z$ written as $(g_{w}C^{(xx)}+\kappa\,\I)\frac{\Delta}{\kappa}=z$:
which shows that the rank-one matrix $\propto\Delta\Delta^{\T}$ modifies
the inverse of the kernel $g_{w}C^{(xx)}$.

\subsubsection*{c) Product of two random matrices}

In the feature learning approach, we have replaced a product of a
random matrix and a random vector by a random Gaussian vector, see
equation \eqref{eq:replaced_by_Gauss}. We here want to check the
validity of this approximation. Consider the product of two random
matrices, which may resemble the connections between adjacent layers
\begin{align}
Y_{ij} & :=\sum_{l}W_{il}V_{lj},\label{eq:replacement_product}\\
\bR^{N\times N}\ni W_{il},V_{lj} & \stackrel{\text{i.i.d.}}{\sim}\N(0,\frac{g}{N}).\nonumber 
\end{align}
Compute the cumulant-generating function of the product \eqref{eq:replacement_product}
and show that to leading order in $N$ the result is
\begin{align*}
Y_{ij} & \stackrel{\text{i.i.d.}}{\sim}\N(0,\frac{g^{2}}{N}).
\end{align*}
You may use that $\ln\det(A)=\tr\ln(A)$ (which can be shown by moving
into the eigenbasis of $A$), and the expansion $\tr\,\ln\big(\I+B)=\tr\,B+\order(B^{2})$.

\subsubsection*{d) Numerics: Predictor statistics and kernels}

We have already seen that we can utilize the theory in Li \& Sompolinksy
for linear networks to obtain the norm of the weight vector after
training. We want to show that we may also obtain results for the
training loss. Check numerically, whether the expressions from the
main text \eqref{eq:avg_training_loss} hold

\begin{align*}
\mathcal{L} & =\frac{1}{2}\mathrm{tr}\left(\langle\Delta_{\alpha}\Delta_{\beta}\rangle\right)\\
\langle\Delta_{\alpha}\Delta_{\beta}\rangle & =\kappa^{2}\left(\left[C^{*}+\kappa\mathbb{I}\right]^{-1}zz^{\top}\left[C^{*}+\kappa\mathbb{I}\right]^{-1}+\frac{\mathbb{I}}{\kappa}-\left[C^{*}+\kappa\mathbb{I}\right]^{-1}\right)
\end{align*}
for a linear network for the form

\begin{align}
h_{\alpha} & =Vx_{\alpha}\quad V_{ij}\stackrel{\text{i.i.d.}}{\sim}\mathcal{N}(0,g_{v}/d)\\
y_{\alpha} & =w^{\T}h_{\alpha}\quad w_{i}\stackrel{\text{i.i.d.}}{\sim}\mathcal{N}(0,g_{w}/N)\\
z_{\alpha} & =y_{\alpha}+\xi_{\alpha}\quad\xi_{\alpha}\stackrel{\text{i.i.d.}}{\sim}\mathcal{N}(0,\kappa)\\
L & =\frac{1}{2}\sum_{\alpha}(y_{\alpha}-z_{\alpha})^{2}
\end{align}
with $C^{*}=QC^{xx}$. Plot the training loss in the feature learning
approximation and the NNGP approximation as a function of $g_{w}$
and compare to numerical results. Choose settings with $P=10$. First
set $g_{w}=1,\kappa=0.01$ and plot the training loss as a function
of the amount of training data $N=10^{1},\ldots,10^{4}$. Are your
results consistent with your expectations? Now keep $N=200$,$P=10$,
$\kappa=0.01$ and scan the loss over the values $g_{w}=0.01,\ldots,0.4$.
Compare the numerical results to the NGGP and the feature learning
results. What is a lower bound for $\mathcal{L}$ (aka. the lowest
value of $\mathcal{L}$ that you would expect to see?)\\
\\
We now want to check, whether the network kernel is in fact only scaled
by a factor $Q$, as predicted by the theory of Li and Sompolinksy.
The program already records the empirical kernel as ``C\_phi\_phi''.
Use a scatter plot of the kernel elements of the theoretically predicted
kernel and the empirical kernel to determine, whether the theory predicts
the correct kernel. Hint: Use the autocorrelation of the loss (which
can be computed by the new method 'compute\_acf') to determine, whether
your sampling distance is appropriate.

\subsubsection*{d) Legendre transform of a Gaussian}

\selectlanguage{english}%
Consider the cumulant generating function of a Gaussian

\begin{align}
W(j) & =j^{\T}\kappa_{1}+\frac{1}{2}j^{\T}\kappa_{2}j\quad j\in\mathbb{R}^{d}\\
j^{\top}\kappa_{1} & :=\sum_{\alpha=1}^{d}j_{\alpha}\kappa_{1\alpha}\\
j^{\T}\kappa_{2}j & :=\sum_{\alpha,\beta=1}^{d}j_{\alpha}\kappa_{2\alpha\beta}j_{\beta}
\end{align}
with a symmetric, invertible, positive definite covariance matrix
$\kappa_{2}\in\mathbb{R}^{d\times d}$ and the mean $\kappa_{1}\in\mathbb{R}^{d}$.
Show that the effective action

\begin{equation}
\Gamma(x)=\sup_{j}\left\{ j^{\T}x-W(j)\right\} 
\end{equation}
is also quadratic. Next show that in general the second derivative
of the cumulant generating function and the second derivative of the
effective action are inverse to each other.

\[
\I=W^{(2)}\Gamma^{(2)}.
\]
To this end, start with $\delta_{ki}=\frac{\partial j_{i}}{\partial j_{k}}$
and use the equation of state $j_{i}=\frac{\partial\Gamma}{\partial x_{i}}$.
What is the property of the Legendre transform that you are exploiting
to achieve this? Does this relation hold for general $W,\Gamma$ ?\selectlanguage{american}%

\chapter{Nomenclature}

We here adapt the nomenclature from the book by Kleinert on path integrals
\citep{Kleinert89}. We denote as $x$ our ordinary random variable
or dynamical variable, depending on the system. Further we use
\begin{itemize}
\item $p(x)$ probability distribution
\item $\langle x^{n}\rangle$ $n$-th moment 
\item $\llangle x^{n}\rrangle$ $n$-th cumulant
\item $x^{\T}y=\sum_{i}x_{i}y_{i}$ scalar product
\item $\N(x|a,b)=\frac{1}{\sqrt{2\pi b}}\,\exp\left(-\frac{1}{2}\frac{(x-a)^{2}}{b}\right)$
Gaussian distribution
\item $S(x)\propto\ln\,p(x)$ action
\item $-\frac{1}{2}x^{\T}Ax$ quadratic action
\item $X\in\bR^{P\times N}$ data matrix
\item $y$ labels
\item $f$ network output
\item $w$ readout weights
\item $W^{(a)}$ network weights of layer $a$
\item $h$ hidden representation, fields, synaptic input
\item $\phi(\circ)$ (non-linear) activation function
\item $C$ kernel, covariance matrix between hidden representations
\item $Z(j)=\langle\exp(j^{\T}x)\rangle$ moment generating function{[}al{]}
(MGF) or partition function
\item $\mathcal{W}(j)=\ln\,Z(j)$ cumulant generating function{[}al{]} (CGF)
or generating function of connected diagrams; (Helmholtz) free energy
\item $\Gamma[x^{\ast}]=\sup_{j}\,j^{\T}x^{\ast}-\mathcal{W}[j]$ generating
function{[}al{]}; Gibbs free energy
\end{itemize}

\end{document}